PAPER

# A unified generalization of inverse regression via adaptive column selection

Yin Jin [1] and Wei Luo [1,*]

[1]Center for Data Science, Zhejiang University, Hangzhou, 310058, Zhejiang, China
[*]Corresponding author. weiluo@zju.edu.cn

## Abstract

A bottleneck of sufficient dimension reduction (SDR) in the modern era is that, among numerous methods, only the sliced inverse regression (SIR) is generally applicable under high-dimensional settings. The higher-order inverse regression methods, which form a major family of SDR methods that are superior to SIR in the population level, suffer from the dimensionality of their intermediate matrix-valued parameters that have an excessive number of columns. In this paper, we propose the generic idea of using a small subset of columns of the matrix-valued parameter for SDR estimation, which breaks the convention of using the ambient matrix for the higher-order inverse regression methods. With the aid of a quick column selection procedure, we then generalize these methods as well as their ensembles under high-dimensional sparse settings, in a uniform manner that resembles sparse SIR without additional assumptions. This is the first promising attempt in the literature to free the higher-order inverse regression methods from their dimensionality, thereby facilitating the application of SDR. The gain of column selection with respect to the efficiency of SDR estimation is also studied under low-dimensional settings. Simulation studies and a real data example are provided at the end.

## 1. Introduction

With the omnipresence of high-dimensional data in supervised learning, sufficient dimension reduction (SDR) has received intensive research interest for decades. Let $X$ be the $p$-dimensional predictor, and let $Y$ be the response. SDR assumes that $X$ affects $Y$ through a low-dimensional linear combination $\beta_0^\top X$; that is,

$$Y \perp\!\!\!\perp X \mid \beta_0^\top X, \tag{1}$$

where $\perp\!\!\!\perp$ denotes independence. Because no parametric assumptions are imposed on $Y|\beta_0^{\mathsf{T}}X$, SDR permits flexible and reliable subsequent modeling of $Y$ using the reduced predictor $\beta_0^{\mathsf{T}}X$. As (1) is valid under any invertible column transformation of $\beta_0$, it characterizes the column space of $\beta_0$. Accordingly, under general conditions on $X$ [42], [9] introduced the central subspace $\mathcal{S}_{Y|X}$ as the parameter of interest for SDR, which is the uniquely smallest subspace of $\mathbb{R}^p$ that satisfies (1). Hereafter, we abuse $\beta_0$ to denote an arbitrary basis matrix of $\mathcal{S}_{Y|X}$, and use $d$ to denote the dimension of $\mathcal{S}_{Y|X}$.

To estimate $\mathcal{S}_{Y|X}$, a major family of existing SDR methods, known as the inverse regression methods, commonly use the moments of $X|Y$ to construct a matrix-valued parameter M called the candidate matrix. Ideally, M satisfies the coverage condition

$$\mathcal{S}(\mathrm{M}) = \mathcal{S}_{Y|X}, \tag{2}$$

where $\mathcal{S}(\mathrm{M})$ denotes the column space of M, so that the leading left singular vectors of a matrix estimator $\widehat{\mathrm{M}}$ span an estimator of $\mathcal{S}_{Y|X}$. For example, suppose $X$ has zero mean and an invertible covariance matrix $\Sigma$, and suppose $Y$ is discrete with support $\{1, \ldots, H\}$; the sliced inverse regression [SIR; 22] constructs

$$\mathrm{M}_{\mathrm{SIR}} = \Sigma^{-1}(\mu_1, \ldots, \mu_H), \tag{3}$$

where $\mu_h$ denotes $E(X|Y = h)$ for $h = 1, \ldots, H$. When $Y$ is continuous, the slicing technique [22] is commonly employed to replace $Y$ with a discrete intermediate response $Y_D$, and (3) still applies. Using the sample moments, a consistent $\widehat{\mathrm{M}}_{\mathrm{SIR}}$ is easily derived when $p$ is fixed as the sample size $n$ grows, and, assuming the sparsity of $\mathcal{S}_{Y|X}$, it is also thoroughly studied under high-dimensional settings [24, 33, 31, 47]. The reduced predictor from SIR is uniquely informative to $Y$, that is, $\mathcal{S}(\mathrm{M}_{\mathrm{SIR}}) \subseteq \mathcal{S}_{Y|X}$, under

(A1) the linearity condition: $E(X \mid \beta_0^{\mathsf{T}}X) = \Sigma\beta_0(\beta_0^{\mathsf{T}}\Sigma\beta_0)^{-1}\beta_0^{\mathsf{T}}X$,

which holds approximately for general $X$ when $p$ is large [12, 16]. However, the full recovery of $\mathcal{S}_{Y|X}$ from SIR, i.e. the coverage condition (2), additionally requires that the effect of $\beta_0^{\mathsf{T}}X$ on $Y$ is asymmetric, and that $Y$ has more than $d$ possible outcomes if discrete. These subtle assumptions cause the risk of insufficient dimension reduction, which is the major limitation that SIR is often criticized for in practice. From (3), this is the intrinsic price of simply using the first-order moment $E(X|Y)$ to characterize (1).

As a remedy of SIR, the other inverse regression methods use the higher-order moments of $X|Y$ and thus are uniformly called the higher-order methods. These include the sliced average variance estimation [11], directional regression [20], and the inverse third moment method [41], etc., whose candidate matrices are respectively

$$\mathrm{M}_{\mathrm{SAVE}} = \Sigma^{-1}(\Sigma - \Sigma_1, \ldots, \Sigma - \Sigma_H),\ \mathrm{M}_{\mathrm{DR}} = \{\mathrm{M}_{\mathrm{SAVE}} + \Sigma^{-1}(\mu_1\mu_1^{\mathsf{T}}, \ldots, \mu_H\mu_H^{\mathsf{T}}), \mathrm{M}_{\mathrm{SIR}}\},$$
$$\mathrm{M}_{\mathrm{TM}} = [\Sigma^{-1}\{E(XX^{\mathsf{T}} \otimes X^{\mathsf{T}}|Y = 1), \ldots, E(XX^{\mathsf{T}} \otimes X^{\mathsf{T}}|Y = H)\}, \mathrm{M}_{\mathrm{SAVE}}, \mathrm{M}_{\mathrm{SIR}}]. \tag{4}$$

Here, $\otimes$ denotes the Kronecker product, $\Sigma_h$ denotes $\mathrm{Var}(X|Y = h)$, and $Y$ is replaced by $Y_D$ again if it is continuous. For $\mathcal{S}(\mathrm{M}) \subseteq \mathcal{S}_{Y|X}$, these methods require both (A1) and

(A2) the constant variance condition: $\mathrm{Var}(X \mid \beta_0^{\mathsf{T}}X) = \Sigma - \Sigma\beta_0(\beta_0^{\mathsf{T}}\Sigma\beta_0)^{-1}\beta_0^{\mathsf{T}}\Sigma$,

and the inverse third moment method additionally requires

(A3) the symmetry condition: $E(XX^{\mathsf{T}} \otimes X^{\mathsf{T}} \mid \beta_0^{\mathsf{T}}X) = \mathbf{0}$,

where $\mathbf{0}$ denotes the origin of the appropriate real space. Similarly to (A1), both (A2) and (A3) are considered nonrestrictive in practice [12, 16, 26]. Because the higher-order methods explore more information of data, they fully recover $\mathcal{S}_{Y|X}$ under fairly general conditions given (A1)-(A3); that is, they can detect symmetric and other subtle effects of $\beta_0^\top X$ on $Y$, and they allow $Y$ to be binary. In this regard, they are more powerful SDR tools than SIR in the population level, and should be used in place of or at least in conjunction with SIR in practice [40].

With the common recognition of their importance, however, the higher-order inverse regression methods are also well known to be more sensitive to the dimensionality of data than SIR. Referring to (4), this is due to their large number of parameters revealed by the number of columns of M, e.g. $pH$ for $\mathrm{M}_{\mathrm{SAVE}}$ and $p^2H$ for $\mathrm{M}_{\mathrm{TM}}$. Consequently, these methods have not been systematically generalized like SIR under high-dimensional sparse settings: except for a forward selection approach that subtly avoids high-dimensional estimation [45], only [31] proposed a sparse sliced average variance estimation under the near sparsity assumptions on both $\Sigma$ and $\Sigma_h$'s, and [36] proposed a sparse principal Hessian direction [21] that is hardly extendable for the other methods. Because the inverse regression methods are the only SDR methods that avoid localization [39, 38], the failure of higher-order methods makes SIR essentially the only choice for conducting SDR under high-dimensional settings. Referring to the aforementioned limitations of SIR, this jeopardizes the applicability of SDR in the modern era of data analysis. Thus, generalizing the higher-order inverse regression methods under high-dimensional settings is an urgent problem, which, as we are aware of, remains unsolved.

At first glance, the issue of dimensionality for the higher-order inverse regression methods is natural and logically inevitable: the estimation errors for the parameters in M accumulate and harm the consistency of $\widehat{\mathrm{M}}$, and subsequently must harm the consistency of the leading left singular vectors of $\widehat{\mathrm{M}}$ in estimating $\mathcal{S}_{Y|X}$. This also complies with the general truth that a more complex statistical model requires a larger sample size. To break this routine, the key observation is that the parameter of interest in SDR is always $\mathcal{S}_{Y|X}$ rather than the candidate matrix M. Therefore, we should directly target at the estimation consistency of $\mathcal{S}(\mathrm{M})$ instead of M, and, other than the universal practice of conducting singular value decomposition on the ambient $\widehat{\mathrm{M}}$, we should be open-minded about how to use $\widehat{\mathrm{M}}$ to estimate $\mathcal{S}(\mathrm{M})$. To this end, note that the dimension of $\mathcal{S}(\mathrm{M})$ is always small under the SDR assumption (1) and the coverage condition (2), regardless of the number of columns of M. Thus, $\mathcal{S}(\mathrm{M})$ must be equivalently spanned by a small subset of columns of M, denoted by $\mathrm{M}_{\mathrm{R}}$; that is,

$$\mathcal{S}_{Y|X} = \mathcal{S}(\mathrm{M}) = \mathcal{S}(\mathrm{M}_{\mathrm{R}}). \tag{5}$$

Subject to the true selection of $\mathrm{M}_{\mathrm{R}}$, one can safely remove the rest of M and regard $\mathrm{M}_{\mathrm{R}}$ as the new candidate matrix to estimate $\mathcal{S}_{Y|X}$. From (3) and (4), $\mathrm{M}_{\mathrm{R}}$ is similar to $\mathrm{M}_{\mathrm{SIR}}$ in both the functional form and the number of parameters, so its estimation is much eased compared with the ambient M and can resemble those for $\mathrm{M}_{\mathrm{SIR}}$ with comparable accuracy. The advantages of the higher-order methods over SIR are still preserved, as they lie in the generality of the coverage condition (2), i.e., the first equation in (5).

In this paper, we develop the generic idea above to uniformly generalize the higher-order inverse regression methods as well as their free ensembles with SIR, e.g., $(\mathrm{M}_{\mathrm{SIR}}, \mathrm{M}_{\mathrm{SAVE}})$ and $(\mathrm{M}_{\mathrm{DR}}, \mathrm{M}_{\mathrm{TM}})$, under high-dimensional sparse settings. The proposed approach has the same level of robustness against dimensionality as the existing sparse SIR-based methods without additional assumptions on data, and, with the aid of a fast and data-driven column selection procedure that ensures the full recovery of $\mathcal{S}(\mathrm{M})$, it preserves the population-level advantages of the original higher-order methods. Under the conventional low-dimensional settings where $p$ is fixed as the sample size $n$ grows, we turn to study the column selection strategy towards the efficiency of SDR estimation, rooted in the fact that removing nonzero columns of M does not always cause information loss but sometimes sharpens the SDR estimation. In summary, the idea of column selection can facilitate SDR in multiple ways, subject to appropriate selection procedures adapted to specific research interests.

The rest of the article is organized as follows. In Section 2, we briefly review the existing sparse SIR-based methods, which will be useful to the high-dimensional study later. The idea of column selection is elaborated in Section 3, where we also discuss how it enhances the efficiency of SDR estimation under low-dimensional settings, and we propose a corresponding column selection procedure. Starting from Section 4, we focus on high-dimensional sparse settings, and introduce a different column selection procedure under the concern of computational feasibility. Some preliminary asymptotic theory is also built in Section 4, and the consistency of the SDR estimation is rigorized in Section 5. The details of implementation are given in Section 6. Section 7 includes the simulation studies and Section 8 includes a real data application. Some possible extensions of the proposed work are discussed at the end. For continuity of text, all the theoretical proofs, as well as some complementary simulation studies, are deferred to the Appendix. `R` code for implementing the proposed methods can be downloaded at https://github.com/yinjinstat/ACS-method.

To ease the presentation, we assume that $X$ has zero mean, and that $Y$ is univariate and continuous but discretized to $Y_D$ by the aforementioned slicing technique [22]; namely, for a prefixed integer $H$ and each $h \in \{1, \ldots, H\}$, $Y_D = h$ if $Y$ falls between its $100\% \times (h-1)/H$ and $100\% \times h/H$ quantiles. We also regard the dimension $d$ of $\mathcal{S}_{Y|X}$ as fixed if $p$ diverges, and discuss its consistent estimation in Section 6. An omitted piece in this article is the convention of standardizing $X$ to $\Sigma^{-1/2}X$ in SDR, under which the inverse regression methods actually take the leading left singular vectors of $\widehat{\Sigma}^{1/2}\widehat{\mathrm{M}}\widehat{\mathrm{W}}$ left-multipled by $\widehat{\Sigma}^{-1/2}$, $\widehat{\Sigma}$ being the sample covariance matrix of $X$ and $\widehat{\mathrm{W}}$ being some matrix transformation of $\widehat{\Sigma}$ [22, 11]. We briefly adjust for this convention at the end of Section 3.

## 2. A review of high-dimensional sparse SIR-based methods

In the literature of high-dimensional statistical analysis [18], the sparsity assumption commonly refers to the existence of a proper subset $A$ of $\{1, \ldots, p\}$, called the active set, such that the corresponding reduced predictor $X_A$ is sufficient for modeling $Y|X$. Under the SDR assumption (1), this is equivalent to

$$\beta_0^i \neq \mathbf{0} \text{ if and only if } i \in A, \tag{6}$$

where $\beta_0^i$ denotes the $i$th row of $\beta_0$ for $i = 1, \ldots, p$. SDR under the sparsity assumption (6) is called the sparse SDR, and its goal is to consistently estimate $\mathcal{S}_{Y|X}$ and pinpoint the index set of nonzero rows of $\beta_0$, i.e., the active set $A$, simultaneously.

With the aid of the rich literature of variable selection and feature screening, SIR has been generalized towards sparsity in various ways. An early approach is the trace pursuit [45], which selects $A$ by a forward selection procedure using the trace of $\widehat{\mathrm{M}}_{\mathrm{SIR}}\widehat{\mathrm{M}}_{\mathrm{SIR}}^{\mathsf{T}}$ as the measure of goodness of fit. The same strategy was used in [44] but with a more delicate estimator of $\mathrm{M}_{\mathrm{SIR}}$, where $\Sigma^{-1}$ in (3) is estimated by $l_1$-minimization under its own sparsity assumption. With slight overestimation of $A$, [24] proposed a SIR-based screening procedure assuming the sparsity of $\Sigma$. The recent progresses [25, 31, 47] commonly reformulate SIR as a least squares-type estimation and then incorporate an appropriate penalty. Namely, let $\Lambda_{\mathrm{SIR}} = (\mu_1, \ldots, \mu_H)$; by (3), $\mathrm{M}_{\mathrm{SIR}}$ solves

$$\Sigma B = \Lambda_{\mathrm{SIR}} \tag{7}$$

or equivalently minimizes $\mathrm{tr}(B^{\mathsf{T}}\Sigma B)/2 - \mathrm{tr}(B^{\mathsf{T}}\Lambda_{\mathrm{SIR}})$ over $B \in \mathbb{R}^{p\times H}$, $\mathrm{tr}(\cdot)$ being the trace of a square matrix. By using $\widehat{\Sigma}$ and the sample moment estimator $\widehat{\Lambda}_{\mathrm{SIR}}$, and including a penalty term $\psi_\lambda(B)$, (7) induces a penalized $l_2$ loss function $\mathcal{L}_{\text{SIR}} : \mathbb{R}^{p\times H} \to \mathbb{R}$ with

$$\mathcal{L}_{\text{SIR}}(B) = \text{tr}(B^{\mathsf{T}}\widehat{\Sigma}B)/2 - \text{tr}(B^{\mathsf{T}}\widehat{\Lambda}_{\text{SIR}}) + \psi_\lambda(B), \tag{8}$$

whose minimizer is a sparse estimator of $\text{M}_{\text{SIR}}$. As this estimator does not require inverting $\widehat{\Sigma}$, it is applicable when $p > n$ without assuming additional structure on $\Sigma$. For example, the sparse SIR-based method proposed in [25] is consistent as long as $p = o(n^2)$, and those proposed in [31] and [47] allow $p$ to diverge in a nearly exponential order of $n$. A similar formulation was proposed in [33].

As seen in (4), M for the general higher-order inverse regression methods also satisfies $\Sigma\text{M} = \Lambda$, where $\Lambda$ is a polynomial of marginal and conditional moments of $X$. Thus, by replacing $\widehat{\Lambda}_{\text{SIR}}$ with the sample-moment estimator $\widehat{\Lambda}$ in (8), these methods could be naturally generalized towards sparsity in the same way as SIR. However, when $p$ is large, the excessive number of columns of M makes each row of $\widehat{\Lambda}$ a long vector whose componentwise estimation errors accumulate, jeopardizing the capability of the penalty term $\psi_\lambda(B)$ in simultaneously truly detecting the zero rows of M and consistently estimating the nonzero rows of M. Consequently, such generalization would be inconsistent under high-dimensional settings. To address this issue, [31] sharpened the estimation of $\text{M}_{\text{SAVE}}$ by assuming its near sparsity, which, as mentioned in the Introduction, requires restrictive assumptions on $\Sigma$ and $\Sigma_h$'s. We instead replace M with some submatrix $\text{M}_{\text{R}}$ that has a small number of columns; see Sections 4 and 5 later.

## 3. The idea of column selection

Before digging into the details, we introduce more notations that will be used throughout the article. Let $X = (X_1, \ldots, X_p)^{\mathsf{T}}$, and let $I_p$ be the $p$-dimensional identity matrix. For an arbitrary matrix M in $\mathbb{R}^{p\times q}$, let $\tau_{\max}(\text{M})$, $\tau_{\min}(\text{M})$, $\text{M}_j^i$, $\text{M}^i$, $\text{M}_j$, and $\Pi(\text{M})$ be its maximal and minimal singular values, its $(i, j)$th entry, $i$th row, and $j$th column for $i = 1, \ldots, p$ and $j = 1, \ldots, q$, and the projection matrix onto $\mathcal{S}(\text{M})$ under the usual Euclidean norm, respectively. Let $\|\text{M}\|_{\text{FB}} = \{\text{tr}(\text{M}\text{M}^{\mathsf{T}})\}^{1/2}$ be the Frobenius norm of M. When M is a vector, $\|\text{M}\|_{\text{FB}}$ is identical to the $l_2$ norm of M. We use $\mathcal{I}_{\text{M}}$ to denote the index set of columns of M, i.e. $\{1, \ldots, q\}$, and, for any $\text{F} \subset \mathcal{I}_{\text{M}}$, we use $\mathcal{C}(\text{F})$ to denote the cardinality of F and use $\text{M}_{\text{F}}$ to denote the submatrix of M consisting of its columns indexed by F. When M has row-wise sparsity, we denote the index set of nonzero rows of M by $\mathcal{A}(\text{M})$. By definition, whenever M spans $\mathcal{S}_{Y|X}$, $\mathcal{A}(\text{M})$ must coincide with the active set $A$ defined in (6). We use $P(\cdot)$ to denote the probability of events induced by $(X, Y)$.

To start with, let $f(X)$ be a general polynomial of variables of $X$ up to order two. The columns of M for all the inverse regression methods can be uniformly written as

$$\Sigma^{-1}[E\{Xf(X)\} - E\{Xf(X - \mu_h) \mid Y_D = h\}], \tag{9}$$

which again falls in $\mathcal{S}_{Y|X}$ under Conditions (A1)-(A3). For example, (9) becomes a column of $\text{M}_{\text{SIR}}$, $\text{M}_{\text{SAVE}}$, and $\text{M}_{\text{TM}}$ if $f(X)$ is a scalar, $X_i$, and $X_iX_j$, respectively. For the most generality of the proposed work, we extend the existing inverse regression methods to regard M as an arbitrarily fixed pool of vectors of form (9), denoted by $(\text{M}_1, \ldots, \text{M}_q)$, with $q > d$ and the functional form of $f(\cdot)$'s in (9) completely known. When $p$ diverges, M becomes a sequence of fixed pools. This includes all the inverse regression methods and their ensembles, e.g., $(\text{M}_{\text{SIR}}, \text{M}_{\text{SAVE}})$, as special cases. To ease the presentation, we assume the coverage condition (2) for $\mathcal{S}(\text{M})$ without mentioning Conditions (A1)-(A3), under which $\beta_0$ is also a basis matrix of $\mathcal{S}(\text{M})$. Intuitively, a larger M better secures this condition but meanwhile complicates the estimation of $\mathcal{S}(\text{M})$. We leave this trade-off aside in this article, and develop the unified theory for any given M.

As mentioned in the Introduction, the proposed work is motivated by the existence of small subsets of columns of M that fully span $\mathcal{S}(\mathrm{M})$. This is formulated as follows. The proof is straightforward by the definition of the rank of a matrix.

**Lemma 1** *Suppose* M *has reduced column rank* $d$. *There exists at least one proper subset* F *of* $\mathcal{I}_{\mathrm{M}}$ *with* $\mathcal{C}(\mathrm{F}) = d$, *such that* $\mathrm{M}_{\mathrm{F}}$ *spans* $\mathcal{S}(\mathrm{M})$ *or equivalently* $\mathcal{S}_{Y|X}$.

Let $\mathcal{F}$ be the collection of all the index sets F such that $\mathrm{M}_{\mathrm{F}}$ spans $\mathcal{S}_{Y|X}$, which includes $\mathcal{I}_{\mathrm{M}}$ as the trivial case as well as numerous small index sets in general. Since M, $\mathcal{F}$, and $\mathcal{S}_{Y|X}$ are unknown, our target is to select an appropriate index set $\mathrm{F} \subseteq \mathcal{I}_{\mathrm{M}}$ such that, first, F falls in $\mathcal{F}$ for all the large samples, second, $\mathrm{M}_{\mathrm{F}}$ delivers a better estimation of $\mathcal{S}_{Y|X}$ than M does, and, finally, the selection procedure is not computationally too expensive. Depending on the research interests, a better estimation of $\mathcal{S}_{Y|X}$ here can refer to different criteria. Before digging into the high-dimensional studies, we first investigate the conventional asymptotic settings where $p$ is fixed as the sample size $n$ grows, under which the dimensionality of M is not worrisome and the research interest is commonly on the asymptotic efficiency of estimating $\mathcal{S}_{Y|X}$. The efficient estimation of $\mathcal{S}_{Y|X}$ is also studied in [10] for SIR, which essentially optimizes $\widehat{\mathrm{M}}_{\mathrm{SIR}}\mathrm{W}$ over $\mathrm{W} \in \mathbb{R}^{H\times H}$. The column selection strategy restricts W to be diagonal with each diagonal entry being one or zero, and targets at the general M of form (9).

For simplicity, we set the number of slices $H$ to be fixed as $n$ grows throughout the rest of the section, which ensures the $n^{1/2}$-consistency and asymptotic normality of $\widehat{\mathrm{M}}$ constructed by the sample moments [27]. For each $\mathrm{F} \in \mathcal{F}$, let $\widehat{\mathrm{M}}_{\mathrm{F}}$ be the estimator of $\mathrm{M}_{\mathrm{F}}$ induced by $\widehat{\mathrm{M}}$, and let $\widehat{\beta}_{\mathrm{F}}$ be the set of its $d$ leading left singular vectors. To measure the accuracy of $\mathcal{S}(\widehat{\beta}_{\mathrm{F}})$ in estimating $\mathcal{S}_{Y|X}$, we use

$$\mathcal{D}(\mathcal{S}(\widehat{\beta}_{\mathrm{F}}), \mathcal{S}_{Y|X}) \equiv nE\{\|\Pi(\widehat{\beta}_{\mathrm{F}}) - \Pi(\beta_0)\|_{\mathrm{FB}}^2\}, \tag{10}$$

which we also denote by $\varphi_n(\mathrm{F})$ if regarded as a function of F. Under the asymptotic normality of $\widehat{\mathrm{M}}$, $\varphi_n(\mathrm{F})$ converges to some $\varphi_0(\mathrm{F}) > 0$; see Lemma 5 in Appendix A. The optimal F that delivers the efficient $\mathcal{S}(\widehat{\beta}_{\mathrm{F}})$ must minimize $\varphi_0(\cdot)$ among all in $\mathcal{F}$.

Because the columns of $\widehat{\mathrm{M}}$ are generally significantly nonzero, one may suspect that using a proper submatrix of $\widehat{\mathrm{M}}$ must cause information loss to the estimation of $\mathcal{S}_{Y|X}$, making the ambient $\widehat{\mathrm{M}}$ or equivalently $\mathrm{F} = \mathcal{I}_{\mathrm{M}}$ the optimal choice under (10). Interestingly, this is not always true, as some less informative columns of $\widehat{\mathrm{M}}$ can indeed be the "liability" to the estimation of $\mathcal{S}_{Y|X}$ and thus are better removed. The following proposition helps clarify this point.

**Proposition 1** *For each* $i \in \mathcal{I}_{\mathrm{M}}$, *write* $\mathrm{M}_i = \beta_0\Gamma_i$ *and* $\widehat{\mathrm{M}}_i = \mathrm{M}_i + \epsilon_i$, *where* $\|\epsilon_i\|_{\mathrm{FB}} = O_P(n^{-1/2})$. *For each* $\mathrm{F} \in \mathcal{F}$, *which means that* $\sum_{i\in\mathrm{F}} \Gamma_i\Gamma_i^{\mathsf{T}}$ *is invertible, let*

$$\widetilde{\beta}_{\mathrm{F}} = \beta_0 + (\textstyle\sum_{i\in\mathrm{F}} \epsilon_i\Gamma_i^{\mathsf{T}})(\sum_{i\in\mathrm{F}} \Gamma_i\Gamma_i^{\mathsf{T}})^{-1}. \tag{11}$$

*Then* $\mathcal{S}(\widehat{\beta}_{\mathrm{F}})$ *is asymptotically equivalent to* $\mathcal{S}(\widetilde{\beta}_{\mathrm{F}})$, *that is,* $\|\Pi(\widehat{\beta}_{\mathrm{F}}) - \Pi(\widetilde{\beta}_{\mathrm{F}})\|_{\mathrm{FB}} = O_P(n^{-1})$.

By Proposition 1, the SDR estimation error from $\widehat{\mathrm{M}}_{\mathrm{F}}$ is roughly a weighted average of the column-wise estimation errors in $\widehat{\mathrm{M}}_{\mathrm{F}}$. Thus, removing a column in $\widehat{\mathrm{M}}$ will enhance the efficiency of SDR estimation if this column is significantly nonzero (i.e., with $\Gamma_i \neq 0$ in (11)) and its estimation error $\epsilon_i$ is either excessively large or strongly dependent with the other $\epsilon_j$'s. As a toy example, suppose $\beta_0$ is a vector, $\mathrm{M} = (\beta_0, \beta_0)$, and $\epsilon_1$ and $\epsilon_2$ have zero mean and $n\{\mathrm{Var}(\epsilon_1), \mathrm{Var}(\epsilon_2), \mathrm{cov}(\epsilon_1, \epsilon_2)\} \to (\Delta, \upsilon^2\Delta, \kappa\upsilon\Delta)$ for some $\upsilon > 0$, $|\kappa| \leq 1$, and positive definite $\Delta$. Then $\widehat{\mathrm{M}}_1$ alone outperforms $\widehat{\mathrm{M}} = (\widehat{\mathrm{M}}_1, \widehat{\mathrm{M}}_2)$ for estimating $\mathcal{S}(\beta_0)$ as long as $\upsilon^2 + 2\kappa\upsilon > 3$, which occurs if $\upsilon$ is substantially large, that is, $\widehat{\mathrm{M}}_2$ is much noisier than $\widehat{\mathrm{M}}_1$, or if $\upsilon$ is larger than one and $\kappa$ is close to one, that is, $\widehat{\mathrm{M}}_2$ completely copies and also amplifies the estimation error in $\widehat{\mathrm{M}}_1$. To explore these in a realistic setting, we next take M from the sliced average variance estimation, i.e., $\mathrm{M}_{\mathrm{SAVE}}$ in (4), assuming that $\mathcal{S}_{Y|X}$ is one-dimensional. For simplicity, we also replace the individual columns with the submatrices $\mathrm{M}_{[i]} \equiv \Sigma^{-1}\{(\Sigma - \Sigma_1)_i, \dots, (\Sigma - \Sigma_H)_i\}$ indexed by $f(X) = X_i$ in (9) as the units of column selection. A candidate $\widehat{\mathrm{M}}_{\mathrm{F}}$ is $(w_1\widehat{\mathrm{M}}_{[1]}, \dots, w_p\widehat{\mathrm{M}}_{[p]})$ with each $w_i \in \{0, 1\}$, and the corresponding $\widehat{\beta}_{\mathrm{F}}$ is denoted by $\widehat{\beta}_w$ where $w = (w_1, \dots, w_p)^{\mathsf{T}}$.

**Proposition 2** *Suppose* $\mathrm{M}_{\mathrm{SAVE}}$ *is of rank one and* $X|(Y_D = h)$ *has a multivariate normal distribution* $N(\mu_h, \Sigma_h)$ *for* $h = 1, \dots, H$. *We have*

$$\lim_{n\to\infty} \mathcal{D}(\mathcal{S}(\widehat{\beta}_w), \mathcal{S}_{Y|X}) = \frac{2\{w^{\mathsf{T}} \odot (\Sigma\beta_0)^{\mathsf{T}}\}[c_1\{I_p - \Pi(\beta_0)\} + \Phi]\{(\Sigma\beta_0) \odot w\}}{w^{\mathsf{T}}(\Sigma\beta_0 \odot \Sigma\beta_0)(\Sigma\beta_0 \odot \Sigma\beta_0)^{\mathsf{T}} w}, \tag{12}$$

*where* $\odot$ *denotes the Hardmard product of two matrices,* $\Phi$ *is a positive semi-definite constant matrix that is invariant of* $w$, *and* $c_1$ *is a nonnegative scalar that is also invariant of* $w$ *with* $c_1 = 0$ *if and only if* $\Sigma_h$ *is invariant of* $h$.

As a special case of Proposition 2, a proper submatrix of $\widehat{\mathrm{M}}_{\mathrm{SAVE}}$ can deliver more efficient SDR estimation than the ambient $\widehat{\mathrm{M}}_{\mathrm{SAVE}}$ when $\mathcal{S}_{Y|X}$ is sparse. For example, if $\mathcal{S}_{Y|X}$ is spanned by $(1, 0, \dots, 0)^{\mathsf{T}}$ and $\Sigma$ conveys the compound symmetry, then $\widehat{\mathrm{M}}_{[1]}$ is the desired submatrix; see more explanations in Appendix B. By contrast, if $\beta_0$ is nowhere sparse, then $\widehat{\mathrm{M}}_{\mathrm{SAVE}}$ may still be the optimal choice, although the issue of dimensionality of $\widehat{\mathrm{M}}_{\mathrm{SAVE}}$ is again overlooked in this conventional asymptotic consideration.

Based on the observations above, the optimal subset of columns of M that delivers the efficient estimation of $\mathcal{S}_{Y|X}$ can vary from small to large, depending on the underlying model. Denote this set by G, which again minimizes $\varphi_0(\cdot)$ among all in $\mathcal{F}$. If multiple sets in $\mathcal{F}$ minimize $\varphi_0(\cdot)$, then we take G as who has the smallest cardinality and assume its uniqueness therein. Because $\varphi_0(\mathrm{F})$ is meaningful only when $\mathrm{F} \in \mathcal{F}$ and $\mathcal{F}$ is unknown, we need to recover $\mathcal{F}$ from all the nonempty subsets of $\mathcal{I}_{\mathrm{M}}$ before selecting G. This leads to a two-step approach as follows.

Step 1. select $\widehat{\mathcal{F}}$ by screening over $\mathcal{G}$, where $\mathcal{G} = \{\mathrm{F} \subset \mathcal{I}_{\mathrm{M}} : \mathrm{F} \text{ is nonempty}\}$;
Step 2. choose $\widehat{\mathrm{G}} \in \widehat{\mathcal{F}}$ that minimizes an approximation of $\varphi_0(\cdot)$ with minimal cardinality.

By definition, $\mathcal{F}$ can be written as $\{\mathrm{F} \in \mathcal{G} : \mathrm{rank}(\mathrm{M}_{\mathrm{F}}) = \mathrm{rank}(\mathrm{M})\}$. Thus, we conduct Step 1 by determining the rank of $\mathrm{M}_{\mathrm{F}}$ based on $\widehat{\mathrm{M}}_{\mathrm{F}}$ for each $\mathrm{F} \in \mathcal{G}$, and selecting those with the maximal rank of $\mathrm{M}_{\mathrm{F}}$ into $\widehat{\mathcal{F}}$. To this end, we apply the predictor augmentation estimator [28] to each $\widehat{\mathrm{M}}_{\mathrm{F}}$, whose theoretical consistency guarantees $\widehat{\mathcal{F}} = \mathcal{F}$ in probability, and whose computational efficiency induces that of $\widehat{\mathcal{F}}$; see Section 6 for details. To simplify the search, we can also automatically include all the supersets of F into $\widehat{\mathcal{F}}$ once we determine $\mathrm{F} \in \widehat{\mathcal{F}}$, based on the fact that $\mathrm{rank}(\mathrm{M}_{\mathrm{F}}) \le \mathrm{rank}(\mathrm{M}_{\mathrm{F}^*})$ for any $\mathrm{F} \subset \mathrm{F}^*$.

Given $\widehat{\mathcal{F}}$, we approximate $\varphi_0(\mathrm{F})$ for each $\mathrm{F} \in \widehat{\mathcal{F}}$ by bootstrap re-sampling; that is, we generate $n$ naive bootstrap samples, and calculate the variation of the resulting estimates $\mathcal{S}(\widehat{\beta}_{\mathrm{F}}^{(1)}), \dots, \mathcal{S}(\widehat{\beta}_{\mathrm{F}}^{(n)})$ measured by $\sum_{i=1}^n \|\Pi(\widehat{\beta}_{\mathrm{F}}^{(i)}) - \Pi(\widehat{\beta}_{\mathrm{F}})\|_{\mathrm{FB}}^2$ that resembles (10). Taking the cardinality of F into account, the objective function is $g : \widehat{\mathcal{F}} \to \mathbb{R}$ with

$$g(\mathrm{F}) = \textstyle\sum_{i=1}^n \|\Pi(\widehat{\beta}_{\mathrm{F}}^{(i)}) - \Pi(\widehat{\beta}_{\mathrm{F}})\|_{\mathrm{FB}}^2 + n^{-c}\mathcal{C}(\mathrm{F}) \tag{13}$$

where $c \in (0, 0.5)$. We take $\widehat{\mathrm{G}}$ as the minimizer of $g(\cdot)$, and estimate $\mathcal{S}_{Y|X}$ by $\mathcal{S}(\widehat{\beta}_{\widehat{\mathrm{G}}})$ induced from $\widehat{\mathrm{M}}_{\widehat{\mathrm{G}}}$. Using the sliced average variance estimation as an example, the consistency of $\widehat{\mathrm{G}}$ is justified as follows. The same result also holds for the general M of form (9).

**Theorem 1** *Suppose* $\mathrm{M} = \mathrm{M}_{\mathrm{SAVE}}$ *and* $E(\|X\|_{\mathrm{FB}}^{8}) < \infty$*. We have* $\widehat{\mathrm{G}} = \mathrm{G}$ *with probability converging to one, which implies the* $n^{1/2}$*-consistency of* $\mathcal{S}(\widehat{\beta}_{\widehat{\mathrm{G}}})$ *and its asymptotic efficiency among* $\{\mathcal{S}(\widehat{\beta}_{\mathrm{F}}) : \mathrm{F} \in \mathcal{F}\}$*; that is, with probability converging to one,*

$$\lim_{n\to\infty} \mathcal{D}(\mathcal{S}(\widehat{\beta}_{\widehat{\mathrm{G}}}), \mathcal{S}_{Y|X}) = \min\{\lim_{n\to\infty} \mathcal{D}(\mathcal{S}(\widehat{\beta}_{\mathrm{F}}), \mathcal{S}_{Y|X}) : \mathrm{F} \in \mathcal{F}\}.$$

By its nature, we call the proposed approach that derives $\mathcal{S}(\widehat{\beta}_{\widehat{\mathrm{G}}})$ the efficiency-adapted column selection approach, which generalizes the inverse regression methods uniformly towards more efficient SDR estimation. The meaning of being adapted is two-fold: first, the selected column set is adapted to the data; second, the criterion for column selection is adapted to the research interest. When M is pooled from multiple inverse regression methods, the selected column set can either be a mix-up across different methods or consistently come from an individual method. For example, suppose $\mathrm{M} = (\mathrm{M}_{\mathrm{SIR}}, \mathrm{M}_{\mathrm{SAVE}})$, then the selected columns are likely from $\mathrm{M}_{\mathrm{SIR}}$ if $X|Y$ conveys a homoscedastic regression model, and they are likely from $\mathrm{M}_{\mathrm{SAVE}}$ if $X|Y$ has zero mean but varying covariance matrix, and they consist of some of $\mathrm{M}_{\mathrm{SIR}}$ and some of $\mathrm{M}_{\mathrm{SAVE}}$ if both patterns exist in the data along different directions of $X$. In this sense, the proposed approach is also an automatic model selection and refining procedure, where a model refers to a single inverse regression method.

Because the implementation of $\widehat{\mathrm{G}}$ requires calculating $g(\cdot)$ exhaustively over $\widehat{\mathcal{F}}$, it quickly slows down as $p$ grows. A compromise is to use the submatrices of M instead of the individual columns as the units of selection, e.g., in Proposition 2. This may reduce the gain of column selection for SDR estimation, particularly if some slices of $Y$ are not as informative, but it also eases the interpretation by avoiding the "selection" of slices of $Y$. The choice of $c$ in (13) affects the cardinality of the selected column set: similar to Theorem 1, using $c > 0.5$ also delivers efficient SDR estimation but with a possibly larger column set, and using $c < 0$ delivers the smallest set in $\mathcal{F}$ regardless of the efficiency of SDR estimation. In practice, we recommend using $c = 0.2$; see Section 7 later.

To comply with the convention of standardizing $X$ mentioned at the end of the Introduction, for each $\mathrm{F} \in \mathcal{G}$, we can first calculate the leading left singular vectors of $\widehat{\Sigma}^{1/2}\widehat{\mathrm{M}}_{\mathrm{F}}\widehat{\mathrm{W}}^{\mathrm{F}}$ and then left-multiply them by $\widehat{\Sigma}^{-1/2}$, where $\widehat{\mathrm{W}}^{\mathrm{F}}$ denotes the submatrix of $\widehat{\mathrm{W}}$ formed by the rows indexed by F. Same as Theorem 1, the modified methods outperform the original inverse regression methods or their ensembles asymptotically. We omit the details as this variation has no theoretical advantage over simply using $\widehat{\mathrm{M}}_{\mathrm{F}}$. Note that the aforementioned optimal $\widehat{\mathrm{M}}_{\mathrm{SIR}}\mathrm{W}$ proposed in [10] differs from this convention.

## 4. High-dimensional column selection

We now apply the generic idea of column selection to high-dimensional sparse settings, where $p$ diverges with $n$ and $\mathcal{S}_{Y|X}$ is assumed to be sparse. As mentioned in the Introduction as well as at the end of Section 2, to preserve the applicability of higher-order inverse regression methods under these settings, the selected subset of columns of M must have a small cardinality subject to full recovery of $\mathcal{S}_{Y|X}$. This conforms to the aforementioned variation of the efficiency-adapted column selection approach with $c < 0$ in (13). However, as neither a consistent $\widehat{\mathrm{M}}$ is available, nor an exhaustive search over $\mathcal{G}$ is computationally feasible, a new column selection procedure must be employed.

Recall from the discussion at the end of Section 2 that, following the literature of sparse SIR-based methods, e.g., (8), the consistent estimation of $\mathrm{M}_{\mathrm{F}}$ is plausible when F has a small cardinality. Accordingly, we generate a consistent estimate for each of the individual columns of M, denoted by $\widetilde{\mathrm{M}}_i$'s, which together serve as the basic units for column selection. Considering the diverging number of columns of M, the computational cost is the primary concern in this stage, under which we use the lasso penalty in generating $\widetilde{\mathrm{M}}_i$'s and propose a quick forward column selection procedure subsequently. The price is that the optimality of the selected set such as the asymptotic efficiency of SDR estimation is not guaranteed. Given the selected set, denoted by R as in the Introduction, we then construct a refined estimate of $\mathrm{M}_{\mathrm{R}}$ using a more delicate penalty function than the lasso penalty in the generalization of (8). Besides estimating $\mathcal{S}_{Y|X}$, we also use the refined estimate of $\mathrm{M}_{\mathrm{R}}$ to recover the active set $A$ defined in (6), so we denote it by $\widehat{\mathrm{M}}_{\mathrm{R}}^{\mathrm{S}}$ where S means sparsity. The process is outlined as Algorithm 1.

**Algorithm 1** High-dimensional sparse inverse regression

1: For each $i \in \mathcal{I}_{\mathrm{M}}$, calculate the initial estimate $\widetilde{\mathrm{M}}_i$ of $\mathrm{M}_i$ using the lasso penalty.
2: Based on $\widetilde{\mathrm{M}}_i$'s, use a forward column selection procedure to select $\mathrm{R} \in \mathcal{G}$.
3: Calculate the refined estimate $\widehat{\mathrm{M}}_{\mathrm{R}}^{\mathrm{S}}$ and its leading $d$ left singular vectors $\widehat{\beta}_{\mathrm{R}}^{\mathrm{S}}$.
4: Estimate $\mathcal{S}_{Y|X}$ by $\mathcal{S}(\widehat{\beta}_{\mathrm{R}}^{\mathrm{S}})$ and recover $A$ by the index set of the nonzero rows of $\widehat{\beta}_{\mathrm{R}}^{\mathrm{S}}$.

Intuitively, the consistency of $\mathcal{S}(\widehat{\beta}_{\mathrm{R}}^{\mathrm{S}})$ requires both $\mathcal{S}(\mathrm{M}_{\mathrm{R}}) = \mathcal{S}_{Y|X}$ in probability from column selection and the consistency of $\widehat{\mathrm{M}}_{\mathrm{R}}^{\mathrm{S}}$ in estimating $\mathrm{M}_{\mathrm{R}}$; the former hinges on the uniform consistency of the initial column estimates $\widetilde{\mathrm{M}}_i$'s. Because both $\widetilde{\mathrm{M}}_i$'s and $\widehat{\mathrm{M}}_{\mathrm{R}}^{\mathrm{S}}$ are of the penalized least squares type that resembles (8), we first propose a preliminary uniform asymptotic result. The forward column selection procedure is elaborated afterwards, and the consistency of SDR estimation is formalized in Section 5. To avoid an extremely "wide" M whose column-wise estimation errors jointly negate any effective column selection, we require the number of columns of M to diverge in a polynomial order of $p$; that is,

$$q \equiv \mathcal{C}(\mathcal{I}_{\mathrm{M}}) = O(p^k), \tag{14}$$

for some positive integer $k$. This is satisfied for all the existing higher-order inverse regression methods as well as their ensembles, with $k \leq 2$. Let $s$ be the cardinality of the active set $A$, which again represents the row-wise sparsity level of M. We allow $s$ to diverge with $p$; see Condition (A5) below for details.

## 4.1. Basic results for a small fixed F

Suppose F is a small fixed subset of $\mathcal{I}_{\mathrm{M}}$. Because $\mathcal{I}_{\mathrm{M}}$ is a sequence of diverging sets as $p$ grows, the rigorous definition of F is a sequence of index sets $\{\mathrm{F}_p : p = 1, \ldots\}$, where each $\mathrm{F}_p$ is a fixed subset of the corresponding $\mathcal{I}_{\mathrm{M}}$ with uniformly upper bounded cardinality. In particular, F can be associated with a set of $f(X)$'s in (9) that are invariant as $p$ grows, which includes each individual column of M as a special case. Clearly, $\mathrm{M}_{\mathrm{F}}$ inherits the row-wise sparsity of M: let $s_{\mathrm{F}}$ be the number of nonzero rows of $\mathrm{M}_{\mathrm{F}}$, then $s_{\mathrm{F}} \leq s$.

Recall from the discussion at the end of Section 2 that $\Lambda$, defined as $\Sigma\mathrm{M}$, is a polynomial of marginal and conditional moments of $X$; that is, $\Lambda = (\Lambda_1, \ldots, \Lambda_q)$ with

$$\Lambda_i = E\{X f_i(X)\} - E\{X f_i(X - E(X|Y_D = h_i)) \mid Y_D = h_i\}, \tag{15}$$

where $f_i(\cdot)$ and $h_i$ are associated with $\mathrm{M}_i$ in its functional form (9). Let $\Lambda_{\mathrm{F}}$ be the subset of columns of $\Lambda$ indexed by F. Since $\Lambda_{\mathrm{F}}$ is equal to $\Sigma\mathrm{M}_{\mathrm{F}}$ and is comparable with $\Lambda_{\mathrm{SIR}}$ in terms of both the functional form and the dimensionality, we can generalize (8) to construct a sparse estimate of $\mathcal{S}(\mathrm{M}_{\mathrm{F}})$. Namely, let $\widehat{\Lambda}_{\mathrm{F}}$ be an estimate of $\Lambda_{\mathrm{F}}$ that replaces $E(\cdot)$ in (15) with the sample mean; we minimize $\mathcal{L}_{\mathrm{F}} : \mathbb{R}^{p\times\mathcal{C}(\mathrm{F})} \to \mathbb{R}$ with

$$\mathcal{L}_{\mathrm{F}}(B) = \mathrm{tr}(B^{\mathsf{T}}\widehat{\Sigma}B)/2 - \mathrm{tr}(B^{\mathsf{T}}\widehat{\Lambda}_{\mathrm{F}}) + \lambda_n \textstyle\sum_{j=1}^{p} v_j \|B^j\|_{\mathrm{FB}}, \tag{16}$$

where $B^j$ again denotes the $j$th row of $B$, and the penalty is applied to the $l_2$ norms of these rows to ensure the row-wise sparsity of the minimizer [6]. In the sparse SIR-based method proposed in [47], the nuclear norm of $B$ is additionally penalized to regulate the rank of the minimizer, which we omit here for simplicity. A naive choice of $v_j$ is to set it constantly at one, which leads to a group lasso estimator $\widetilde{\mathrm{M}}_{\mathrm{F}}$ that includes the aforementioned initial column estimate $\widetilde{\mathrm{M}}_i$ as a special case. If there is a need for polishing the estimation, we can use $\widetilde{\mathrm{M}}_{\mathrm{F}}$ as the initial value and follow the spirit of [50] to make $v_j$ adaptive to $\|\widetilde{\mathrm{M}}_{\mathrm{F}}^j\|_{\mathrm{FB}}$, that is,

$$v_j = \begin{cases} \|\widetilde{\mathrm{M}}_{\mathrm{F}}^j\|_{\mathrm{FB}}^{-\rho/2} & \text{if } \widetilde{\mathrm{M}}_{\mathrm{F}}^j \neq \mathbf{0} \\ K & \text{if } \widetilde{\mathrm{M}}_{\mathrm{F}}^j = \mathbf{0} \end{cases} \tag{17}$$

where $\rho > 0$ and $K$ is large enough to force the corresponding estimate of $\mathrm{M}_{\mathrm{F}}^j$ to be zero. The adaptive $v_j$ diminishes the estimation bias caused by penalization and relaxes the irrepresentable condition that is typically adopted for the lasso regression [50]. A similar penalty has also been used in [6] and [31]. Denote the resulting minimizer of $\mathcal{L}_{\mathrm{F}}(\cdot)$ by $\widehat{\mathrm{M}}_{\mathrm{F}}^{\mathrm{S}}$, which conforms to $\widehat{\mathrm{M}}_{\mathrm{R}}^{\mathrm{S}}$ mentioned at the beginning of this section and differs from $\widehat{\mathrm{M}}_{\mathrm{F}}$ previously used under low-dimensional settings.

For both choices of $v_j$, $\mathcal{L}_{\mathrm{F}}(\cdot)$ can be readily minimized following the rich literature of penalized least squares estimation. In particular, $\widetilde{\mathrm{M}}_{\mathrm{F}}$ can be implemented by the standard $l_1$ minimization and thus is computationally efficient; see Section 6 for details. Same as for $\mathcal{L}_{\mathrm{SIR}}(\cdot)$ in (8), there is no need to invert $\widehat{\Sigma}$ in minimizing $\mathcal{L}_{\mathrm{F}}(\cdot)$, which means that both $\widetilde{\mathrm{M}}_{\mathrm{F}}$ and $\widehat{\mathrm{M}}_{\mathrm{F}}^{\mathrm{S}}$ are applicable when $p > n$ without additional restrictions on $\Sigma$ or other moments. For the consistency of these estimators, we adopt the following regularity conditions. Hereafter, for any sequences of positive scalars $\{a_n : n = 1, \ldots\}$ and $\{b_n : n = 1, \ldots\}$, we write $a_n = \Omega(b_n)$ if $b_n = O(a_n)$, that is, if $a_n/b_n$ is uniformly bounded below from zero.

(A4) $\max\{E(X_i^2) : i = 1, \ldots, p\} = O(1)$, and, for any $\epsilon > 0$, there exists $C > 0$ such that $\max\{P(|X_i| > \epsilon) : i = 1, \ldots, p\} < \exp(-\epsilon^2/2C)$.

(A5) $H^2/p = O(1)$ and there exists $\zeta > 0$ such that $(Hs + s^2)\log(p)/n = O(n^{-2\zeta})$.

(A6) $\max\{\|\mathrm{M}_i\|_{\mathrm{FB}} : i \in \mathcal{I}_{\mathrm{M}}\} = O(1)$.

(A7) For each small fixed $\mathrm{F} \subset \mathcal{I}_{\mathrm{M}}$, there exists $c_2, c_3, \nu > 0$ such that, with probability less than $c_3 p^{-\nu}/\mathcal{C}(\mathcal{I}_{\mathrm{M}})$, $\max\{\|\widehat{\Lambda}_{\mathrm{F}}^i - \Lambda_{\mathrm{F}}^i\|_{\mathrm{FB}} : i = 1, \ldots, p\} > c_2\{H\log(p)/n\}^{1/2}$.

(S1) For $\zeta$ defined in (A5) and each small fixed $\mathrm{F} \subset \mathcal{I}_{\mathrm{M}}$, there exists $\phi < \zeta$ such that $\min\{\|\mathrm{M}_{\mathrm{F}}^i\|_{\mathrm{FB}} : i \in \mathcal{A}(\mathrm{M}_{\mathrm{F}})\} = \Omega(n^{-\phi})$, where $\mathcal{A}(\mathrm{M}_{\mathrm{F}})$ indexes the nonzero rows of $\mathrm{M}_{\mathrm{F}}$.

Condition (A4) regulates the individual variables of $X$ to be uniformly sub-Gaussian, which is typically assumed in the literature of high-dimensional covariance matrix analysis [37, 3]. Condition (A5) allows $p$ to diverge in a nearly exponential order of $n$, and allows slightly larger $H$ and $s$ than the sparse SIR-based methods [31, 47]; the latter is due to the fixed number of columns of $\mathrm{M}_{\mathrm{F}}$ in contrast to $H$ for $\mathrm{M}_{\mathrm{SIR}}$ that may diverge. For a specific M, Conditions (A6) and (A7) can be replaced with more explicit conditions on both the multicollinearity of $X$ and the stability of $X|Y_D$. This is illustrated below for the sliced average variance estimation and directional regression. Following the literature [24, 31], (A6) can also be implied by restricting $\Sigma$ to have bounded $\tau_{\max}(\Sigma)$ and $\tau_{\min}(\Sigma)$. As mentioned before, we choose not to do so because it requires $\Sigma$ to be nearly diagonal and thus would harm the applicability of SDR.

**Proposition 3** *Suppose Conditions (A4)-(A5) hold. If we have, as p diverges, (i)* $\max\{\tau_{\max}(\Sigma^{-1}\Sigma_h) : h = 1, \ldots, H\} = O(1)$, *(ii)* $\max\{\mu_h^{\top}\Sigma^{-1}\mu_h : h = 1, \ldots, H\} = O(1)$, *(iii) for any* $\epsilon > 0$, $\max\{P(|X_i| > \epsilon|Y_D = h) : i = 1, \ldots, p, h = 1, \ldots, H\} < \exp(-\epsilon^2/2C)$ *for some* $C > 0$, *then both* $\mathrm{M}_{\mathrm{SAVE}}$ *and* $\mathrm{M}_{\mathrm{DR}}$ *satisfy (A6) and (A7) with* $\nu = 1$ *in (A7).*

Condition (S1) is labeled differently as it regulates the signal strength. Namely, it requires a (vanishing) lower bound for both the overall magnitude of $\mathrm{M}_{\mathrm{F}}$ and the contribution of each active variable of $X$ to $\mathcal{S}(\mathrm{M}_{\mathrm{F}})$; see a detailed explanation in Section 5 later. Its necessity is intuitive for the variable selection consistency of $\widehat{\mathrm{M}}_{\mathrm{F}}^{\mathrm{S}}$, as for which the magnitude of each nonzero row of $\mathrm{M}_{\mathrm{F}}$ (indexed by $\mathcal{A}(\mathrm{M}_{\mathrm{F}})$) must dominate the corresponding estimation error in $\widehat{\mathrm{M}}_{\mathrm{F}}^{\mathrm{S}}$. Under Condition (A6), $\min\{\|\mathrm{M}_{\mathrm{F}}^{i}\|_{\mathrm{FB}} : i \in \mathcal{A}(\mathrm{M}_{\mathrm{F}})\}$ must be $O(s_{F}^{-1/2})$. Thus, (S1) implicitly requires $s_{\mathrm{F}} = o(n^{2\zeta})$. Since $s_{\mathrm{F}} \leq s$, this can be implied either by a sparse enough $Y|X$ that delivers a slowly diverging $s$, or by Condition (A5) if $s$ and $\log(p)$ are "jointly" small enough to make $\zeta > 1/6$; the latter is justified in the proof of Theorem 2 below in Appendix B.

**Theorem 2** *Suppose Conditions (A4)-(A7) hold and* F *is a small fixed subset of* $\mathcal{I}_{\mathrm{M}}$*. Let* $\lambda_n$ *in (16) be* $\widetilde{\lambda}_n = c_4\{(H + s_{\mathrm{F}})\log(p)/n\}^{1/2}$ *and* $\widehat{\lambda}_n = c_5 n^{-\rho\phi}\widetilde{\lambda}_n$ *for* $v_j$ *being equal and being specified in (17), respectively, where* $c_4$ *is a sufficiently large positive constant,* $\rho$ *is defined in (17), and* $\phi$ *is defined in (S1). For* $\nu$ *defined in (A7) and sufficiently large positive constants* $c_6$ *and* $C$ *that increase with* $\mathcal{C}(\mathrm{F})$ *but otherwise are invariant of* F*, we have, with probability greater than* $1 - c_6 p^{-\nu}/\mathcal{C}(\mathcal{I}_{\mathrm{M}})$*,*

*(i)* $\|\widetilde{\mathrm{M}}_{\mathrm{F}} - \mathrm{M}_{\mathrm{F}}\|_{\mathrm{FB}} \leq C(H^{1/2}s_{\mathrm{F}}^{1/2} + s_{\mathrm{F}})\{\log(p)/n\}^{1/2} = o(1)$,
*(ii)* $\|\widehat{\mathrm{M}}_{\mathrm{F}}^{\mathrm{S}} - \mathrm{M}_{\mathrm{F}}\|_{\mathrm{FB}} \leq C(H^{1/2}s_{\mathrm{F}}^{1/2} + s_{\mathrm{F}})\{\log(p)/n\}^{1/2} = o(1)$ *and* $\mathcal{A}(\widehat{\mathrm{M}}_{\mathrm{F}}^{\mathrm{S}}) = \mathcal{A}(\mathrm{M}_{\mathrm{F}})$ *under (S1).*

By Theorem 2, for any small fixed F, both $\widetilde{\mathrm{M}}_{\mathrm{F}}$ and $\widehat{\mathrm{M}}_{\mathrm{F}}^{\mathrm{S}}$ consistently estimate $\mathrm{M}_{\mathrm{F}}$, and $\widehat{\mathrm{M}}_{\mathrm{F}}^{\mathrm{S}}$ truly selects $\mathcal{A}(\mathrm{M}_{\mathrm{F}})$. A special case is the consistency of the initial column estimate $\widetilde{\mathrm{M}}_i$ for each $i \in \mathcal{I}_{\mathrm{M}}$. By relaxing $s_{\mathrm{F}}$ to $s$ as discussed at the beginning of the subsection and applying Boole's Inequality, the initial column estimates together satisfy

$$\max\{\|\widetilde{\mathrm{M}}_i - \mathrm{M}_i\|_{\mathrm{FB}} : i \in \mathcal{I}_{\mathrm{M}}\} \leq C(H^{1/2}s^{1/2} + s)\{\log(p)/n\}^{1/2} \tag{18}$$

with probability greater than $1 - c_6 p^{-\nu}$; that is, their estimation errors are uniformly $o_P(1)$ under Condition (A5). Intuitively, (18) is necessary for the consistent estimation of $\mathcal{S}_{Y|X}$ in any form based on an estimate of M: if it fails, then, for a non-negligible probability, there will exist some column of M that is estimated with significant bias; as this column cannot be identified based on data, its bias will inevitably pass to the estimation of $\mathcal{S}_{Y|X}$. In addition, (18) is weaker than any type of convergence of $\widetilde{\mathrm{M}} = (\widetilde{\mathrm{M}}_1, \ldots, \widetilde{\mathrm{M}}_q)$ to M, including $\tau_{\max}(\widetilde{\mathrm{M}} - \mathrm{M}) = o_P(1)$ that is generally necessary for the consistency of using the leading left singular vectors of $\widetilde{\mathrm{M}}$ to estimate $\mathcal{S}_{Y|X}$ [43]. Because (18) plays the central role in the theoretical studies later (see Subsection 4.2 and Section 5 below), its generality in comparison to the consistent estimation of the ambient M reveals the essential superiority of the column selection strategy over the original inverse regression methods.

## 4.2. The forward column selection procedure

Given the initial column estimates $\widetilde{\mathrm{M}}_i$'s, we select the elements of R one after another, using the $l_2$ norm of $\widetilde{\mathrm{M}}_i$'s as a simple measure of their importance. To address the possibility that the dominating columns of M fall into a proper subspace of $\mathcal{S}_{Y|X}$, we adopt an iterative projection strategy such that, after each column is selected, we project the rest of the columns onto its orthogonal complement, and pick the next column as whose remainder has the largest $l_2$ norm. The procedure is summarized in Algorithm 2, where we temporally assume that an upper bound $D$ of $d$ is known in order to set the stopping rule. The selection of $D$ is deferred to Section 6.

Had the true M been known and used in the selection, the orthogonality between the selected columns would immediately imply $\mathcal{S}(\mathrm{M}_{\mathrm{R}}) = \mathcal{S}_{Y|X}$. To achieve this through $\widetilde{\mathrm{M}}_i$'s, any direction in $\mathcal{S}_{Y|X}$ must deliver a strong enough signal in $\mathrm{M}_i$'s to dominate

**Algorithm 2** The forward column selection procedure for Step 2 in Algorithm 1

1: Given $\widetilde{\mathrm{M}}_i$'s that minimize (16) with the lasso penalty and $\widetilde{\lambda}_n$ specified in Theorem 2, select $\mathrm{S}(1) \in \mathcal{I}_{\mathrm{M}}$ as who maximizes $\|\widetilde{\mathrm{M}}_i\|_{\mathrm{FB}}$ for $i \in \mathcal{I}_{\mathrm{M}}$. Let $k = 1$ and $\mathrm{F}(1) = \{\mathrm{S}(1)\}$.
2: Given $k$ and $\mathrm{F}(k)$, let $\mathcal{E}_k = \{\|\widetilde{\mathrm{M}}_i - \Pi(\widetilde{\mathrm{M}}_{\mathrm{F}(k)})\widetilde{\mathrm{M}}_i\|_{\mathrm{FB}} : i \in \mathcal{I}_{\mathrm{M}}, i \notin \mathrm{F}(k)\}$; select $\mathrm{S}(k+1)$ that indexes the largest element in $\mathcal{E}_k$, update $\mathrm{F}(k)$ to $\mathrm{F}(k+1) = \mathrm{F}(k) \cup \{\mathrm{S}(k+1)\}$, and update $k$ to $k+1$.
3: Stop when $\mathcal{E}_k$ only includes zero or $k$ reaches a prefixed $D$. Let $\mathrm{R} = \mathrm{F}(k)$.

the corresponding estimation errors. Therefore, in relation to Condition (A5) and the convergence rate of estimation errors justified in Theorem 2, we adopt

(S2) $\min\{\max\{|\beta^{\mathsf{T}}\mathrm{M}_j| : j \in \mathcal{I}_{\mathrm{M}}\} : \beta \in \mathcal{S}_{Y|X}, \beta^{\mathsf{T}}\beta = 1\} = \Omega(n^{-\xi})$ for some $\xi < \zeta/\min(d, 2)$.

This condition differs from (S1) in that it is built for the exhaustive recovery of $\mathcal{S}_{Y|X}$ rather than detecting the variable-wise signals, and it is for the ambient M rather than a specific submatrix $\mathrm{M}_{\mathrm{F}}$. It is equivalent to the existence of a small submatrix $\mathrm{M}_{\mathrm{R}^*}$ that spans $\mathcal{S}_{Y|X}$ with $\tau_{\min}(\mathrm{M}_{\mathrm{R}^*}) = \Omega(n^{-\xi})$, and therefore a regulation on top of Lemma 1 in Section 3. The generality of (S2) can be seen in the case $\mathrm{M} = \mathrm{M}_{\mathrm{SIR}}$ with $H$ fixed as $n$ grows: if we strengthen (S2) to $\xi = 0$, then it complies with Condition (7) of [25] and Condition (A5) of [31] that require $\tau_{\min}(\mathrm{M}_{\mathrm{SIR}})$ to have a positive lower limit. In this sense, (S2) is a reformulation as well as a relaxation of the commonly adopted assumptions for the higher-order inverse regression methods. For a specific M, it can be implied by regulating the moments of $\beta_0^{\mathsf{T}}X$ on $Y$, details omitted.

**Theorem 3** *Under the coverage condition (2) and Conditions (A4)-(A7) and (S2), for any $k < d$, the maximal element of $\mathcal{E}_k$ defined in Step 2 of Algorithm 2 is $\Omega(n^{-\xi})$ with probability converging to one. Consequently, $\mathcal{S}(\mathrm{M}_{\mathrm{R}})$ derived in Algorithm 2 coincides with $\mathcal{S}_{Y|X}$ with probability converging to one.*

Given the coincidence between $\mathcal{S}(\mathrm{M}_{\mathrm{R}})$ and $\mathcal{S}_{Y|X}$ justified in Theorem 3, Theorem 2 can be applied again to justify the consistency of $\widehat{\mathrm{M}}_{\mathrm{R}}^{\mathrm{S}}$ in estimating $\mathcal{S}_{Y|X}$. The result will be rigorized in Section 5. As raised by a Referee, it is interesting to compare column selection with the aforementioned variable selection and feature screening procedures [45, 13, 48], which can be regarded as row selection of M due to the correspondence between the active set $A$ and the nonzero rows of M. A key difference is that, instead of finding the unique active set $A$, we only need the selected column set R to be one of the numerous small sets in $\mathcal{F}$ that ensure exhaustive recovery of $\mathcal{S}_{Y|X}$, and we allow it to be randomly distributed among these sets. Intuitively, this makes the consistency of column selection more general than that of variable selection or feature screening. More theoretical and numerical comparisons toward this direction are deferred to future.

## 5. High-dimensional sparse inverse regression

We now rigorously build the consistency of $\mathcal{S}(\widehat{\beta}_{\mathrm{R}}^{\mathrm{S}})$ derived in Algorithm 1 under high-dimensional sparse settings. Based on the discussions so far, the only issue we need to address is the potential rambling of R. Accordingly, we replace Condition (S1) with

(S3) $\min\{\max\{|\mathrm{M}_j^i| : j \in \mathcal{I}_{\mathrm{M}}\} : i \in A\} = \Omega(n^{-\phi})$ for some $\phi < \zeta$,

where the same $\phi$ as in (S1) is abused. As (S3) can be rewritten as

$$\min\{\max\{\|\mathrm{M}_{\mathrm{F}}^i\|_{\mathrm{FB}} : \mathrm{F} \subset \mathcal{I}_{\mathrm{M}}, \mathcal{C}(\mathrm{F}) \leq D\} : i \in A\} = \Omega(n^{-\phi}),$$

it relaxes (S1) to that the latter holds for at least one small fixed F. Intuitively, this relaxation is due to the construction of R that makes $\mathrm{M}_{\mathrm{R}}$ among the most informative submatrices of M. Similar to (S1), (S3) regulates the variable-wise signal strength in M. As seen next, given (S2) that ensures appropriate signal strength of $\mathcal{S}_{Y|X}$ in M, (S3) is also a regulation of the generic variable-wise signal strength in $\mathcal{S}_{Y|X}$, regardless of the choice of M.

**Proposition 4** *Under Conditions (A5) and (S2), Condition (S3) holds if there exists $\omega < \zeta - \xi$, where $\zeta$ is defined in (A5) and $\xi$ is defined in (S2), such that*

$$\min\{\|\beta_0^i\|_{\mathrm{FB}} : i \in A\} = \Omega(n^{-\omega}). \tag{19}$$

In the literature, a counterpart of (19) is commonly required for the consistency of the sparse SIR-based methods [25, 31, 47], an analogue of which is also widely adopted in general high-dimensional analysis; see, for example, Condition 3 in [13], Condition A3 in [48], and Condition (9) in [29]. By Proposition 4, (S3) resembles these conditions and therefore does not restrict the applicability of SDR. Since $\min\{\|\beta_0^i\|_{\mathrm{FB}} : i \in A\} = O(s^{-1/2})$ due to the constraint $\|\beta_0\|_{\mathrm{FB}} = d$, (19) additionally requires $s$, the cardinality of the active set $A$, to satisfy $s = o(n^{2\zeta})$. This complies with the discussion about (S1) in Section 4, and its necessity is intuitive as a large number of active variables in $X$ usually means that their effects are mostly weak, which are undetectable from limited data.

Given Conditions (S2) and (S3), we now combine the results in Theorem 2 and Theorem 3 to justify the consistency of $\widehat{\mathrm{M}}_{\mathrm{R}}^{\mathrm{S}}$ in estimating $\mathrm{M}_{\mathrm{R}}$, in recovering $\mathcal{S}_{Y|X}$, and in selecting the active set $A$.

**Theorem 4** *Suppose Conditions (A4)-(A7) and (S2)-(S3) hold, and $\widehat{\lambda}_n$ in Theorem 2 is used in (16). We have, with $\zeta$, $\xi$, and $\phi$ defined in (A5), (S2), and (S3), respectively,*

*(i)* $\|\widehat{\mathrm{M}}_{\mathrm{R}}^{\mathrm{S}} - \mathrm{M}_{\mathrm{R}}\|_{\mathrm{FB}} = O_P[(H^{1/2}s^{1/2} + s)\{\log(p)/n\}^{1/2}] = O_P(n^{-\zeta})$;

*(ii)* $\|\Pi(\widehat{\beta}_{\mathrm{R}}^{\mathrm{S}}) - \Pi(\beta_0)\|_{\mathrm{FB}} = O_P[(H^{1/2}s^{1/2} + s)\{\log(p)/n\}^{1/2}n^{\xi}] = o_P(1)$;

*(iii)* $\mathcal{A}(\widehat{\beta}_{\mathrm{R}}^{\mathrm{S}}) = A$ *with probability converging to one.*

We call the proposed approach that derives $\mathcal{S}(\widehat{\beta}_{\mathrm{R}}^{\mathrm{S}})$ the sparsity-adapted column selection approach, which is a uniform generalization of inverse regression methods towards sparse SDR. For the following two reasons, it is a substantial progress in the SDR research that enables free applications of higher-order inverse regression methods under high-dimensional sparse settings.

First, except for the general moment conditions, our approach does not require additional assumptions on either $X$ or $X|Y$. This owes to column selection that reduces the working number of parameters for the SDR estimation. By contrast, as [31] use the ambient $\mathrm{M}_{\mathrm{SAVE}}$ to estimate $\mathcal{S}_{Y|X}$, they assume the near sparsity of $\mathrm{M}_{\mathrm{SAVE}}$ to reduce the working number of parameters. Accordingly, [31] require the near sparsity of both $\Sigma$ and $\Sigma_h$, which are strong independence assumptions on the variables of $X$ that can be easily violated in practice.

Second, same as the efficiency-adapted column selection approach proposed in Section 3, the sparsity-adapted column selection approach is also an automatic model selection and refining procedure if M is pooled from multiple inverse regression methods. For different choices of M, its implementation is unified as shown in Algorithm 1 and Algorithm 2, and, by Theorem 4, its convergence rate is also invariant as long as the number of columns of M diverges in a polynomial order of $p$ and the signal strength of $\mathcal{S}_{Y|X}$ in M is regulated by Condition (S2). When $H$ diverges, this uniform convergence rate is slightly faster than that of the existing sparse SIR-based methods [31, 47], which is typically $(Hs^{1/2} + s)\{\log(p)/n\}^{1/2}$. Referring to the discussion about Condition (A5) in Section 4, this is not surprising as $\mathrm{M}_{\mathrm{SIR}}$ has $H$ columns and $\mathrm{M}_{\mathrm{R}}$ is designed to have up to $D$ columns, that is, $\mathrm{M}_{\mathrm{R}}$ has a smaller number of parameters asymptotically.

In the column selection stage, we may end up selecting a little more columns than necessary. This will not jeopardize the consistency of the subsequent SDR estimation, but it motivates a column re-selection procedure applied to $\mathrm{M}_{\mathrm{R}}$ that may further polish the estimation. Namely, we can apply the best subset selection in Section 3 to $\mathrm{M}_{\mathrm{R}}$ towards efficient SDR estimation, which potentially reduces R to its proper subset. The criterion $g(\cdot)$ in (13) needs to be adjusted due to the change of convergence rate of the matrix estimation. Because the cardinality of R is at most $D$, the computational cost is acceptable. More theoretical studies are deferred to future.

## 6. Details of implementation

We now address the implementation of the proposed sparsity-adapted column selection approach outlined as Algorithm 1 and Algorithm 2 in Section 4. Recall that both the initial column estimate $\widetilde{\mathrm{M}}_i$ and the refined estimate $\widehat{\mathrm{M}}_{\mathrm{R}}^{\mathrm{S}}$ can be unified as the minimizer of (16), with F being the singleton $\{i\}$ and R, respectively. Thus, the implementation is centred on an effective minimization of (16). In addition, we will discuss how to determine the dimension $d$ of $\mathcal{S}_{Y|X}$ and its upper bound $D$ involved in the stopping rule of Algorithm 2. The implementation of the efficiency-adapted column selection approach, which is designed for low-dimensional settings, is straightforward and therefore omitted.

### 6.1. Minimization algorithm

Motivated by the existing sparse SIR-based methods [25, 31, 47], we minimize (16) by reformulating it into a penalized least squares problem. First, referring to (9), we rewrite $\mathrm{M}_{\mathrm{F}}$ as $\Sigma^{-1}E\{Xf_C(X,Y_D)\}$, where the subscript C means centred. For example, the form of $f_C(X,Y_D)$ for each column of $\mathrm{M}_{\mathrm{SAVE}}$ is

$$X_i - H\delta(Y_D = h)\{X_i - E(X_i|Y_D = h)\},$$

where $\delta(\cdot)$ is the indicator function. Let $\widetilde{f}_C(X,Y_D)$ be the estimator of $f_C(X,Y_D)$ using the sample moments, and let $\mathbb{X} \in \mathbb{R}^{n\times p}$ and $\mathbb{Y} \in \mathbb{R}^{n\times C(\mathrm{F})}$ be the matrices whose rows are the sample copies of $X$ and $\widetilde{f}_C(X,Y_D)$, respectively. Up to an intercept, (16) is equal to

$$\frac{1}{2n}\sum_{i\in\mathrm{F}} \|\mathbb{Y}_i - \mathbb{X}B_i\|_{\mathrm{FB}}^2 + \lambda_n \sum_{j=1}^{p} v_j \|B^j\|_{\mathrm{FB}}, \tag{20}$$

which is a typical penalized $l_2$ loss function whose minimization has been extensively studied in the literature. For example, the `R` package `glmnet` [15] gives a fast and approximate minimizer of (20), and the ADMM algorithm is slower but more precise.

Under the concern of computational cost, we use `glmnet` to calculate the initial column estimates, and, after column selection, we use the ADMM algorithm to re-estimate each selected column and apply the forward column selection procedure again to check and remove any redundant columns. Given the resulting R, we use the ADMM algorithm to calculate the refined estimate $\widehat{\mathrm{M}}_{\mathrm{R}}^{\mathrm{S}}$. For each estimation, we follow the sparse SIR-based methods [31, 47] to tune $\lambda_n$ in (20) by cross validation.

### 6.2. Order determination

The predictor augmentation estimator [28] determines the rank of a general matrix-valued parameter M based on a general $\widehat{\mathrm{M}}$. Its consistency requires either the asymptotic normality of $n^{1/2}(\widehat{\mathrm{M}}-\mathrm{M})$ under low-dimensional settings, or simply a vanishing $\tau_{\max}(\widehat{\mathrm{M}}-\mathrm{M})$ under the additional assumption that $X$ is normally distributed. Based on the former, we apply this estimator to determine the rank of each $\mathrm{M}_{\mathrm{F}}$ in Step 1 of implementing the efficiency-adapted column selection approach, which guarantees $\widehat{\mathcal{F}} = \mathcal{F}$ in probability as discussed above Theorem 1. Based on the latter, we apply this estimator to determine $d$ based on $\widehat{\mathrm{M}}_{\mathrm{R}}^{\mathrm{S}}$, whose consistency is implied by Theorem 4 subject to the normality of $X$ and $D \geq d$. Because $\mathrm{rank}(\mathrm{M}_{\mathrm{R}}) < D$ if and only if $D > d$, we determine $D$ and $d$ iteratively, as elaborated next. Subject to a sufficiently large $D$, other ways to determine $d$ include applying the Bayesian information criterion [49] to the singular values of $\widehat{\mathrm{M}}_{\mathrm{R}}^{\mathrm{S}}$, and, as mentioned below (16), penalizing the nuclear norm of $B$ in (16) to make the estimate of $\mathrm{M}_{\mathrm{R}}$ have rank $d$. Both approaches allow non-normal $X$, details omitted.

Since $d \leq 15$ in most SDR applications, we start with setting $D = 15$, which gives an initial $\widehat{\mathrm{M}}_{\mathrm{R}}^{\mathrm{S}}$. Let $\widehat{A}$ be the index set of nonzero rows of $\widehat{\mathrm{M}}_{\mathrm{R}}^{\mathrm{S}}$, and let $X_{\widehat{A}}$ be the corresponding subvector of $X$. To apply the predictor augmentation estimator, we generate a five-dimensional noise $U$ from the standard multivariate normal distribution and augment $X_{\widehat{A}}$ to $X^{*} = (X_{\widehat{A}}^{\mathsf{T}}, U^{\mathsf{T}})^{\mathsf{T}}$, and then calculate $\widehat{\mathrm{M}}_{\mathrm{R}}^{*}$ for $(Y, X^{*})$ in the same way as the non-sparse $\widehat{\mathrm{M}}_{\mathrm{R}}$, regarding R as known. For $\ell = 1, \ldots, \mathcal{C}(\mathrm{R})$, let $\widehat{\beta}_{(\ell)}$ be the left singular vector of $\widehat{\mathrm{M}}_{\mathrm{R}}^{*}$ associated with its $\ell$th largest singular value $\widehat{\tau}_{\ell}$, and let $\widehat{\beta}_{(\ell),U}$ be the subvector of $\widehat{\beta}_{(\ell)}$ corresponding to $U$. We minimize $\eta : \{0, \ldots, \mathcal{C}(\mathrm{R})\} \to \mathbb{R}$ with

$$\eta(\ell) = \textstyle\sum_{i=0}^{\ell} \|\widehat{\beta}_{(i),U}\|_{\mathrm{FB}}^{2} + \widehat{\tau}_{\ell+1}^{2}/(1 + \sum_{i=1}^{\ell+1} \widehat{\tau}_{i}^{2}), \tag{21}$$

where both $\widehat{\beta}_{(0),U}$ and $\widehat{\tau}_{\mathcal{C}(\mathrm{R})+1}$ are set to be zero. For more stable results, [28] recommended repeating the predictor augmentation process for multiple times and using the average of (21). Denote the minimizer of (21) by $\widehat{d}$, which consistently estimates the rank of $\mathrm{M}_{\mathrm{R}}$. As mentioned above, to ensure $D > d$, we check whether $D > \widehat{d}$: if not, then we will enlarge $D$, update $\widehat{\mathrm{M}}_{\mathrm{R}}^{\mathrm{S}}$ and $\widehat{d}$, and repeat the process until the inequality is satisfied.

## 7. Simulation studies

We now evaluate the effectiveness of the proposed column selection approaches using simulation models. First, we apply the efficiency-adapted column selection approach to the sliced average variance estimation under low-dimensional settings. As mentioned in the Introduction, the sliced average variance estimation here means conducting singular value decomposition on $\widehat{\mathrm{M}}_{\mathrm{SAVE}}$ without involving the standardization of $X$, although our experience suggests that the two versions perform similarly. We then evaluate the sparsity-adapted column selection approach under high-dimensional settings.

### 7.1. Efficiency-adapted column selection under low-dimensional settings

For simplicity, we set $c = 0.2$ in (13) and $H = 5$ in $\mathrm{M}_{\mathrm{SAVE}}$, and follow Proposition 2 to use the $p$ submatrices of $\mathrm{M}_{\mathrm{SAVE}}$ indexed by the variables of $X$ as the units of column selection. The latter lightens the computational burden but again may reduce the gain of column selection. We generate $X$ from $N(\mathbf{0}, \Sigma)$, where $\Sigma$ has the compound symmetry with diagonal elements being one and off-diagonal elements being some scalar $a$. To simulate different degrees of dependence between the variables of $X$, we let $a$ be each of 0.2 and 0.8. The corresponding $\Sigma$ is denoted by $\Sigma_2$ and $\Sigma_8$, respectively. Given $X$, we generate $Y$ from each of the following models, where $\epsilon$ is randomly generated from $N(0, 0.2^2)$.

Model I: $Y = \exp\{X_1^2\} + \epsilon$;
Model II: $Y = \log\{(X_1 + \ldots + X_p)^2 + 5\} + \epsilon$;
Model III: $Y = 0.4X_1^2 + 3|X_2|^{1/2} + \epsilon$;
Model IV: $Y = 3\sin(X_1/4) + 0.4X_2^2 + \epsilon$.

By simple algebra, $\mathcal{S}(\mathrm{M}_{\mathrm{SAVE}})$ coincides with $\mathcal{S}_{Y|X}$ in all the four models, so its dimension is one, one, two, and two, respectively. To address the relative complexity of Models III and IV, we set $n = 200$ for Models I and II, and raise it to 500 for Models III and IV. To evaluate how the gain of column selection changes with the dimensionality, we set $p$ at each of six and ten.

Referring to the discussion below Proposition 2, the optimal choice of column selection for all the large samples in Model I is always the submatrix of $\mathrm{M}_{\mathrm{SAVE}}$ associated with $X_1$; see (4) and (9) for details. Based on an omitted simulation study that evaluates $\varphi_0(\cdot)$ defined below (10) with $\widehat{\mathrm{M}} = \widehat{\mathrm{M}}_{\mathrm{SAVE}}$, the optimal choice of column selection is the ambient $\mathrm{M}_{\mathrm{SAVE}}$ in Model II, the set of the two submatrices of $\mathrm{M}_{\mathrm{SAVE}}$ associated with $X_1$ and $X_2$ in Model IV, and the set of those associated with $X_1$ and $X_2$ for the case of $\Sigma = \Sigma_2$ and that with $X_2$ for the case of $\Sigma = \Sigma_8$ in Model III. Because the sliced average variance estimation is theoretically consistent in estimating $\mathcal{S}_{Y|X}$ for all the four models, we only record how its accuracy changes after applying column selection, that is,

$$\widehat{\mathcal{D}}(\mathcal{S}(\widehat{\beta}_{\widehat{\mathrm{G}}}), \mathcal{S}_{Y|X}) \,/\, \widehat{\mathcal{D}}(\mathcal{S}(\widehat{\beta}), \mathcal{S}_{Y|X}), \tag{22}$$

where $\mathcal{S}(\widehat{\beta})$ and $\mathcal{S}(\widehat{\beta}_{\widehat{\mathrm{G}}})$ are the estimates of $\mathcal{S}_{Y|X}$ based on the ambient $\widehat{\mathrm{M}}_{\mathrm{SAVE}}$ and the selected submatrix $\widehat{\mathrm{M}}_{\widehat{\mathrm{G}}}$, respectively, and $\widehat{\mathcal{D}}(\cdot, \mathcal{S}_{Y|X})$ is a bootstrap estimate of $\mathcal{D}(\cdot, \mathcal{S}_{Y|X})$ defined in (10) assuming $\mathcal{S}_{Y|X}$ known *a priori*. A smaller value of (22) means more gain of column selection. With $\widehat{\mathrm{G}}$ replaced by the optimal column set G, the latter specified in the previous paragraph, an oracle value of (22) is also recorded to serve as the benchmark. To comprehend the selected column set $\widehat{\mathrm{G}}$, we record both its cardinality $\mathcal{C}(\widehat{\mathrm{G}})$ and its approximation to G measured by the true selection rate $\mathcal{C}(\widehat{\mathrm{G}} \cap \mathrm{G})/\mathcal{C}(\mathrm{G})$. Based on 200 independent runs, these terms are summarized in Table 1. For reference, the ambient $\mathrm{M}_{\mathrm{SAVE}}$ has 30 columns when $p = 6$ and 50 columns when $p = 10$.

When $\mathcal{S}_{Y|X}$ is sparse, i.e., in Models I, III, and IV, Table 1 suggests that the proposed column selection approach sharpens the sliced average variance estimation. The gain slightly increases with $p$ when $\Sigma = \Sigma_2$, indicating that the proposed approach also enhances the robustness of SDR against dimensionality in this case. As seen from the summary of the corresponding $\widehat{\mathrm{G}}$, this is due to the consistent selection of the optimal column set G, whose cardinality is invariant of $p$ in these models. When $\Sigma = \Sigma_8$, a similar number of columns are selected, but their "quality" is lowered as revealed by the reduced percentage of true column selection, resulting in compromised SDR estimation compared with the oracle method. A possible reason is that the large correlation between the variables of $X$ in this case causes the similarity of $\mathcal{S}(\widehat{\beta}_{\mathrm{F}})$ between different choices of $\mathrm{F} \in \mathcal{F}$, so a larger $n$ is required for their bootstrap sample variances in (13) to be notably distinct from each other.

When $\mathcal{S}_{Y|X}$ is non-sparse, i.e., in Model II, Table 1 indicates the slight sub-optimality of the proposed column selection approach with respect to the original sliced average variance estimation. From an omitted simulation study, this is due to the instability of the bootstrap sample variance of $\mathcal{S}(\widehat{\beta}_{\mathrm{F}})$ in (13) when F is large, and it is alleviated when a larger $n$ is used.

## 7.2. Sparsity-adapted column selection under high-dimensional settings

We now turn to high-dimensional settings where $n$ is set at 200 and $p$ is raised to each of 200 and 1000, and we apply the proposed sparsity-adapted column selection approach to both the sliced average variance estimation and its ensemble with SIR, that is, with M being $\mathrm{M}_{\mathrm{SAVE}}$ and $(\mathrm{M}_{\mathrm{SIR}}, \mathrm{M}_{\mathrm{SAVE}})$, respectively. The aforementioned sparse SIR-based methods proposed in [31] and [47] and the sparse sliced average variance estimation proposed in [31] are included in the comparison, and, in favor of the former two methods, we assume $d$ to be known in their implementation. For simplicity, we fix $H = 2$ in $\mathrm{M}_{\mathrm{SAVE}}$ and $H = 5$ in $\mathrm{M}_{\mathrm{SIR}}$ for all the sparse SDR methods when $Y$ is continuous. The results for some other choices of $H$, as well as a run time comparison between the sparse SDR methods, can be found in Appendix C.

We still generate $X$ from $N(\mathbf{0}, \Sigma)$, and consider two new structures of $\Sigma$: $\Sigma_a = (.5^{|i-j|})$ that is known as the autoregressive covariance and frequently adopted in the literature of high-dimensional covariance matrix estimation [4, 3], and $\Sigma_b = (\sigma_{ij})$ with $\sigma_{ij} = 0.5^{|i-j|}$ if either $i = j$ or one of $i, j$ falls in the active set $A$ and $\sigma_{ij} = 0.5$ otherwise. Only $\Sigma_a$ satisfies the near sparsity assumption required for the sparse sliced average variance estimation proposed in [31], and $\Sigma_b$ means that $X$ includes numerous highly dependent variables. Given $X$, we generate $Y$ from each of the following models. For any positive integer $r < p$, we use $\beta_{(r)}$ to denote the vector in $\mathbb{R}^p$ with the first $r$ entries being one and the rest being zero, and use $\beta_{(-r)}$ to denote that with the last $r$ entries being one and the rest being zero.

Model V: $Y \sim \text{Bernoulli}[\pi(X)/\{1 + \pi(X)\}]$ where $\pi(X) = \exp\{(\beta_{(5)}^{\mathsf{T}} X)^2 - 4.4\}$;
Model VI: $Y = 0.4(\beta_{(3)}^{\mathsf{T}} X)^2 + 3|\beta_{(-3)}^{\mathsf{T}} X|^{1/2} + \epsilon$;
Model VII: $Y = 0.2 \exp(\beta_{(2)}^{\mathsf{T}} X) + \epsilon$;
Model VIII: $Y = \{2\delta(\beta_{(-2)}^{\mathsf{T}} X > 0) - 1\}\{(\beta_{(2)}^{\mathsf{T}} X)^{1/3} + 0.5\} + \epsilon$.

Since the effect of $X$ on $Y$ is symmetric in Models V and VI and is monotone in Models VII and VIII, $\text{M}_{\text{SIR}}$ spans the entire $\mathcal{S}_{Y|X}$ only in the latter two models and meanwhile conveys a stronger signal than $\text{M}_{\text{SAVE}}$ therein. Accordingly, we apply the sparse SIR-based methods to Models VII and VIII only, where they serve as the benchmark of the comparison. For a sparse SDR estimator $\mathcal{S}(\widetilde{\beta})$ with the selected active set $\widehat{A}$, we measure its consistency in estimating $\mathcal{S}_{Y|X}$ by the (absolute) spectral loss

$$\tau_{\max}\{\Pi(\widetilde{\beta}) - \Pi(\beta_0)\} \tag{23}$$

ranged in $(0, 1)$, and we measure its variable selection consistency by both $\mathcal{C}(\widehat{A} \cap A^{\mathsf{C}})$ and $\mathcal{C}(\widehat{A}^{\mathsf{C}} \cap A)$, i.e., the number of falsely selected inactive variables of $X$ and that of falsely unselected active variables of $X$. The results based on 200 independent runs are summarized in Table 2, where we also record the average number of selected columns in the proposed approach.

From Table 2, the consistency of the sparse sliced average variance estimation proposed in [31] heavily relies on the near sparsity of $\Sigma$: it is overall consistent in the case of $\Sigma_a$ but completely fails in the case of $\Sigma_b$. By contrast, with the aid of column selection, our sparse sliced average variance estimation is consistent regardless of the near sparsity of $\Sigma$, and it can outperform that proposed in [31] in both SDR estimation (Models V and VIII) and variable selection (Models VI and VII) even when $\Sigma$ is nearly sparse. It is also comparable with the better of the sparse SIR-based methods for Model VIII, where the latter serves as the benchmark as mentioned above. These phenomena can be observed for both $p = 200$ and $p = 1000$, suggesting that the proposed column selection approach gives a consistently more reliable generalization of the sliced average variance estimation than [31] under high-dimensional sparse settings.

With M enriched from $\text{M}_{\text{SAVE}}$ to $(\text{M}_{\text{SIR}}, \text{M}_{\text{SAVE}})$, the proposed column selection approach is not adversely affected by the additional column estimation for Models V and VI, where $\text{M}_{\text{SIR}}$ is uninformative, and it is enhanced and becomes equally consistent as the better of the sparse SIR-based methods (i.e., the benchmark) for Models VII and VIII, where $\text{M}_{\text{SIR}}$ is exclusively informative. The same phenomena can be observed for both $p = 200$ and $p = 1000$, suggesting that the proposed approach also has the same level of robustness against dimensionality as the sparse SIR-based methods. These together illustrate the effectiveness of the proposed sparsity-adapted column selection approach for conducting sparse SDR on high-dimensional complex data, where it preserves the theoretical advantages of all the individual inverse regression methods involved in M.

## 8. Real data analysis

We now apply the sparsity-adapted column selection approach to the Iconix dataset [8, 30] available at https://www.ncbi.nlm.nih.gov/geo/query/acc.cgi?acc=GSE24417, which was collected to assess how the gene expression of Sprague-Dawley rats affects the probability that they develop tumor, after short term exposure of non-genotoxic chemicals. The data consist of $n = 216$ male rats who were treated for five days with one of 76 non-genotoxic carcinogens. The response variable $Y$ is a binary indicator of whether a rat is tested positive (73 cases) or negative (143 cases) for hepatic tumor after the treatment, and the predictor $X$ consists of $p = 10,752$ genomic variables generated using the Amersham Codelink Uniset Rat 1 Bioarray. A validation set was also included in the original data, but it is suspiciously heterogeneous from the rest of the data [17, 19, 46] and thus is removed from the current analysis. Following [7], we impute the missing values in the data by the $K$-nearest neighborhood algorithm with $K = 10$. To avoid the impact of extreme outliers of $X$, we apply the arc-tangent transformation to each variable of $X$ and then standardize it to have zero mean and unit variance.

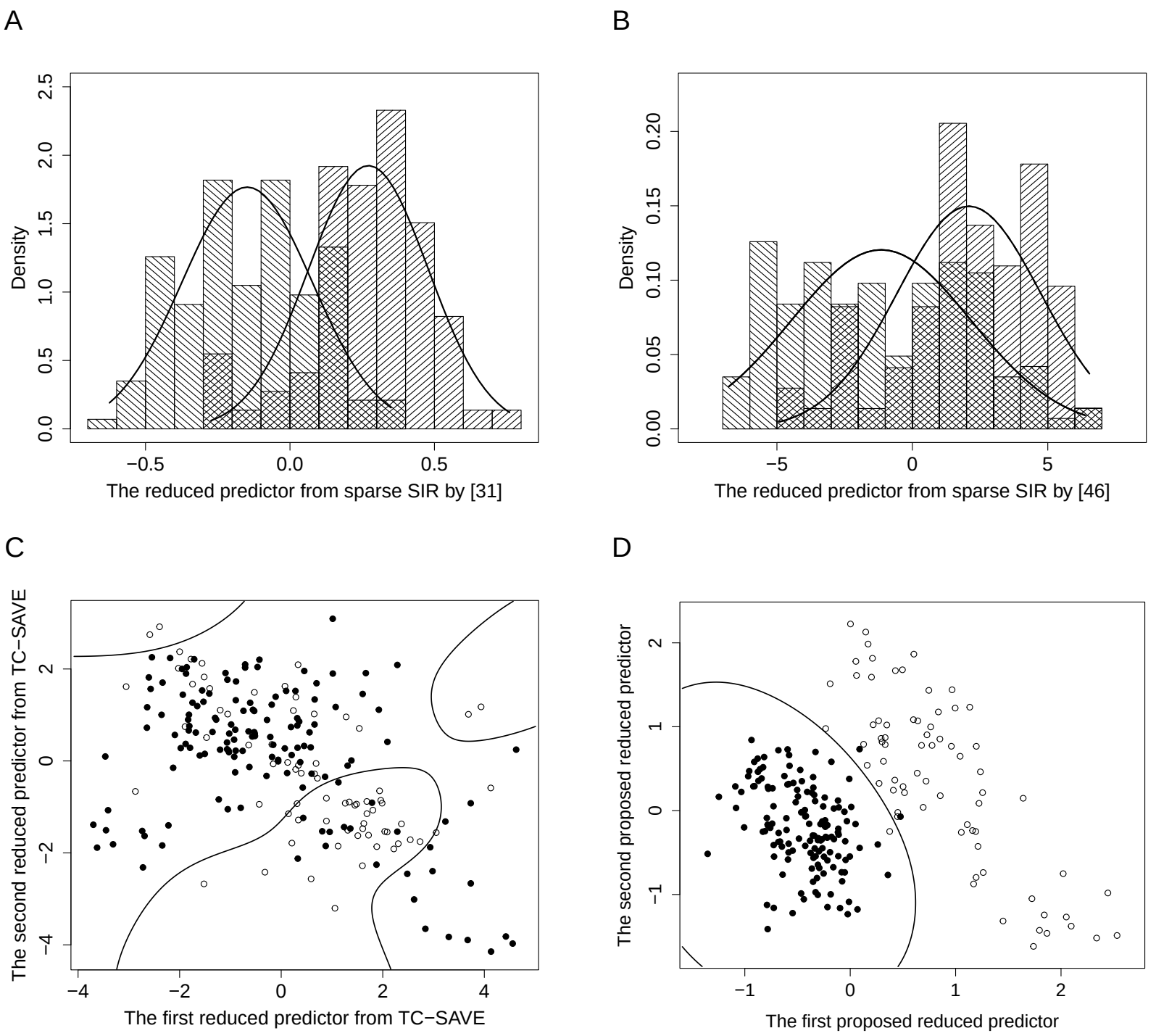

**Fig. 1.** Panel A displays the histograms per class of the univariate reduced predictor derived from the sparse SIR-based method proposed in [31], with the bars filled with "\" for the class of positive cases and those with "/" for the negative cases. Panel B displays the same for the sparse SIR-based method proposed in [47]. Panel C displays the scatter plot of the bivariate reduced predictor from the sparse sliced average variance estimation proposed in [31] (abbreviated as TC-SAVE), with "∘" for the positive cases and "•" for the negative cases. Panel D displays the same for the proposed sparsity-adapted column selection approach applied to directional regression.

To start with, we conduct SDR on the dataset by applying the two sparse SIR-based methods proposed in [31] and [47] as well as the sparse sliced average variance estimation proposed in [31], where $H$ is set at two automatically due to the binary nature of $Y$. The former two methods deliver univariate reduced predictors as $H = 2$, whose histograms specified per class of $Y$ are depicted in panels A and B of Figure 1; the latter delivers a bivariate reduced predictor whose dimension is determined by cross validation, and the corresponding scatter plot is displayed in panel C of Figure 1 where the dot types represent the classes of $Y$. The moderate overlap between the two classes in these plots suggests room for possible improvement of SDR.

Using $\mathrm{M} = \mathrm{M}_{\mathrm{DR}}$ from directional regression, we then apply the proposed sparsity-adapted column selection approach to the dataset. The reduced predictor is binary as determined by the predictor augmentation estimator, whose scatter plot is displayed in panel D of Figure 1 with the dot types again representing the classes. Clearly, the two classes are now distinguished to a much better extent, which suggests the sufficiency of our reduced predictor for the classification analysis. The two classes also differ in both the mean and the covariance matrix of the reduced predictor, where the latter pattern is intrinsically undetectable by SIR. These illustrate the necessity of effectively generalizing the higher-order inverse regression methods for the reliability of SDR under high-dimensional settings, which is achieved by the proposed sparsity-adapted column selection approach.

To confirm that the findings from Figure 1 are not illusions caused by over-fitting, we use the leave-one-out cross validation to calculate the classification error of quadratic discriminant analysis (QDA) after applying each of the SDR methods mentioned above. The appropriateness of QDA is supported by the compactness of the reduced predictors shown in Figure 1. For each training set of size $n - 1$, the reduced predictor from SDR is re-estimated before fitting QDA. The resulting classification error is 81/216 for the sparse SIR-based method proposed in [31], 66/216 for that proposed in [47], 78/216 for the sparse sliced average variance estimation proposed by [31], and sharply lowered to 1/216 for the proposed approach. These comply with the findings from Figure 1, which again suggest the unique sufficiency of the proposed reduced predictor. Among all the 10752 genomic variables, 119 of them are detected to have an impact on the chemical-tumor association, which can be further investigated in the future genetic analysis.

## 9. Discussion

In this paper, we propose the idea of column selection for the candidate matrices of the higher-order inverse regression methods, which breaks the convention in these methods that uses the ambient candidate matrices to estimate $\mathcal{S}_{Y|X}$. With the aid of appropriate column selection procedures, it improves the estimation efficiency of the higher-order inverse regression methods under low-dimensional settings, and, more importantly, it delivers a uniform generalization of these methods that has comparable accuracy with SIR under high-dimensional sparse settings. The latter facilitates the application of SDR in the modern era.

As mentioned in Section 4, under high-dimensional sparse settings, it is interesting to compare column selection with row selection of the candidate matrices, where the latter can be achieved by the variable selection and feature screening technique. In addition, column selection can be applied under high-dimensional non-sparse settings, again to free the higher-order inverse regression methods from the dimensionality of their candidate matrices. Under these settings, the consistency of SIR has been justified when $p = o(n)$ [24] and further relaxed to $p > n$ under a factor model assumption on $X$ [14]. With the aid of column selection, similar results seem promising for the higher-order inverse regression methods. Beyond SDR, column selection may also facilitate the principal component analysis and other statistical analyses where the research interest is to recover the low-dimensional column space of a matrix-valued parameter.

In terms of methodology, the proposed work can be refined or generalized as follows.

First, the forward column selection procedure for high-dimensional settings is proposed under the concern of computational cost. There may exist more delicate column selection procedures that can detect a weaker signal than that regulated in Condition (S2), and/or take the efficiency of SDR estimation into consideration.

Second, as mentioned in the Introduction, the practice of column selection is motivated by the more general spirit that one should play smart when using a matrix estimator $\widehat{\mathrm{M}}$ to estimate $\mathcal{S}_{Y|X}$. Namely, the principle is to use $\widehat{\mathrm{M}}\mathrm{W}$ for a carefully picked weight matrix W, and, while optimizing a free W is infeasible under high-dimensional settings, column selection simplifies the problem by restricting W to be diagonal with its diagonal entries being one or zero. A reasonably larger set of candidate W may deliver more efficient SDR estimation.

Third, the proposed work regards M as a given pool of vectors of form (9). If we instead regard it as open and unknown, column selection will be changed to a data-driven search for an appropriate set of vectors of form (9). This permits more freedom in choosing the polynomial $f(\cdot)$ in (9) and thus room for further improvement.

## References

1. E. Altınışık. On the smallest singular value in the class of unit lower triangular matrices with entries in [- a, a]. *Special Matrices*, 9(1):297–304, 2021.
2. P. J. Bickel and D. A. Freedman. Some asymptotic theory for the bootstrap. *The annals of statistics*, 9(6):1196–1217, 1981.
3. P. J. Bickel and E. Levina. Covariance regularization by thresholding. *The Annals of Statistics*, 36(6):2577–2604, 2008.
4. P. J. Bickel and E. Levina. Regularized estimation of large covariance matrices. *The Annals of Statistics*, 36(1):199–227, 2008.
5. T. T. Cai and A. Zhang. Rate-optimal perturbation bounds for singular subspaces with applications to high-dimensional statistics. 2018.
6. X. Chen, C. Zou, and R. Cook. Coordinate-independent sparse suffcient dimension reduction and variable selection. *The Annals of Statistics*, 6:3696–3723, 2010.
7. C.-C. Chiu, S.-Y. Chan, C.-C. Wang, and W.-S. Wu. Missing value imputation for microarray data: a comprehensive comparison study and a web tool. *BMC systems biology*, 7(6):1–13, 2013.
8. M. Consortium et al. The microarray quality control (maqc)-ii study of common practices for the development and validation of microarray-based predictive models. *Nature biotechnology*, 28(8):827, 2010.
9. R. D. Cook. *Regression Graphics*. Wiley, New York, 1998.
10. R. D. Cook and L. Ni. Sufficient dimension reduction via inverse regression: A minimum discrepancy approach. *Journal of the American Statistical Association*, 100(470):410–428, 2005.
11. R. D. Cook and S. Weisberg. Discussion of “Sliced inverse regression for dimension reduction”. *Journal of the American Statistical Association*, 86:316–342, 1991.
12. P. Diaconis and D. Freedman. Asymptotics of graphical projection pursuit. *The annals of statistics*, pages 793–815, 1984.
13. J. Fan and J. Lv. Sure independence screening for ultrahigh dimensional feature space. *Journal of the Royal Statistical Society: Series B (Statistical Methodology)*, 70(5):849–911, 2008.
14. X. L. Fan, J. and J. Yao. Sufficient forecasting using factor models. *Journal of Econometrics*, 201(2):292–306, 2017.
15. J. Friedman, T. Hastie, and R. Tibshirani. Regularization paths for generalized linear models via coordinate descent. *Journal of statistical software*, 33(1):1, 2010.

16. P. Hall and K.-C. Li. On almost linearity of low dimensional projections from high dimensional data. *The Annals of Statistics*, 47(5):867–889, 1993.
17. S. Ham, H. Qin, and D. Yu. An improved ratio-based (irb) batch effects removal algorithm for cancer data in a co-analysis framework. In *2014 IEEE International Conference on Bioinformatics and Bioengineering*, pages 212–219. IEEE, 2014.
18. T. Hastie, R. Tibshirani, J. H. Friedman, and J. H. Friedman. *The elements of statistical learning: data mining, inference, and prediction*, volume 2. Springer, 2009.
19. H. Huang, H. Qin, S. Yoo, and D. Yu. Physics-based anomaly detection defined on manifold space. *ACM Transactions on Knowledge Discovery from Data (TKDD)*, 9(2):1–39, 2014.
20. B. Li and S. Wang. On directional regression for dimension reduction. *Journal of the American Statistical Association*, 102(479):997–1008, 2007.
21. K. Li. On principal hessian directions for data visualization and dimension reduction: another application of Stein's lemma. *Journal of the American Statistical Association*, 87:1025–1039, 1992.
22. K. C. Li. Sliced inverse regression for dimension reduction (with discussion). *Journal of the American Statistical Association*, 86:316–342, 1991.
23. Y. Li and L.-X. Zhu. Asymptotics for sliced average variance estimation. *The Annals of Statistics*, 35:41–69, 2007.
24. Q. Lin, Z. Zhao, and J. S. Liu. On consistency and sparsity for sliced inverse regression in high dimensions. *The Annals of Statistics*, 46(2):580–610, 2018.
25. Q. Lin, Z. Zhao, and J. S. Liu. Sparse sliced inverse regression via lasso. *Journal of the American Statistical Association*, 114(528):1726–1739, 2019.
26. W. Luo. On the second-order inverse regression methods for a general type of elliptical predictors. *Statistica Sinica*, 28(3):1415–1436, 2018.
27. W. Luo and B. Li. Combining eigenvalues and variation of eigenvectors for order determination. *Biometrika*, 103(4):875–887, 2016.
28. W. Luo and B. Li. On order determination by predictor augmentation. *Biometrika*, 108:557–574, 2021.
29. D. Nandy, F. Chiaromonte, and R. Li. Covariate information number for feature screening in ultrahigh-dimensional supervised problems. *Journal of the American Statistical Association*, pages 1–14, 2021.
30. V. Popovici, W. Chen, B. D. Gallas, C. Hatzis, W. Shi, F. W. Samuelson, Y. Nikolsky, M. Tsyganova, A. Ishkin, T. Nikolskaya, et al. Effect of training-sample size and classification difficulty on the accuracy of genomic predictors. *Breast Cancer Research*, 12(1):1–13, 2010.
31. W. Qian, S. Ding, and R. D. Cook. Sparse minimum discrepancy approach to sufficient dimension reduction with simultaneous variable selection in ultrahigh dimension. *Journal of the American Statistical Association*, 114(527):1277–1290, 2019.
32. G. W. Stewart. Perturbation theory for the singular value decomposition. Technical report, 1998.
33. K. M. Tan, Z. Wang, T. Zhang, H. Liu, and R. D. Cook. A convex formulation for high-dimensional sparse sliced inverse regression. *Biometrika*, 105(4):769–782, 2018.
34. R. Tibshirani, J. Bien, J. Friedman, T. Hastie, N. Simon, J. Taylor, and R. J. Tibshirani. Strong rules for discarding predictors in lasso-type problems. *Journal of the Royal Statistical Society Series B: Statistical Methodology*, 74(2):245–266, 2012.
35. D. E. Tyler. Asymptotic inference for eigenvectors. *The Annals of Statistics*, 9(4):725–736, 1981.
36. C. Wang, B. Jiang, and L. Zhu. Penalized interaction estimation for ultrahigh dimensional quadratic regression. *Statistica Sinica*, 2021.

37. W. Wang and J. Fan. Asymptotics of empirical eigenstructure for high dimensional spiked covariance. *Annals of statistics*, 45(3):1342, 2017.
38. Y. Xia. A constructive approach to the estimation of dimension reduction directions. *The Annals of Statistics*, pages 2654–2690, 2007.
39. Y. Xia, H. Tong, W. Li, and L.-X. Zhu. An adaptive estimation of dimension reduction space. *Journal of the Royal Statistical Society: Series B (Statistical Methodology)*, 64(3):363–410, 2002.
40. Z. Ye and R. E. Weiss. Using the bootstrap to select one of a new class of dimension reduction methods. *Journal of the American Statistical Association*, 98:968–979, 2003.
41. X. Yin and R. D. Cook. Estimating central subspaces via inverse third moments. *Biometrika*, 90(1):113–125, 2003.
42. X. Yin, B. Li, and R. D. Cook. Successive direction extraction for estimating the central subspace in a multiple-index regression. *Journal of Multivariate Analysis*, 99(8):1733–1757, 2008.
43. Y. Yu, T. Wang, and R. J. Samworth. A useful variant of the davis–kahan theorem for statisticians. *Biometrika*, 102:315–323, 2015.
44. Z. Yu, Y. Dong, and J. Shao. On marginal sliced inverse regression for ultrahigh dimensional model-free feature selection. *The Annals of Statistics*, 44(6):2594–2623, 2016.
45. Z. Yu, Y. Dong, and L.-X. Zhu. Trace pursuit: A general framework for model-free variable selection. *Journal of the American Statistical Association*, 111(514):813–821, 2016.
46. J. Zahoor and K. Zafar. Classification of microarray gene expression data using an infiltration tactics optimization (ito) algorithm. *Genes*, 11(7):819, 2020.
47. J. Zeng, Q. Mai, and X. Zhang. Subspace estimation with automatic dimension and variable selection in sufficient dimension reduction. *Journal of the American Statistical Association*, pages 1–13, 2022.
48. T. Zhou, L. Zhu, and R. Li. Model-free forward regression via cumulative divergence. *Journal of the American Statistical Association*, 2019.
49. L. Zhu, B. Miao, and H. Peng. On sliced inverse regression with high-dimensional covariates. *Journal of the American Statistical Association*, 101:630–642, 2006.
50. H. Zou. The adaptive lasso and its oracle properties. *Journal of the American statistical association*, 101(476):1418–1429, 2006.

**Table 1.** Performance of the efficiency-adapted column selection applied to the sliced average variance estimation: columns 2-6 record the case $\Sigma = \Sigma_2$ and columns 7-11 are for $\Sigma = \Sigma_8$; in each case, $a(b)$ in Column "Efficiency" records the estimate $(a)$ and its bootstrap standard error $(b)$ of the relative efficiency of the proposed column selection approach with respect to the original sliced average variance estimation, column "# Col." records the average number of columns selected by the proposed efficiency-adapted approach, and column "TSR" records the ratio between the average number of desired columns actually selected and the number of all the desired columns, all based on 500 independent runs; column "Opt" records the limiting value of the relative efficiency of column selection in the oracle case that $\widehat{\mathrm{G}}$ is replaced with G.

| Model | $\Sigma_2$ | | | | | $\Sigma_8$ | | | | |
|---|---|---|---|---|---|---|---|---|---|---|
| | $p$ | Efficiency | # Col. | TSR | Opt | $p$ | Efficiency | # Col. | TSR | Opt |
| I | 6 | 0.83(0.009) | 5.3 | 1.00 | 0.83 | 6 | 0.92(0.009) | 8.1 | 0.91 | 0.92 |
| | 10 | 0.76(0.008) | 5.5 | 1.00 | 0.76 | 10 | 0.93(0.007) | 9.9 | 0.77 | 0.92 |
| II | 6 | 1.07(0.009) | 25.3 | 0.84 | 1.00 | 6 | 1.10(0.007) | 8.0 | 0.27 | 1.00 |
| | 10 | 1.18(0.009) | 26.7 | 0.53 | 1.00 | 10 | 1.03(0.002) | 8.3 | 0.17 | 1.00 |
| III | 6 | 0.87(0.009) | 10.3 | 1.00 | 0.87 | 6 | 0.92(0.008) | 8.2 | 0.86 | 0.87 |
| | 10 | 0.84(0.008) | 13.1 | 1.00 | 0.82 | 10 | 0.94(0.004) | 10.4 | 0.76 | 0.89 |
| IV | 6 | 0.86(0.011) | 10.3 | 1.00 | 0.84 | 6 | 0.88(0.011) | 10.9 | 0.80 | 0.83 |
| | 10 | 0.83(0.013) | 10.1 | 1.00 | 0.82 | 10 | 0.94(0.011) | 13.8 | 0.79 | 0.84 |

**Table 2.** Performance of the sparsity-adapted column selection approach: $a(b)$ in Column "Eff." records the sample average $(a)$ and the standard error $(b)$ of the spectral loss (23) of the SDR estimator, "ACv." records the average number of falsely unselected active variables, "ICv." records the average number of falsely selected inactive variables, and "Col." records the average number of columns used in the sparsity-adapted column selection approach, all based on 200 independent runs. "TC-SAVE", "SC-SIR", "SEAS-SIR", "SCS-SAVE", and "SCS-ENS" stand for the sparse sliced average variance estimation proposed in [31], the sparse SIR-based method proposed in [31], the sparse SIR-based method proposed in [47], and the proposed sparsity-adapted column selection approach applied to the sliced average variance estimation and to the ensemble of that and SIR, respectively.

| Model | $p$ | Method | $\Sigma_a$ | | | | $\Sigma_b$ | | | |
|---|---|---|---|---|---|---|---|---|---|---|
| | | | Eff. | ACv. | ICv. | Col. | Eff. | ACv. | ICv. | Col. |
| V | 200 | TC-SAVE | 0.684(0.037) | 2.85 | 3.91 | – | 0.999(0.000) | 3.23 | 115.46 | – |
| | | SCS-SAVE | 0.292(0.011) | 0.00 | 0.60 | 4.36 | 0.280(0.011) | 0.00 | 0.58 | 4.78 |
| | | SCS-ENS | 0.290(0.012) | 0.00 | 0.72 | 4.46 | 0.280(0.010) | 0.05 | 0.60 | 5.00 |
| | 1000 | TC-SAVE | 0.697(0.035) | 2.80 | 3.41 | – | 0.996(0.000) | 1.86 | 591.00 | – |
| | | SCS-SAVE | 0.328(0.014) | 0.01 | 2.04 | 9.83 | 0.335(0.015) | 0.04 | 1.70 | 9.26 |
| | | SCS-ENS | 0.327(0.015) | 0.02 | 2.05 | 10.03 | 0.337(0.014) | 0.04 | 1.70 | 9.88 |
| VI | 200 | TC-SAVE | 0.500(0.026) | 0.08 | 2.23 | – | 0.999(0.000) | 1.72 | 151.14 | – |
| | | SCS-SAVE | 0.439(0.014) | 0.04 | 0.64 | 5.73 | 0.465(0.015) | 0.15 | 1.00 | 6.94 |
| | | SCS-ENS | 0.397(0.012) | 0.10 | 2.16 | 6.28 | 0.459(0.014) | 0.12 | 2.01 | 7.12 |
| | 1000 | TC-SAVE | 0.567(0.035) | 1.02 | 73.82 | – | 0.999(0.000) | 0.24 | 930.48 | – |
| | | SCS-SAVE | 0.445(0.017) | 0.24 | 3.42 | 8.08 | 0.483(0.017) | 0.25 | 3.82 | 7.30 |
| | | SCS-ENS | 0.476(0.020) | 0.32 | 3.48 | 8.21 | 0.482(0.017) | 0.25 | 3.47 | 8.60 |
| VII | 200 | SC-SIR | 0.081(0.005) | 0.00 | 4.99 | – | 0.076(0.004) | 0.00 | 5.62 | – |
| | | SEAS-SIR | 0.095(0.004) | 0.00 | 0.80 | – | 0.094(0.005) | 0.00 | 1.24 | – |
| | | TC-SAVE | 0.131(0.008) | 0.00 | 0.70 | – | 0.998(0.000) | 0.33 | 153.05 | – |
| | | SCS-SAVE | 0.118(0.010) | 0.06 | 0.12 | 1.81 | 0.122(0.012) | 0.04 | 0.18 | 1.98 |
| | | SCS-ENS | 0.090(0.005) | 0.00 | 0.01 | 3.95 | 0.089(0.008) | 0.00 | 0.04 | 3.22 |
| | 1000 | SC-SIR | 0.086(0.006) | 0.00 | 8.48 | – | 0.078(0.004) | 0.00 | 11.11 | – |
| | | SEAS-SIR | 0.096(0.008) | 0.00 | 0.91 | – | 0.093(0.008) | 0.00 | 0.81 | – |
| | | TC-SAVE | 0.153(0.009) | 0.00 | 1.78 | – | 0.995(0.000) | 0.06 | 882.92 | – |
| | | SCS-SAVE | 0.151(0.015) | 0.10 | 0.30 | 2.21 | 0.124(0.015) | 0.00 | 0.15 | 2.26 |
| | | SCS-ENS | 0.095(0.018) | 0.00 | 0.03 | 4.82 | 0.096(0.018) | 0.01 | 0.05 | 3.55 |
| VIII | 200 | SC-SIR | 0.429(0.018) | 0.02 | 3.53 | – | 0.377(0.016) | 0.08 | 5.86 | – |
| | | SEAS-SIR | 0.252(0.008) | 0.00 | 3.40 | – | 0.253(0.008) | 0.00 | 3.52 | – |
| | | TC-SAVE | 0.467(0.024) | 0.04 | 0.61 | – | 0.995(0.000) | 0.09 | 174.92 | – |
| | | SCS-SAVE | 0.262(0.008) | 0.02 | 0.71 | 4.16 | 0.271(0.008) | 0.04 | 0.78 | 4.10 |
| | | SCS-ENS | 0.255(0.008) | 0.00 | 0.55 | 5.16 | 0.257(0.008) | 0.00 | 0.63 | 4.80 |
| | 1000 | SC-SIR | 0.525(0.028) | 0.28 | 6.49 | – | 0.505(0.024) | 0.28 | 12.93 | – |
| | | SEAS-SIR | 0.262(0.008) | 0.03 | 1.83 | – | 0.261(0.011) | 0.07 | 1.55 | – |
| | | TC-SAVE | 0.592(0.029) | 0.06 | 0.65 | – | 0.999(0.000) | 0.46 | 857.04 | – |
| | | SCS-SAVE | 0.285(0.009) | 0.04 | 0.08 | 4.90 | 0.292(0.010) | 0.00 | 0.22 | 4.53 |
| | | SCS-ENS | 0.267(0.008) | 0.01 | 0.76 | 7.48 | 0.275(0.008) | 0.00 | 0.44 | 7.16 |

# Appendix A Some preliminary theoretical results

## A.1 An overview of preliminary results

We first introduce some lemmas and review some existing results that will be useful in the proof of the proposed theorems later. In Section A.2, Lemma 2 justifies the generality of regarding the dimension of $\mathcal{S}_{Y|X}$ as known *a priori*, by which we can safely assume $d$ to be known in all the proofs later for simplicity. Section A.3 includes three lemmas for proving Theorem 1 under low-dimensional settings, and Section A.4 includes three lemmas for proving the propositions and theorems under high-dimensional settings. In Section A.5, we review the classical Wedin's Theorem and Weyl's Theorem. More details about these results can be found at the beginning of each subsection.

To ease the presentation, we introduce a probabilistic version of $\Omega(1)$; that is, a sequence of random variables $\{Z_n : n = 1, \ldots, \}$ is written as $Z_n = \Omega_P(1)$ if there exists $C > 0$ such that $P(Z_n > C) \to 1$, which means that $Z_n$ is $\Omega(1)$ with probability converging to one. Equivalently, $Z_n = \Omega_P(1)$ if $P(Z_n > \epsilon_n) \to 1$ for any vanishing constant sequence $\epsilon_n = o(1)$. We refer to [27] for more details about this concept. Similarly to $\tau_{\min}(\cdot)$ and $\tau_{\max}(\cdot)$ defined at the beginning of Section 3 of the main text, for any matrix $\mathrm{M} \in \mathbb{R}^{p\times q}$, we use $\tau_1(\mathrm{M}) \geq \tau_2(\mathrm{M}) \geq \ldots \geq \tau_q(\mathrm{M})$ to denote the singular values of M and allow them to be zero, and use $\beta_1, \ldots, \beta_q$ to denote the corresponding left singular vectors. Considering the multiplicities of $\tau_i(\mathrm{M})$'s, we also use $\tau_{[1]}(\mathrm{M}) > \tau_{[2]}(\mathrm{M}) \ldots > \tau_{[r]}(\mathrm{M}) \geq 0$ to denote the distinct singular values of M, and use $B_1(\mathrm{M}), \ldots, B_r(\mathrm{M})$ to denote the corresponding subsets of $\beta_i(\mathrm{M})$'s; that is, $B_i(\mathrm{M})$ spans the $i$th eigenspace of $\mathrm{M}^{\otimes 2}$, where $\mathrm{M}^{\otimes 2}$ denotes $\mathrm{M}\mathrm{M}^\mathsf{T}$ for simplicity. The Moore–Penrose inverse of M is denoted by $\mathrm{M}^\dagger$, and the max norm of M, i.e., $\max_{i,j} |\mathrm{M}_{ij}|$, is denoted by $\|\mathrm{M}\|_\infty$.

## A.2 Justification of assuming the dimension of $\mathcal{S}_{Y|X}$ known

The following lemma justifies the equivalence of the asymptotic study for any statistic that involves $d$, the dimension of $\mathcal{S}_{Y|X}$, if we replace $d$ with its arbitrary consistent estimator $\widehat{d}$ in the statistic. The essence is that $\widehat{d}$ has discrete outcomes, which makes its convergence rate to $d$ arbitrarily high in the conventional sense of convergence in probability.

**Lemma 2** *For any statistic $S_n$ that involves $\widehat{d}$, where $\widehat{d}$ is a consistent estimator of $d$ in terms of $P(\widehat{d} = d) \to 1$, let $R_n$ be constructed identically as $S_n$ but with $\widehat{d}$ replaced by $d$. We have the asymptotic equivalence between $R_n$ and $S_n$; that is, $S_n - R_n = o_P(n^{-r})$ for any $r > 0$.*

*Proof* For any $r > 0$ and $\delta > 0$, we have

$$P(|S_n - R_n| > n^{-r}\delta) = P(|S_n - R_n| > n^{-r}\delta, \hat{d} \neq d) \leq Pr(\hat{d} \neq d) \to 0 \tag{24}$$

The equation above is due to the fact $S_n = R_n$ whenever $\hat{d} = d$. Thus, we have $S_n - R_n = o_P(n^{-r})$ for any $r > 0$ as the desired result. □

## A.3 Lemmas for the proof of Theorem 1

Regarding Theorem 1 where $\mathrm{M} = \mathrm{M}_{\mathrm{SAVE}}$, Lemma 3 connects $\Pi(\widehat{\beta}_{\mathrm{F}})$ that appears in $g(\mathrm{F})$ in (13) with the corresponding $\widehat{\mathrm{M}}_{\mathrm{F}}$, which is easier to study; Lemma 4 justifies the asymptotic properties of $\widehat{\mathrm{M}}_{\mathrm{F}}$ based on the full sample and based on the bootstrap sub-sample, respectively. Building upon these lemmas, Lemma 5 gives the key results to the proof of Theorem 1. Lemma 3 is a slight modification of Lemma 4.1 in [35].

Because bootstrap re-sampling is involved in the proposed efficiency-adapted column selection approach, which introduces two layers of randomness, we first follow [2] and [27] (see Section 1 of their supplementary material) to formulate a specific asymptotic setting. Let $(\Omega, \mathbf{E}, P)$ be a probability space and $\mathbb{N} = \{1, 2, \ldots\}$. Let $\Omega_S$ be the collection of all the sequences $\{a_n : n \in \mathbb{N}\}$ where $a_n \in \mathbb{R}^{p+1}$, and let $\mathbf{E}_S$ be the Borel $\sigma-$field on $\Omega_S$. Let $S : \Omega \to \Omega_S$ be a random element that is measurable with respect to $\mathbf{E}/\mathbf{E}_S$; that is, for each $\omega$, $S(\omega)$ is a sequence of vectors in $\mathbb{R}^{p+1}$. For each $n \in \mathbb{N}$, let $S_n(\omega)$ be the $n$th element of $S(\omega)$. We can also regard $S$ as a sequence of $(p+1)$-dimensional random vectors $\{S_n : n \in \mathbb{N}\}$. Let $P_S = P \circ S^{-1}$ be the probability on $(\Omega_S, \mathbf{E}_S)$ induced by $S$. Then $(\Omega_S, \mathbf{E}_S, P_S)$ is the probability space of the sequence of random vectors $\{S_n : n \in \mathbb{N}\}$. We now specify each $S_n$ to consist of $(X_n, Y_n)$, where $X_n$ is a $p$-dimensional random vector representing the predictor and $Y_n$ is a random variable representing the response, and we assume that $\{S_n : n \in \mathbb{N}\}$ are independent and identically distributed (i.i.d). In this way, we identify a random sample of size $n$ with the first $n$ elements in a random sequence $S$, and identify a random sequence $S$ with a sequence of random samples with increasing sample sizes.

Let $F$ be the distribution of $S_1$, and let $F_n$ be the empirical distribution based on $\{S_1, \ldots, S_n\}$, which is again the set of the first $n$ elements of $S$. Recall that $\widehat{\mathrm{M}}_{\mathrm{SAVE}}$, $\widehat{\mathrm{M}}_{\mathrm{F}}$, and $\widehat{\Sigma}$ are the estimators based on $F_n$. We denote the corresponding bootstrap estimators, which are based on an i.i.d. sample of $F_n$, by $\widehat{\mathrm{M}}^*_{\mathrm{SAVE}}$, $\widehat{\mathrm{M}}^*_{\mathrm{F}}$, and $\widehat{\Sigma}^*$, respectively. Given an outcome $s$ of $S$, a general full-sample estimator $\widehat{\mathrm{M}}$ based on $F_n$ becomes a fixed sequence of matrices, and the corresponding bootstrap estimator $\widehat{\mathrm{M}}^*$ is still a random sequence of matrices. We formulate the asymptotic property of $\widehat{\mathrm{M}}^*$ as the probabilistic limit of $n^{-r}(\widehat{\mathrm{M}}^* - \widehat{\mathrm{M}})$ for some $r > 0$ (given $S = s$) almost surely $P_S$. For consistency, we also formulate the asymptotic property of $\widehat{\mathrm{M}}$ in the same way; that is, we discuss the deterministic limit of $n^{-t}(\widehat{\mathrm{M}} - \mathrm{M})$ for some $t > 0$ (given $S = s$) almost surely $P_S$. Referring to the connection between $(\Omega_S, \mathbf{E}_S, P_S)$ and $(\Omega, \mathbf{E}, P)$ above, the convergence of $\widehat{\mathrm{M}}$ in this sense is identical to the convergence of $\widehat{\mathrm{M}}$ almost surely, which is stronger than the usual convergence in probability for $\widehat{\mathrm{M}}$.

**Lemma 3** *Suppose* $\mathrm{M} \in \mathbb{R}^{p\times q}$ *has rank $d$, and $B$ is the set of its left singular vectors. Let* $\widehat{\mathrm{M}}$ *be an estimator of* $\mathrm{M}$ *that satisfies* $\|\widehat{\mathrm{M}} - \mathrm{M}\|_{\mathrm{FB}} = o(1)$ *almost surely $P_S$, and let $\widehat{B}$ be the set of its leading $d$ left singular vectors. We have, as $n \to \infty$,*

$$vec\{\Pi(\widehat{B}) - \Pi(B)\} = h(\mathrm{M})vec(\widehat{\mathrm{M}} - \mathrm{M}) + \epsilon_n,$$

*where $h(\cdot)$ is a non-stochastic and continuous mapping of* $\mathrm{M}$*, and* $\|\epsilon_n\|_{\mathrm{FB}} \le c_{\mathrm{M}}\|\widehat{\mathrm{M}}-\mathrm{M}\|^2_{\mathrm{FB}}$ *almost surely $P_S$ for some positive constant $c_{\mathrm{M}}$ specific for* $\mathrm{M}$*.*

*Proof* The proof resembles that of Lemma 4.1 in [35], with $M$, $M_n$, $a_n$, $P_0$, $A$, $\Gamma_n$ and $\Gamma$ in the equation there replaced by $\mathrm{M}^{\otimes 2}$, $\widehat{\mathrm{M}}^{\otimes 2}$, $1/2$, $\Pi(B)$, $B$, $I_p$ and $I_p$, respectively, and with $vec(\cdot)$ applied to both sides of the equation. Again, $\mathrm{M}^{\otimes 2}$ here denotes $\mathrm{M}\mathrm{M}^{\mathsf{T}}$. In their lemma, $\|\widehat{\mathrm{M}}-\mathrm{M}\|^2_{\mathrm{FB}}$ is assumed to be upper bounded by a constant, under which $\epsilon_n$ satisfies $|\epsilon_n| \le c_{\mathrm{M}}\|\widehat{\mathrm{M}} - \mathrm{M}\|^2_{\mathrm{FB}}$ for some constant $c_{\mathrm{M}} > 0$ specific for $\mathrm{M}$. This assumption is satisfied in our case conditional on almost all $s \in \Omega_S$, as $\widehat{\mathrm{M}}$ satisfies $\|\widehat{\mathrm{M}} - \mathrm{M}\|_{\mathrm{FB}} = o(1)$ almost surely $P_S$. The functional form of $h(\mathrm{M})$ can be easily derived from the major term of the right-hand side of the equation in Lemma 4.1 of [35], which is

$$\begin{aligned}\textstyle\sum_{k=1}^{r} \Big[&\Pi\{B_k(\mathrm{M})\} \otimes \{\mathrm{M}^{\otimes 2} - \tau^2_{[k]}(\mathrm{M})I_p\}^{\dagger} \\ &+ \{\mathrm{M}^{\otimes 2} - \tau^2_{[k]}(\mathrm{M})I_p\}^{\dagger} \otimes \Pi\{B_k(\mathrm{M})\}\Big]\, vec(\widehat{\mathrm{M}}^{\otimes 2} - \mathrm{M}^{\otimes 2}),\end{aligned} \tag{25}$$

where $vec(\widehat{\mathrm{M}}^{\otimes 2} - \mathrm{M}^{\otimes 2})$ can be written as

$$vec(\widehat{\mathrm{M}}^{\otimes 2} - \mathrm{M}^{\otimes 2}) = \{I_p \otimes \mathrm{M} + (\mathrm{M} \otimes I_p)K_{(p,q)}\}vec(\widehat{\mathrm{M}} - \mathrm{M}) + O(\|\widehat{\mathrm{M}} - \mathrm{M}\|^2_{\mathrm{FB}}). \tag{26}$$

Here, $B_k(\mathrm{M})$, $\tau_{[k]}(\mathrm{M})$, and $\mathrm{M}^{\dagger}$ in (25) are defined at the end of Section A.1 above, and $K_{(p,q)}$ in (26) transforms $vec(\mathrm{M})$ to $vec(\mathrm{M}^{\mathsf{T}})$. To see the continuity of $h(\cdot)$, suppose $\{\mathrm{M}_n : n \in \mathbb{N}\}$ is a sequence of matrices with $\|\mathrm{M}_n - \mathrm{M}\|_{\mathrm{FB}} = o(1)$. By (25) and (26), we have

$$\begin{aligned}\|h(\mathrm{M}_n) - h(\mathrm{M})\|_{\mathrm{FB}} = O\big[\,&\|\mathrm{M}_n - \mathrm{M}\|_{\mathrm{FB}} + \textstyle\sum_{k=1}^{r} \|\Pi\{B_k(\mathrm{M}_n)\} - \Pi\{B_k(\mathrm{M})\}\|_{\mathrm{FB}} \\ &+ \textstyle\sum_{i=1}^{q} |\tau_i(\mathrm{M}_n) - \tau_i(\mathrm{M})|\,\big].\end{aligned} \tag{27}$$

By Weyl's Theorem (see Subsection A.5 below), we have $|\tau_i(\mathrm{M}_n)-\tau_i(\mathrm{M})| = O(\|\mathrm{M}_n-\mathrm{M}\|_{\mathrm{FB}})$ for $i = 1, \ldots, q$. By the strict inequalities between $\tau_{[1]}(\mathrm{M}), \ldots, \tau_{[r]}(\mathrm{M})$ and Wedin's Theorem (again see Subsection A.5 below), we have $\|\Pi\{B_k(\mathrm{M}_n)\} - \Pi\{B_k(\mathrm{M})\}\|_{\mathrm{FB}} = O(\|\mathrm{M}_n - \mathrm{M}\|_{\mathrm{FB}})$ for $k = 1, \ldots, r$. Together with (27), these imply

$$\|h(\mathrm{M}_n) - h(\mathrm{M})\|_{\mathrm{FB}} = O(\|\mathrm{M}_n - \mathrm{M}\|_{\mathrm{FB}}),$$

which justifies the continuity of $h(\cdot)$. This completes the proof. □

**Lemma 4** *Suppose $E(\|X\|_{\mathrm{FB}}^8) < \infty$, and denote $\log\{\log(n)\}$ by $c_n$. We have, for almost surely $P_S$,*

*(i) $n^{1/2}vec(\widehat{\mathrm{M}}_{\mathrm{SAVE}} - \mathrm{M}_{\mathrm{SAVE}}) = n^{-1/2}\sum_{i=1}^n \phi(X_n, Y_n) + O(n^{-1/2}c_n)$, where $\phi(X, Y)$ is a matrix-valued measurable function of $(X, Y)$ with zero mean and finite covariance matrix; this implies $n^{1/2}vec(\widehat{\mathrm{M}}_{\mathrm{F}} - \mathrm{M}_{\mathrm{F}}) = n^{-1/2}\sum_{i=1}^n \phi_{[\mathrm{F}]}(X_n, Y_n) + O(n^{-1/2}c_n)$ for any $\mathrm{F} \subset \mathcal{I}_{\mathrm{M}}$, where $\phi_{[\mathrm{F}]}(X, Y)$ is the corresponding submatrix of $\phi(X, Y)$;*

*(ii) $n^{1/2}vec(\widehat{\mathrm{M}}^*_{\mathrm{SAVE}} - \widehat{\mathrm{M}}_{\mathrm{SAVE}}) = n^{-1/2}\sum_{i=1}^n \phi(X_i^*, Y_i^*) + O_P(n^{-1/2})$, and, for any $\mathrm{F} \subset \mathcal{I}_{\mathrm{M}}$, $n^{1/2}vec(\widehat{\mathrm{M}}^*_{\mathrm{F}} - \widehat{\mathrm{M}}_{\mathrm{F}}) = n^{-1/2}\sum_{i=1}^n \phi_{[\mathrm{F}]}(X_i^*, Y_i^*) + O_P(n^{-1/2})$, where $\{(X_i^*, Y_i^*) : i = 1, \ldots, n\}$ is an i.i.d bootstrap sample from $F_n$;*

*(iii) $E_n[\{n^{1/2}vec(\widehat{\mathrm{M}}^*_{\mathrm{F}} - \widehat{\mathrm{M}}_{\mathrm{F}})\}^{\otimes 2}] = E\{\phi^{\otimes 2}_{[\mathrm{F}]}(X, Y)\} + O(n^{-1/2}c_n^{1/2})$ for any $\mathrm{F} \subset \mathcal{I}_{\mathrm{M}}$.*

*Proof* $(i)$ Let $E_a = E_1 \cap E_2 \cap E_3 \cap E_4$, with the events $E_1, E_2, E_3, E_4 \in \mathbf{E}_S$ being

$$E_1 = \{\widehat{\Sigma} = \Sigma + O(n^{-1/2}c_n^{1/2})\},$$

$$E_2 = \{\widehat{\Sigma}_h = \Sigma_h + O(n^{-1/2}c_n^{1/2}) \text{ for all } h = 1, ..., H\},$$

$$E_3 = \{\widehat{\mu} = \mu + O(n^{-1/2}c_n^{1/2})\},$$

$$E_4 = \{\widehat{\mu}_h = \mu_h + O(n^{-1/2}c_n^{1/2})\} \text{ for all } h = 1, ..., H\},$$

where $\mu_h$ and $\Sigma_h$ are defined in the Introduction of the main text. By the law of iterated logarithm and the assumption $E(\|X\|_{\mathrm{FB}}^8) < \infty$, we have $P(E_a) = 1$. Recall the functional form of $\mathrm{M}_{\mathrm{SAVE}}$ in Equation (4) of the main text, we have

$$n^{1/2}(\widehat{\mathrm{M}}_{\mathrm{SAVE}} - \mathrm{M}_{\mathrm{SAVE}}) = \{n^{1/2}(\widehat{\Sigma}^{-1}\widehat{\Sigma}_h - \Sigma^{-1}\Sigma_h)\}_{h=1,\ldots,H}.$$

For each $h = 1, \ldots, H$, we also have, conditional on any $s \in E_a$,

$$n^{1/2}(\widehat{\Sigma}^{-1}\widehat{\Sigma}_h - \Sigma^{-1}\Sigma_h) = n^{1/2}\{\Sigma^{-1}(\widehat{\Sigma}_h - \Sigma_h) + \Sigma^{-1}(\widehat{\Sigma} - \Sigma)\Sigma^{-1}\Sigma_h\} + O(n^{-1/2}c_n).$$

These together imply $n^{1/2}vec(\widehat{\mathrm{M}}_{\mathrm{SAVE}} - \mathrm{M}_{\mathrm{SAVE}}) = n^{-1/2}\sum_{i=1}^n \phi(X_n, Y_n) + O(n^{-1/2}c_n)$ given any $s \in E_a$, where $\phi(X, Y) = [vec\{\phi_h(X, Y)\}]_{h=1,\ldots,H}$ with each $\phi_h(X, Y)$ being:

$$\Sigma^{-1}[\{XI(Y = h) - \mu_h\}^{\otimes 2} - \Sigma_h + \mu_h\{XI(Y = h) - \mu_h\}^{\mathsf{T}} + \{XI(Y = h) - \mu_h\}\mu_h^{\mathsf{T}}] + \Sigma^{-1}\{(X - \mu)^{\otimes 2} - \Sigma + \mu(X - \mu)^{\mathsf{T}} + (X - \mu)\mu^{\mathsf{T}}\}\Sigma^{-1}\Sigma_h. \tag{28}$$

The proof of the statement for $\widehat{\mathrm{M}}_F$ follows immediately.

$(ii)$ Conditioning on any $s \in \Omega_S$, the central limit theorem implies $\widehat{\Sigma}^* = \widehat{\Sigma} + O_P(n^{-1/2})$ and similarly the $n^{1/2}$-convergence (in probability) of $\widehat{\Sigma}_h^*$, $\widehat{\mu}^*$, and $\widehat{\mu}_h^*$. The asymptotic linearity of $n^{1/2}vec(\widehat{\mathrm{M}}^*_{\mathrm{SAVE}} - \widehat{\mathrm{M}}_{\mathrm{SAVE}})$ follows the same reasoning as the counterpart in the proof of $(i)$ above, where $\phi(X, Y)$ is now replaced with $\phi(X^*, Y^*)$ where $(X^*, Y^*)$ has the empirical distribution $F_n$. This completes the proof of $(ii)$. To ease the proof of $(iii)$ below, we now strengthen $(ii)$ to that the remainder term, denoted by

$$\varepsilon_{[\mathrm{F}]} \equiv n^{1/2}vec(\widehat{\mathrm{M}}^*_{\mathrm{F}} - \widehat{\mathrm{M}}_{\mathrm{F}}) - n^{-1/2}\sum_{i=1}^n \phi_{[\mathrm{F}]}(X_i^*, Y_i^*), \tag{29}$$

further satisfies $E_n(\varepsilon^{\otimes 2}_{[\mathrm{F}]}) = O(n^{-1})$ almost surely $P_S$. Since $E_n\{(\widehat{\mu}^*)^{\otimes 2}\} = n^{-1}\widehat{\Sigma}$ and $E_n\{(\widehat{\mu}_h^*)^{\otimes 2}\} = n^{-1}\widehat{\Sigma}_h$, both terms are $O(n^{-1})$ conditioning on any $s \in E_1 \cap E_2$, i.e., almost surely $P_S$. Under $E(\|X\|_{\mathrm{FB}}^8) < \infty$, we can similarly prove $E_n[\{vec(\widehat{\Sigma}^*)\}^{\otimes 2}] = O(n^{-1})$ and $E_n[\{vec(\widehat{\Sigma}_h^*)\}^{\otimes 2}] = O(n^{-1})$ almost surely $P_S$ by constructing appropriate $E_5, E_6 \in \mathbf{E}_S$ using higher-order moments of $X$. By the same algebra as in the proof of *(i)* above, these together imply $E_n(\varepsilon_{[\mathrm{F}]}^{\otimes 2}) = O(n^{-1})$ almost surely $P_S$.

(*iii*) Recall $\varepsilon_{[\mathrm{F}]}$ defined in (29) with $E_n(\varepsilon_{[\mathrm{F}]}^{\otimes 2}) = O(n^{-1})$ almost surely $P_S$. By the Cauchy-Schwarz Inequality, this implies

$$E_n[\{n^{1/2} vec(\widehat{\mathrm{M}}_{\mathrm{F}}^{*} - \widehat{\mathrm{M}}_{\mathrm{F}})\}^{\otimes 2}] = E_n\{\phi_{[\mathrm{F}]}^{\otimes 2}(X^*, Y^*)\} + O(n^{-1/2}) \tag{30}$$

almost surely $P_S$, where $(X^*, Y^*)$ has the empirical distribution $F_n$. By the law of iterated logarithm and $E(\|X\|_{\mathrm{FB}}^{8}) < \infty$, we also have

$$E_n\{\phi_{[\mathrm{F}]}^{\otimes 2}(X^*, Y^*)\} = E\{\phi_{[\mathrm{F}]}^{\otimes 2}(X, Y)\} + O(n^{-1/2} c_n^{1/2})$$

almost surely $P_S$. Plugging this into (30) readily implies (*iii*). This completes the proof. □

The next lemma gives the key results to the proof of Theorem 1; that is, for almost surely $P_S$, the first term of $g(\mathrm{F})$ converges to a positive constant specific for F with the error being $O_P(n^{-1/2} c_n^{1/2})$, and $\mathcal{D}(\mathcal{S}(\widehat{\beta}_{\mathrm{F}}), \mathcal{S}_{Y|X})$ also converges to the same constant with the error being $O(n^{-1/2} c_n^{3/2})$ if $\mathrm{F} \in \mathcal{F}$.

**Lemma 5** *Suppose $E(\|X\|_{\mathrm{FB}}^{8}) < \infty$. We have, for any $\mathrm{F} \in \mathcal{I}_{\mathrm{M}}$ and almost surely $P_S$,*

*(i)* $\sum_{i=1}^{n} \|\Pi(\widehat{\beta}_{\mathrm{F}}^{(i)}) - \Pi(\widehat{\beta}_{\mathrm{F}})\|_{\mathrm{FB}}^2 = \mathrm{tr}\left[h(\mathrm{M}_{\mathrm{F}}) E\{\phi_{[\mathrm{F}]}^{\otimes 2}(X, Y)\} h^{\mathsf{T}}(\mathrm{M}_{\mathrm{F}})\right] + O_P(n^{-1/2} c_n^{1/2})$ *where $h(\cdot)$ is defined in Lemma 3 and $\phi_{[\mathrm{F}]}(\cdot)$ is defined in Lemma 4(i),*

*(ii)* $\mathcal{D}(\mathcal{S}(\widehat{\beta}_{\mathrm{F}}), \mathcal{S}_{Y|X}) = \mathrm{tr}\left[h(\mathrm{M}_{\mathrm{F}}) E\{\phi_{[\mathrm{F}]}^{\otimes 2}(X, Y)\} h^{\mathsf{T}}(\mathrm{M}_{\mathrm{F}})\right] + O_P(n^{-1/2} c_n^{3/2})$ *if $\mathrm{F} \in \mathcal{F}$.*

*Proof (i)* Similarly to $\widehat{\beta}_{\mathrm{F}}^{(i)}$ in (13), we use $\widehat{\mathrm{M}}_{\mathrm{F}}^{(i)}$ to denote the $i$th bootstrap estimator of $\mathrm{M}_{\mathrm{F}}$. Let $\widehat{\mathrm{M}}$ and M in Lemma 3 be $\widehat{\mathrm{M}}_{\mathrm{F}}^{(i)}$ and $\widehat{\mathrm{M}}_{\mathrm{F}}$, respectively. We denote the corresponding remainder term $\epsilon_n$ by $\epsilon^{(i)}$, which satisfies $\|\epsilon^{(i)}\|_{\mathrm{FB}} < c_{\widehat{\mathrm{M}}} \|\widehat{\mathrm{M}}_{\mathrm{F}}^{(i)} - \widehat{\mathrm{M}}_{\mathrm{F}}\|_{\mathrm{FB}}^2$. By the definition of $\|\cdot\|_{\mathrm{FB}}$ and the Cauchy-Schwarz Inequality, we have

$$\begin{aligned}
\sum_{i=1}^{n} \|\Pi(\widehat{\beta}_{\mathrm{F}}^{(i)}) - \Pi(\widehat{\beta}_{\mathrm{F}})\|_{\mathrm{FB}}^2 &= \sum_{i=1}^{n} \|h(\widehat{\mathrm{M}}_{\mathrm{F}}) vec(\widehat{\mathrm{M}}_{\mathrm{F}}^{(i)} - \widehat{\mathrm{M}}_{\mathrm{F}}) + \epsilon^{(i)}\|_{\mathrm{FB}}^2 \\
&= \sum_{i=1}^{n} \|h(\widehat{\mathrm{M}}_{\mathrm{F}}) vec(\widehat{\mathrm{M}}_{\mathrm{F}}^{(i)} - \widehat{\mathrm{M}}_{\mathrm{F}})\|_{\mathrm{FB}}^2 + 2\sum_{i=1}^{n} \|h(\widehat{\mathrm{M}}_{\mathrm{F}}) vec(\widehat{\mathrm{M}}_{\mathrm{F}}^{(i)} - \widehat{\mathrm{M}}_{\mathrm{F}})\|_{\mathrm{FB}} \|\epsilon^{(i)}\|_{\mathrm{FB}} \\
&\quad + \sum_{i=1}^{n} \|\epsilon^{(i)}\|_{\mathrm{FB}}^2 \\
&\equiv \mathrm{I} + \mathrm{II} + \mathrm{III},
\end{aligned}$$

We next show $\mathrm{I} = \mathrm{tr}\left[h(\mathrm{M}_{\mathrm{F}}) E\{\phi_{[\mathrm{F}]}^{\otimes 2}(X, Y)\} h^{\mathsf{T}}(\mathrm{M}_{\mathrm{F}})\right] + O_P(n^{-1/2} c_n^{1/2})$, $\mathrm{II} = O_P(n^{-1/2} c_n^{1/2})$, and $\mathrm{III} = O_P(n^{-1/2} c_n^{1/2})$ almost surely $P_S$, which readily imply statement ($i$). Conditioning on any $s \in \Omega_S$, $\widehat{\mathrm{M}}_{\mathrm{F}}^{(1)}, \ldots, \widehat{\mathrm{M}}_{\mathrm{F}}^{(n)}$ are i.i.d., so the central limit theorem implies

$$\sum_{i=1}^{n} \|h(\widehat{\mathrm{M}}_{\mathrm{F}}) vec(\widehat{\mathrm{M}}_{\mathrm{F}}^{(i)} - \widehat{\mathrm{M}}_{\mathrm{F}})\|_{\mathrm{FB}}^2 = E_n(\|h(\widehat{\mathrm{M}}_{\mathrm{F}}) n^{1/2} vec(\widehat{\mathrm{M}}_{\mathrm{F}}^{(i)} - \widehat{\mathrm{M}}_{\mathrm{F}})\|_{\mathrm{FB}}^2) + O_P(n^{-1/2}). \tag{31}$$

By Lemma 4(*iii*), we have $E_n\{n^{1/2} vec(\widehat{\mathrm{M}}_{\mathrm{F}}^{(i)} - \widehat{\mathrm{M}}_{\mathrm{F}})\}^{\otimes 2} = E_n\{\phi_{[\mathrm{F}]}^{\otimes 2}(X, Y)\} + O(n^{-1/2} c_n^{1/2})$ almost surely $P_S$. By Lemma 4($i$) and $E_n\{\phi_{[\mathrm{F}]}(X^*, Y^*)\} = O(n^{-1/2} c_n^{1/2})$ almost surely $P_S$ inherited from the law of iterated logarithm, we also have

$$\|\widehat{\mathrm{M}}_{\mathrm{F}} - \mathrm{M}_{\mathrm{F}}\|_{\mathrm{FB}} = \|E_n\{\phi_{[\mathrm{F}]}(X^*, Y^*)\}\|_{\mathrm{FB}} + O(n^{-1} c_n) = O(n^{-1/2} c_n^{1/2}) \tag{32}$$

almost surely $P_S$, which implies $h(\widehat{\mathrm{M}}_{\mathrm{F}}) - h(\mathrm{M}_{\mathrm{F}}) = O(n^{-1/2} c_n^{1/2})$ almost surely $P_S$ by the continuity of $h(\cdot)$ shown in Lemma 3. By plugging these into (31), we have

$$\mathrm{I} = \mathrm{tr}\left[h(\mathrm{M}_{\mathrm{F}}) E\{\phi_{[\mathrm{F}]}^{\otimes 2}(X, Y)\} h^{\mathsf{T}}(\mathrm{M}_{\mathrm{F}})\right] + O_P(n^{-1/2} c_n^{1/2})$$

almost surely $P_S$. Since $\|\epsilon^{(i)}\|_{\mathrm{FB}} < c_{\widehat{\mathrm{M}}}\|\widehat{\mathrm{M}}_{\mathrm{F}}^{(i)} - \widehat{\mathrm{M}}_{\mathrm{F}}\|_{\mathrm{FB}}^2$, we have

$$\mathrm{II} = O(\textstyle\sum_{i=1}^n \|\widehat{\mathrm{M}}_{\mathrm{F}}^{(i)} - \widehat{\mathrm{M}}_{\mathrm{F}}\|_{\mathrm{FB}}^3), \quad \mathrm{III} = O(\textstyle\sum_{i=1}^n \|\widehat{\mathrm{M}}_{\mathrm{F}}^{(i)} - \widehat{\mathrm{M}}_{\mathrm{F}}\|_{\mathrm{FB}}^4) \tag{33}$$

almost surely $P_S$. Since $E(\|X\|_{\mathrm{FB}}^8) < \infty$, by simple algebra, we have

$$E\{n^2\|vec(\widehat{\mathrm{M}}_{\mathrm{F}}^{(i)} - \widehat{\mathrm{M}}_{\mathrm{F}})\|_{\mathrm{FB}}^4\} < \infty,$$

which, by Jensen's inequality, also implies $E\{n^{3/2}\|vec(\widehat{\mathrm{M}}_{\mathrm{F}}^{(i)} - \widehat{\mathrm{M}}_{\mathrm{F}})\|_{\mathrm{FB}}^3\} < \infty$. By the strong law of large numbers, these indicate

$$E_n\{n^{3/2}\|vec(\widehat{\mathrm{M}}_{\mathrm{F}}^{(i)} - \widehat{\mathrm{M}}_{\mathrm{F}})\|_{\mathrm{FB}}^3\} = O(1), \quad E_n\{n^2\|vec(\widehat{\mathrm{M}}_{\mathrm{F}}^{(i)} - \widehat{\mathrm{M}}_{\mathrm{F}})\|_{\mathrm{FB}}^4\} = O(1), \tag{34}$$

almost surely $P_S$. Since for sequence of non-negative random variables $Z_n$, we always have $Z_n = O_P(E(Z_n))$ by Markov's Inequality, (34) implies

$$\textstyle\sum_{i=1}^n \|\widehat{\mathrm{M}}_{\mathrm{F}}^{(i)} - \widehat{\mathrm{M}}_{\mathrm{F}}\|_{\mathrm{FB}}^3 = O_P(n^{-1/2}), \quad \sum_{i=1}^n \|\widehat{\mathrm{M}}_{\mathrm{F}}^{(i)} - \widehat{\mathrm{M}}_{\mathrm{F}}\|_{\mathrm{FB}}^4 = O_P(n^{-1})$$

almost surely $P_S$, which, by (33) and $c_n = \Omega(1)$, implies $\mathrm{II} = O_P(n^{-1/2}c_n^{1/2})$ and $\mathrm{III} = O_P(n^{-1/2}c_n^{1/2})$ almost surely $P_S$. This completes the proof of statement $(i)$.

*(ii)* Since $\mathrm{F} \in \mathcal{F}$, we have $\mathcal{S}(\beta_{\mathrm{F}}) = \mathcal{S}(\beta_0)$. Thus, $\mathcal{D}(\mathcal{S}(\widehat{\beta}_{\mathrm{F}}), \mathcal{S}_{Y|X})$ in this case reduces to $nE(\|\Pi(\widehat{\beta}_{\mathrm{F}}) - \Pi(\beta_F)\|_{\mathrm{FB}}^2)$. Let $\widehat{\mathrm{M}}$ and M in Lemma 3 be $\widehat{\mathrm{M}}_{\mathrm{F}}$ and $\mathrm{M}_{\mathrm{F}}$, respectively; then the remainder term $\epsilon_n$ satisfies $\epsilon_n = O(\|\widehat{\mathrm{M}}_{\mathrm{F}} - \mathrm{M}_{\mathrm{F}}\|_{\mathrm{FB}}^2)$ almost surely $P_S$. Therefore, similar to the proof of statement $(i)$ above, we have

$$\begin{aligned} &nE(\|\Pi(\widehat{\beta}_{\mathrm{F}}) - \Pi(\beta_F)\|_{\mathrm{FB}}^2) \\ &\quad = nE\{\|h(\mathrm{M}_{\mathrm{F}})vec(\widehat{\mathrm{M}}_{\mathrm{F}} - \mathrm{M}_{\mathrm{F}})\|_{\mathrm{FB}}^2\} + nO\{E(\|\widehat{\mathrm{M}}_{\mathrm{F}} - \mathrm{M}_{\mathrm{F}}\|_{\mathrm{FB}}^3) + E(\|\widehat{\mathrm{M}}_{\mathrm{F}} - \mathrm{M}_{\mathrm{F}}\|_{\mathrm{FB}}^4)\} \\ &\quad = nE\{\|h(\mathrm{M}_{\mathrm{F}})vec(\widehat{\mathrm{M}}_{\mathrm{F}} - \mathrm{M}_{\mathrm{F}})\|_{\mathrm{FB}}^2\} + O(n^{-1/2}c_n^{3/2}) \\ &\quad = [E\{\phi_{[\mathrm{F}]}^{\otimes 2}(X,Y)\} + O(n^{-1/2}c_n)] + O(n^{-1/2}c_n^{3/2}), \end{aligned}$$

where the second equation is implied by (32) and $E(n^2\|\widehat{\mathrm{M}}_{\mathrm{F}} - \mathrm{M}_{\mathrm{F}}\|_{\mathrm{FB}}^4) = O(E(\|X\|_{\mathrm{FB}}^8))$, and the last equation is implied by Lemma 4$(i)$ and the Cauchy-Schwarz Inequality. This completes the proof of statement $(ii)$. □

## A.4 Lemmas for the theory under high-dimensional settings

This subsection consists of three lemmas that help establish the theoretical results under high-dimensional settings. Lemma 6 explores the convergence rate of the sample covariance matrix $\widehat{\Sigma}$ with the aid of the sub-Gaussian property of $X$ in Condition (A4). It is essentially a modification of Equation (10) of [3], and it helps validate Condition (A7) in Proposition 3 and contributes to the proof of Theorem 2. Lemma 7 resembles Lemma 6 and is also used to validate Condition (A7) in Proposition 3, and it explores the convergence rate of the submatrices of the sample conditional covariance matrix $\widehat{\Sigma}_h$ that have fixed numbers of columns. Its proof is modified from that of Lemma 4 of [31]. Lemma 8 is a mirror image of a part of Theorem 2 in [1] and gives the lower bound of $\tau_{\min}(\mathrm{M})$ when M is an upper-triangular matrix, which will appear in the proof of Theorem 3.

**Lemma 6** *Let $\widehat{\Sigma}$ be sample covariance matrix of $X$ that estimates $\Sigma$. Under Conditions (A4)-(A5), for any positive integer $k$, there exists $c_{(k)} > 0$ such that*

$$P(\|\widehat{\Sigma} - \Sigma\|_\infty < c_{(k)}\sqrt{\log(p)/n}) = 1 - O(p^{-k}). \tag{35}$$

*Proof* Under Condition (A4), we have $\|\Sigma\|_\infty = O(1)$. Equation (10) of [3] then justifies

$$P(\|\hat{\Sigma} - \Sigma\|_\infty \geq t) \leq p^2 q e^{-wnt^2} \tag{36}$$

for any $|t| < \delta$, where $q, w, \delta$ are fixed constants determined by an upper bound of $\|\Sigma\|_\infty$. Under Condition (A5), we have $log(p)/n = o(1)$. Thus, for any positive integer $k$ and for all the large $n$, we can take $t = c_{(k)}\sqrt{\log(p)/n}$ with $c_{(k)} = \sqrt{(k+2)/w}$. (36) then implies

$$P(\|\hat{\Sigma} - \Sigma\|_\infty \geq c_{(k)}\sqrt{\log(p)/n}) \leq p^{2-wc^2_{(k)}} q = p^{-k} q,$$

which readily implies (35). □

**Lemma 7** *For each fixed* $\mathrm{F} \subseteq \{1, \ldots, pH\}$*, let* $\Sigma_\mathrm{F} = (\sigma_{\mathrm{F}ij})$ *be the submatrix of* $\Sigma_\Omega \equiv (\Sigma_1, \ldots, \Sigma_H)$ *consisting of its columns indexed by* F*, and let* $\widehat{\Sigma}_\mathrm{F} = (\widehat{\sigma}_{\mathrm{F}ij})$ *be the estimator of* $\Sigma_\mathrm{F}$ *induced from* $\widehat{\Sigma}_\Omega$*. Under Condition (iii) in Proposition 3, there exists* $c_7 > 0$ *such that, for all the large* $n$*,*

$$P(\|\widehat{\Sigma}_\mathrm{F} - \Sigma_\mathrm{F}\|_\infty \leq c_7\sqrt{H\log(p)/n}) \geq 1 - 12\mathcal{C}(\mathrm{F})p^{-3}. \tag{37}$$

*Proof* By Equations (A.53) and (A.54) in [31], there exists a uniform $c_7 > 0$ such that $P(|\widehat{\sigma}_{\mathrm{F}ij} - \sigma_{\mathrm{F}ij}| \leq c_7\sqrt{H\log(p)/n}) \geq 1 - 12p^{-4}$ for all $i = 1, \ldots, p$ and $j = 1, \ldots, \mathcal{C}(\mathrm{F})$. By the definition of $\|\cdot\|_\infty$, this implies

$$\begin{aligned}
&P(\|\widehat{\Sigma}_\mathrm{F} - \Sigma_\mathrm{F}\|_\infty \geq c_7\sqrt{H\log(p)/n}) \\
&\quad = P(\cup_{i=1,\ldots,p,j=1,\ldots,\mathcal{C}(\mathrm{F})}|\widehat{\sigma}_{\mathrm{F}ij} - \sigma_{\mathrm{F}ij}| \geq c_7\sqrt{H\log(p)/n}) \\
&\quad \leq p\mathcal{C}(\mathrm{F}) \max_{i=1,\ldots,p,j=1,\ldots,\mathcal{C}(\mathrm{F})} P(|\widehat{\sigma}_{\mathrm{F}ij} - \sigma_{\mathrm{F}ij}| \geq c_7\sqrt{H\log(p)/n}) \leq 12\mathcal{C}(\mathrm{F})p^{-3}.
\end{aligned}$$

which readily implies (37). □

**Lemma 8** *For any* $a \geq 1$ *and positive integer* $d$*, let* $K_d(a)$ *be the set of* $d \times d$*-dimensional upper triangular matrices with all the off-diagonal entries falling in* $[-a, a]$ *and all the diagonal entries being one. We have*

$$\min\{\tau_{\min}(U) : U \in K_d(a)\} \geq \left\{\frac{2d}{a+2} + \frac{(a+1)^{2d} - 1}{(a+2)^2}\right\}^{-1/2}.$$

*Remark.* This lemma immediately implies that, for any fixed $d$, any sequence of $d \times d$-dimensional upper triangular matrices with diagonal entries being one and off-diagonal entries being $O_P(1)$ must have its minimum singular value being $\Omega_P(1)$. This result will be used in the proof of Theorem 3 later.

*Proof* This lemma is an immediate result of Theorem 2 of [1] (see the lower bound in the inequality there), as long as we change their $\min\{\tau_{\min}(U) : U^\top \in K_d(a)\}$ to $\min\{\tau_{\min}(U) : U \in K_d(a)\}$.

## A.5 Review of some classical results

This subsection reviews Wedin's Theorem and Weyl's Theorem. The presentation below is based on Section 2.3 of [5] and Theorem 1 of [32]. For Wedin's Theorem, the $\sin\Theta$ distance is used to measure the difference between two semi-orthogonal matrices; that is, for $V, \widehat{V} \in \mathbb{R}^{p\times q}$ with $V^\mathsf{T} V = I_q$ and $\widehat{V}^\mathsf{T}\widehat{V} = I_q$, let

$$\sin\Theta(V, \widehat{V}) = \mathrm{diag}[\{(1-\sigma_1)^{1/2}, \ldots, (1-\sigma_r)^{1/2}\}],$$

where $\mathrm{diag}(v)$ denotes the diagonal matrix whose diagonal is $v$ and $\sigma_1, \ldots, \sigma_r$ are the eigenvalues of $V^\mathsf{T}\widehat{V}\widehat{V}^\mathsf{T}V^\mathsf{T}$ that always fall in the interval $(0, 1)$. Since $\sin\Theta(V, \widehat{V})$ is invariant of orthogonal column transformations of $V$ and $\widehat{V}$, it indeed measures the deviance between $\mathcal{S}(V)$ and $\mathcal{S}(\widehat{V})$, or equivalently between $\Pi(\mathrm{M})$ and $\Pi(\widehat{\mathrm{M}})$ for any M and $\widehat{\mathrm{M}}$ span $\mathcal{S}(V)$ and $\mathcal{S}(\widehat{V})$, respectively. This connects Wedin's Theorem with the proposed methods, and the corollaries that elaborate this connection, as presented in the remark below the theorem, will be used in the proofs of Proposition 1, Theorem 3, and Theorem 4. Weyl's Theorem will be applied in the proof of Proposition 1 and the proof of Theorem 4.

*Wedin's Theorem* [5] Suppose a matrix M and its perturbation $\widehat{\mathrm{M}} = R + Z$ have the singular value decomposition

$$\mathrm{M} = \begin{bmatrix} U_1 & U_2 \end{bmatrix} \begin{bmatrix} \Sigma_1 & 0 \\ 0 & \Sigma_2 \end{bmatrix} \begin{bmatrix} V_1^\mathsf{T} \\ V_2^\mathsf{T} \end{bmatrix}, \widehat{\mathrm{M}} = \begin{bmatrix} \widehat{U}_1 & \widehat{U}_2 \end{bmatrix} \begin{bmatrix} \widehat{\Sigma}_1 & 0 \\ 0 & \widehat{\Sigma}_2 \end{bmatrix} \begin{bmatrix} \widehat{V}_1^\mathsf{T} \\ \widehat{V}_2^\mathsf{T} \end{bmatrix}, \tag{38}$$

respectively, where we allow $\Sigma_2$ and $\widehat{\Sigma}_2$ to be singular so that $[U_1 U_2]$ and $[\widehat{U}_1\widehat{U}_2]$ have the same number of columns as $R$. If $\delta \equiv \tau_{\min}(\widehat{\Sigma}_1) - \tau_{\max}(\Sigma_2) > 0$, then we have

$$\max\{\|\sin\Theta(U_1, \widehat{U}_1)\|_{\mathrm{FB}},\ \|\sin\Theta(V_1, \widehat{V}_1)\|_{\mathrm{FB}}\} \le \delta^{-1}\max\{\|Z\widehat{V}_1\|_{\mathrm{FB}},\ \|\widehat{U}_1^\mathsf{T} Z\|_{\mathrm{FB}}\}.$$

*Remark.* Lemma 1 of [5] shows $\|\sin\Theta(U_1, \widehat{U}_1)\|_{\mathrm{FB}} = \|\Pi(U_1) - \Pi(\widehat{U}_1)\|_{\mathrm{FB}}/\sqrt{2}$. Thus, an immediate corollary of Wedin's Theorem is

$$\|\Pi(U_1) - \Pi(\widehat{U}_1)\|_{\mathrm{FB}} \le \sqrt{2}\delta^{-1}\|Z\|_{\mathrm{FB}}, \tag{39}$$

where the right-hand side is derived from $\|Z\widehat{V}_1\|_{\mathrm{FB}} \le \{\|Z\widehat{V}_1\|_{\mathrm{FB}}^2 + \|Z\widehat{V}_2\|_{\mathrm{FB}}^2\}^{1/2} = \|Z\|_{\mathrm{FB}}$ and likewise $\|\widehat{U}_1^\mathsf{T} Z\|_{\mathrm{FB}} \le \|Z\|_{\mathrm{FB}}$. This corollary gives an upper bound of the working measure of distance $\|\Pi(U_1) - \Pi(\widehat{U}_1)\|_{\mathrm{FB}}$, which will be useful in the proofs of Proposition 1 and Theorem 4 later. Furthermore, suppose $\Sigma_2 = \widehat{\Sigma}_2 = \mathbf{0}$, which means $\delta = \tau_{\min}(\widehat{\Sigma}_1)$, $\mathrm{M} = U_1\Sigma_1V_1^\mathsf{T}$, and $\widehat{\mathrm{M}} = \widehat{U}_1\widehat{\Sigma}_1\widehat{V}_1^\mathsf{T}$. Then (39) implies

$$\|\Pi(\mathrm{M}) - \Pi(\widehat{\mathrm{M}})\|_{\mathrm{FB}} = \|\Pi(U_1) - \Pi(\widehat{U}_1)\|_{\mathrm{FB}} \le \sqrt{2}\tau_{\min}^{-1}(\widehat{\mathrm{M}})\|\widehat{\mathrm{M}} - \mathrm{M}\|_{\mathrm{FB}}. \tag{40}$$

This will be useful in the proof of Theorem 3 later.

*Weyl's Theorem* [32] Suppose that M is a matrix from $\mathbb{R}^{p\times q}$ and $\widehat{\mathrm{M}}$ is a perturbation of M. For each $i \in \{1, \ldots, \min(p, q)\}$, we have

$$|\tau_i(\widehat{\mathrm{M}}) - \tau_i(\mathrm{M})| \le \tau_{\max}(\widehat{\mathrm{M}} - \mathrm{M}),$$

where $\tau_i(\cdot)$ is defined in Subsection A.1 above.

*Remark.* We will relax the upper bound $\tau_{\max}(\widehat{\mathrm{M}} - \mathrm{M})$ in this theorem to $\|\widehat{\mathrm{M}} - \mathrm{M}\|_{\mathrm{FB}}$ in the proofs later.

# Appendix B Proofs of the theorems and propositions

We now present in order the proofs of all the theorems and propositions in the main text, including Proposition 1, Proposition 2, Proposition 3, Theorem 2, Theorem 3, Proposition 4, and Theorem 4. The proof of Lemma 1 is omitted as it is straightforward by the definition of the rank of a matrix.

## B.1 Proof of Proposition 1

Write $\mathrm{M}_{\mathrm{F}} = (v_i)_{i\in\mathrm{F}}$ and $\widehat{\mathrm{M}}_{\mathrm{F}} = (\widehat{v}_i)_{i\in\mathrm{F}}$. The left singular vectors of $\widehat{\mathrm{M}}_{\mathrm{F}}$ are the same as the eigenvectors of $\widehat{\mathrm{M}}_{\mathrm{F}}^{\otimes 2}$ associated with nonzero eigenvalues. Thus, by simple algebra, we have

$$\begin{aligned}\widehat{\mathrm{M}}_{\mathrm{F}}^{\otimes 2} &= \textstyle\sum_{i\in\mathrm{F}} \widehat{v}_i^{\otimes 2} = \sum_{i\in\mathrm{F}}(\beta_0\Gamma_i^{\otimes 2}\beta_0^{\mathsf{T}} + \epsilon_i\Gamma_i^{\mathsf{T}}\beta_0^{\mathsf{T}} + \beta_0\Gamma_i\epsilon_i^{\mathsf{T}} + \epsilon_i^{\otimes 2})\\ &= [\{\beta_0 + (\textstyle\sum_{i\in\mathrm{F}}\epsilon_i\Gamma_i)^{\mathsf{T}}(\sum_{i\in\mathrm{F}}\Gamma_i^{\otimes 2})^{-1}\}(\sum_{i\in\mathrm{F}}\Gamma_i^{\otimes 2})^{1/2}]^{\otimes 2} + O_P(n^{-1})\\ &= \{\widetilde{\beta}_{\mathrm{F}}(\textstyle\sum_{i\in\mathrm{F}}\Gamma_i^{\otimes 2})^{1/2}\}^{\otimes 2} + O_P(n^{-1}).\end{aligned} \tag{41}$$

Denote $\widetilde{\beta}_{\mathrm{F}}(\sum_{i\in\mathrm{F}}\Gamma_i^{\otimes 2})^{1/2}$ by $\widetilde{\mathrm{M}}_{\mathrm{F}}$, which spans the same column space as $\widetilde{\beta}_{\mathrm{F}}$, i.e., with $\Pi(\widetilde{\mathrm{M}}_{\mathrm{F}}) = \Pi(\widetilde{\beta}_{\mathrm{F}})$, as $(\sum_{i\in\mathrm{F}}\Gamma_i^{\otimes 2})^{1/2}$ is invertible. Recall Wedin's theorem reviewed in Subsection A.5 of Appendix A above. By setting $U_1$, $\widehat{\mathrm{R}}$, $\mathrm{R}$, $\widehat{U}_1$ in (38) to be $\widehat{\beta}_{\mathrm{F}}$, $\widetilde{\mathrm{M}}_{\mathrm{F}}^{\otimes 2}$, $\widehat{\mathrm{M}}_{\mathrm{F}}^{\otimes 2}$, and the leading $d$ left singular vectors of $\widetilde{\mathrm{M}}_{\mathrm{F}}$, by (39), we have

$$\|\Pi(\widetilde{\beta}_{\mathrm{F}}) - \Pi(\widehat{\beta}_{\mathrm{F}})\|_{\mathrm{FB}}^2 \le 2\{\tau_d^2(\widetilde{\mathrm{M}}_{\mathrm{F}}) - \tau_{d+1}^2(\widehat{\mathrm{M}}_{\mathrm{F}})\}^{-2}\|\widehat{\mathrm{M}}_{\mathrm{F}}^{\otimes 2} - \widetilde{\mathrm{M}}_{\mathrm{F}}^{\otimes 2}\|_{\mathrm{FB}}^2 \tag{42}$$

with probability converging to one, From (41), we also have $\|\widehat{\mathrm{M}}_{\mathrm{F}}^{\otimes 2} - \widetilde{\mathrm{M}}_{\mathrm{F}}^{\otimes 2}\|_{\mathrm{FB}}^2 = O_P(n^{-1})$. Thus, to derive $\|\Pi(\widetilde{\beta}_{\mathrm{F}}) - \Pi(\widehat{\beta}_{\mathrm{F}})\|_{\mathrm{FB}} = O_P(n^{-1})$, it suffices to show $\tau_d^2(\widetilde{\mathrm{M}}_{\mathrm{F}}) - \tau_{d+1}^2(\widehat{\mathrm{M}}_{\mathrm{F}}) = \Omega_P(1)$. By Weyl's Theorem reviewed at the end of Section A.5, we have

$$\tau_d(\widetilde{\mathrm{M}}_{\mathrm{F}}) \ge \tau_d(\mathrm{M}_{\mathrm{F}}) + O_P(n^{-1/2}) = \Omega_P(1), \quad \tau_{d+1}(\widehat{\mathrm{M}}_{\mathrm{F}}) \le \tau_{d+1}(\mathrm{M}_{\mathrm{F}}) + O_P(n^{-1/2}) = o_P(1),$$

which together imply $\tau_d^2(\widetilde{\mathrm{M}}_{\mathrm{F}}) - \tau_{d+1}^2(\widehat{\mathrm{M}}_{\mathrm{F}}) = \Omega_P(1)$. This completes the proof. □

## B.2 Proof of Proposition 2

Let $\mathrm{M}_{[h]}$ be $\Sigma^{-1}(\Sigma - \Sigma_h)$ , and let $\mathrm{M}_w = (\mathrm{M}_{[h]}D_w^{1/2})_{h=1,\ldots,H}$ where $D_w$ is the diagonal matrix with $w$ being its diagonal. Let $\widetilde{\beta}_w$ be $\widetilde{\beta}_{\mathrm{F}}$ in Proposition 1 with M being $\mathrm{M}_w$ and F being $\mathcal{I}_{\mathrm{M}}$. Since $\mathrm{M}_{[h]}$ is a column transformation of $\beta_0$ and $\mathrm{M}_{[h]}\Sigma^{-1}$ is symmetric, there exists a scalar $a_h$ such that $\mathrm{M}_{[h]} = a_h\beta_0\beta_0^{\mathsf{T}}\Sigma$, and, for $\Gamma_i$'s defined in Proposition 1, we have $\Gamma_{[h]} \equiv (\Gamma_{(h-1)p+1}, \ldots, \Gamma_{hp})^{\mathsf{T}} = a_h\Sigma\beta_0$. Since $\mathrm{M}_{[h]}$ is a column transformation of $\beta_0$ and $\mathrm{M}_{[h]}\Sigma^{-1}$ is symmetric, there exists a scalar $a_h$ such that $\mathrm{M}_{[h]} = a_h\beta_0\beta_0^{\mathsf{T}}\Sigma$, and, for $\Gamma_i$'s defined in Proposition 1, we have $\Gamma_{[h]} \equiv (\Gamma_{(h-1)p+1}, \ldots, \Gamma_{hp})^{\mathsf{T}} = a_h\Sigma\beta_0$. Let $\epsilon_w$ be the estimation error of $\widetilde{\beta}_w$ as defined in Proposition 1, i.e.

$$\epsilon_w = (\textstyle\sum_{h=1}^H \epsilon_{[h]}D_w\Gamma_{[h]})/(\sum_{h=1}^H \Gamma_{[h]}^{\mathsf{T}}D_w\Gamma_{[h]}),$$

where $\epsilon_{[h]} \equiv \widehat{\mathrm{M}}_{[h]} - \mathrm{M}_{[h]}$ is the estimation error of the sample moment estimator $\widehat{\mathrm{M}}_{[h]}$. Clearly, $\epsilon_w = O_P(n^{-1/2})$. Let $Q(\beta) = I_p - \beta(\beta^{\mathsf{T}}\beta)^{-1}\beta^{\mathsf{T}}$, which reduces to $I_p - \beta\beta^{\mathsf{T}}$ if $\beta$ is semi-orthogonal. We have

$$\begin{aligned}&E(n\|\Pi(\widetilde{\beta}_w) - \Pi(\beta_0)\|_{\mathrm{FB}}^2) = nE[\mathrm{tr}\{\widetilde{\beta}_w(\widetilde{\beta}_w^{\mathsf{T}}\widetilde{\beta}_w)^{-1}\widetilde{\beta}_w^{\mathsf{T}} - \beta_0\beta_0^{\mathsf{T}}\}^2]\\ &= -2nE[\mathrm{tr}\{\widetilde{\beta}_w(\widetilde{\beta}_w^{\mathsf{T}}\widetilde{\beta}_w)^{-1}\widetilde{\beta}_w^{\mathsf{T}}\beta_0\beta_0^{\mathsf{T}}\} - 1] = -2nE\{(\beta_0^{\mathsf{T}}\widetilde{\beta}_w)^2/(\widetilde{\beta}_w^{\mathsf{T}}\widetilde{\beta}_w) - 1\}\\ &= -2nE\{(\beta_0^{\mathsf{T}}\epsilon_w + 1)^2/(\epsilon_w^{\mathsf{T}}\epsilon_w + 2\beta_0^{\mathsf{T}}\epsilon_w + 1) - 1\} = 2E\{n\epsilon_w^{\mathsf{T}}Q(\beta_0)\epsilon_w\} + o(1)\end{aligned}$$

$$= \frac{2\sum_{h,j=1,\dots,H}\Gamma_{[h]}^{\mathsf{T}}D_w E\{n\epsilon_{[h]}^{\mathsf{T}}Q(\beta_0)\epsilon_{[j]}\}D_w\Gamma_{[j]}}{(\sum_{h=1}^{H}\Gamma_{[h]}^{\mathsf{T}}D_w\Gamma_{[h]})^2} + o(1). \tag{43}$$

To calculate $E\{n\epsilon_{[h]}^{\mathsf{T}}Q(\beta_0)\epsilon_{[j]}\}$ for each $h, j = 1, \dots, H$, we first show

$$\begin{aligned} S_{hj} &\equiv (n/H)E\{(\widehat{\Sigma}_h - \Sigma_h)\Sigma^{-1}Q(\beta_0)\Sigma^{-1}(\widehat{\Sigma}_j - \Sigma_j)\} \\ &\to \delta(h = j)[Q(\beta_0) + \Sigma_h \mathrm{tr}\{Q(\beta_0)\Sigma^{-1}\}]. \end{aligned} \tag{44}$$

When $h \neq j$, this is obvious since the biases of $\widehat{\Sigma}_h$ and $\widehat{\Sigma}_j$ are $O(H/n)$ and the sample moments from different slices are mutually independent. Now suppose $h = j$. Let $E_{n,h}(\cdot)$ be the sample mean within the $h$th slice. By simple algebra, we have $\widehat{\Sigma}_h = E_{n,h}(X - \mu_h)^{\otimes 2} + O_P(n^{-1})$. Let $Z$ be a random vector following the $p$-dimensional standard multivariate normal distribution. For each $i, j = 1, \dots, p$, let $\delta_{ij}$ be the indicator of $i = j$. For $i, j, k, \ell = 1, \dots, p$, we have

$$E\{(Z_iZ_j - \delta_{ij})(Z_kZ_\ell - \delta_{k\ell})\} = \begin{cases} 1 & \text{if } i \neq \ell, \text{ and } j = \ell, k = i, \\ 1 & \text{if } i = \ell, \text{ and } j \neq i, k = j, \\ 2 & \text{if } i = \ell, \text{ and } j = k = i, \\ 0 & \text{otherwise.} \end{cases}$$

This means $E\{(Z^{\otimes 2} - I_p)B(Z^{\otimes 2} - I_p)\} = B + \mathrm{tr}(B)I_p$ for any $B \in \mathbb{R}^{p\times p}$, or equivalently

$$\begin{aligned} S_{hh} &\to \Sigma_h^{1/2}E\{(Z^{\otimes 2} - I_p)\Sigma_h^{1/2}\Sigma^{-1}Q(\beta_0)\Sigma^{-1}\Sigma_h^{1/2}(Z^{\otimes 2} - I_p)\}\Sigma_h^{1/2} \\ &= \Sigma_h^{1/2}\{\Sigma_h^{1/2}\Sigma^{-1}Q(\beta_0)\Sigma^{-1}\Sigma_h^{1/2} + I_p\mathrm{tr}(\Sigma_h^{1/2}\Sigma^{-1}Q(\beta_0)\Sigma^{-1}\Sigma_h^{1/2})\}\Sigma_h^{1/2} \\ &= \Sigma_h\Sigma^{-1}Q(\beta_0)\Sigma^{-1}\Sigma_h + \Sigma_h\mathrm{tr}\{\Sigma_h\Sigma^{-1}Q(\beta_0)\Sigma^{-1}\}. \end{aligned} \tag{45}$$

Because $\mathcal{S}(\Sigma^{-1}(\Sigma - \Sigma_h)) \subseteq \mathcal{S}(\beta_0)$, we have $Q(\beta_0)\Sigma^{-1}(\Sigma - \Sigma_h) = \mathbf{0}$. This means $\Sigma_h\Sigma^{-1}Q(\beta_0) = Q(\beta_0)$, which, together with (45), implies (44).

Recall that $E_n(\cdot)$ is the sample mean of a random element. Since $\widehat{\Sigma} = E_nX^{\otimes 2} + O_P(n^{-1})$ and $E_n(X^{\otimes 2}) = H^{-1}\sum_{h=1}^{H}E_{n,h}(X^{\otimes 2})$, we have

$$\widehat{\Sigma} - \Sigma = H^{-1}\textstyle\sum_{h=1}^{H}(\widehat{\Sigma}_h - \Sigma_h) + H^{-1}\sum_{h=1}^{H}(\widehat{\mu}_h^{\otimes 2} - \mu_h^{\otimes 2}) + O_P(n^{-1}). \tag{46}$$

Together with (44), the independence between $\widehat{\Sigma}_h$ and $\widehat{\mu}_h$, the mutual independence between the sample moments from different slices, and that $Q(\beta_0)\Sigma^{-1}\mu_h = \mathbf{0}$ from the consistency of SIR, (46) indicates

$$\begin{aligned} S_h &\equiv nE\{(\widehat{\Sigma}_h - \Sigma_h)\Sigma^{-1}Q(\beta_0)\Sigma^{-1}(\widehat{\Sigma} - \Sigma)\} \to Q(\beta_0) + \Sigma_h\mathrm{tr}\{Q(\beta_0)\Sigma^{-1}\}. \\ S &\equiv nE\{(\widehat{\Sigma} - \Sigma)\Sigma^{-1}Q(\beta_0)\Sigma^{-1}(\widehat{\Sigma} - \Sigma)\} \\ &= Q(\beta_0) + H^{-1}\textstyle\sum_{h=1}^{H}\Sigma_h\mathrm{tr}\{Q(\beta_0)\Sigma^{-1}\} \\ &\quad + nH^{-2}\textstyle\sum_{h=1}^{H}E\{(\widehat{\mu}_h^{\otimes 2} - \mu_h^{\otimes 2})\Sigma^{-1}Q(\beta_0)\Sigma^{-1}(\widehat{\mu}_h^{\otimes 2} - \mu_h^{\otimes 2})\} + O_P(n^{-1}) \\ &\to Q(\beta_0) + H^{-1}\textstyle\sum_{h=1}^{H}\Sigma_h\mathrm{tr}\{Q(\beta_0)\Sigma^{-1}\} + H^{-1}\sum_{h=1}^{H}\mu_h\mu_h^{\mathsf{T}}\mathrm{tr}\{\Sigma_h\Sigma^{-1}Q(\beta_0)\Sigma^{-1}\} \\ &= Q(\beta_0) + \Sigma\mathrm{tr}\{Q(\beta_0)\Sigma^{-1}\}. \end{aligned}$$

These together with (44) imply

$$\begin{aligned} &E\{n\epsilon_{[h]}^{\mathsf{T}}Q(\beta_0)\epsilon_{[j]}\} = E\{n(\widehat{\Sigma}^{-1}\widehat{\Sigma}_h - \Sigma^{-1}\Sigma_h)Q(\beta_0)(\widehat{\Sigma}^{-1}\widehat{\Sigma}_j - \Sigma^{-1}\Sigma_j)\} \\ &= E[n\{(\widehat{\Sigma}^{-1} - \Sigma^{-1})\widehat{\Sigma}_h + \Sigma^{-1}(\widehat{\Sigma}_h - \Sigma_h)\}^{\mathsf{T}}Q(\beta_0)\{(\widehat{\Sigma}^{-1} - \Sigma^{-1})\widehat{\Sigma}_j + \Sigma^{-1}(\widehat{\Sigma}_j - \Sigma_j)\}] \end{aligned}$$

$$
\begin{aligned}
&= E[n\{-\Sigma^{-1}(\widehat{\Sigma}-\Sigma)\Sigma^{-1}\Sigma_h+\Sigma^{-1}(\widehat{\Sigma}_h-\Sigma_h)\}^{\mathsf{T}}Q(\beta_0)\\
&\qquad\qquad\{-\Sigma^{-1}(\widehat{\Sigma}-\Sigma)\Sigma^{-1}\Sigma_j+\Sigma^{-1}(\widehat{\Sigma}_j-\Sigma_j)\}]+o(1)\\
&= E\{n\Sigma_h\Sigma^{-1}(\widehat{\Sigma}-\Sigma)\Sigma^{-1}Q(\beta_0)\Sigma^{-1}(\widehat{\Sigma}-\Sigma)\Sigma^{-1}\Sigma_j\}\\
&\qquad -E\{n\Sigma_h\Sigma^{-1}(\widehat{\Sigma}-\Sigma)\Sigma^{-1}Q(\beta_0)\Sigma^{-1}(\widehat{\Sigma}_j-\Sigma_j)\}\\
&\qquad -E\{n(\widehat{\Sigma}_h-\Sigma_h)\Sigma^{-1}Q(\beta_0)\Sigma^{-1}(\widehat{\Sigma}-\Sigma)\Sigma^{-1}\Sigma_j\}\\
&\qquad +E\{n(\widehat{\Sigma}_h-\Sigma_h)\Sigma^{-1}Q(\beta_0)\Sigma^{-1}(\widehat{\Sigma}_j-\Sigma_j)\}\\
&= \Sigma_h\Sigma^{-1}S\Sigma^{-1}\Sigma_j-\Sigma_h\Sigma^{-1}S_j-S_h\Sigma^{-1}\Sigma_j+HS_{hj}\\
&= \{\delta(h=j)H-1\}Q(\beta_0)+\Phi_{hj}.
\end{aligned}
$$

where $\Phi_{hj}=\mathrm{tr}\{Q(\beta_0)\Sigma^{-1}\}\{H\delta(h=j)\Sigma_h-\Sigma_h\Sigma^{-1}\Sigma_j\}$. By plugging this into (43), we have

$$
\begin{aligned}
&n\mathcal{D}(\mathcal{S}(\widetilde{\beta}_w),\mathcal{S}_{Y|X})\\
&\quad=\frac{2\sum_{h,j=1,\dots,H}\Gamma_{[h]}^{\mathsf{T}}D_w[\{H\delta(h=j)-1\}Q(\beta_0)+\Phi_{hj}]D_w\Gamma_{[j]}}{(\sum_{h=1}^{H}\Gamma_{[h]}^{\mathsf{T}}D_w\Gamma_{[h]})^2}\\
&\quad=\frac{2\beta_0^{\mathsf{T}}\Sigma D_w\{(a^{\mathsf{T}}\Delta_H a)Q(\beta_0)+\sum_{h,j=1,\dots,H}a_ha_j\Phi_{hj}\}D_w\Sigma\beta_0}{\{(\sum_{h=1}^{H}a_h^2)\beta_0^{\mathsf{T}}\Sigma D_w\Sigma\beta_0\}^2}\\
&\quad\equiv\frac{(w^{\mathsf{T}}\odot(\Sigma\beta_0)^{\mathsf{T}})\{c_1Q(\beta_0)+\Phi\}\{(\Sigma\beta_0)\odot w\}}{[w^{\mathsf{T}}\{(\Sigma\beta_0)\odot(\Sigma\beta_0)\}]^2},
\end{aligned}
\tag{47}
$$

where $a=(a_1,\dots,a_H)^{\mathsf{T}}$ and $\Delta_H$ is the $H$-dimensional symmetric matrix with diagonal entries being $H-1$ and off-diagonal entries being $-1$. Since the eigenvalues of $\Delta_H$ are $H,\dots,H,0$, $c_1$ is always nonnegative and it is zero if and only if $a_1=\dots=a_H$ or equivalently $\Sigma_1=\dots=\Sigma_H$. When $\mu_h$'s are zero, then $c_1=0$ means $\Sigma_h=\Sigma$ for all $h$ and subsequently $\mathrm{M}_{\mathrm{SAVE}}=\mathbf{0}$.

We now expand $\Phi$. Let $\Psi=\Sigma\beta_0\beta_0^{\mathsf{T}}\Sigma$. Then the denominator of the last term in (47) is $\{w^{\mathsf{T}}\odot(\Sigma\beta_0)^{\mathsf{T}}\}\Psi\{w\odot(\Sigma\beta_0)\}$. Thus, to prove (12) from (47), we only need to show $\Phi=c_1\mathrm{tr}\{Q(\beta_0)\Sigma^{-1}\}\Sigma+c\Psi$ for some scalar $c$. Since $\Sigma_h=\Sigma-a_h\Sigma\beta_0\beta_0^{\mathsf{T}}\Sigma$, we have

$$
\begin{aligned}
(\textstyle\sum_{h=1}^{H}a_h^2)\Phi &= \textstyle\sum_{h,j=1,\dots,H}a_ha_j\Phi_{hj}\\
&= \mathrm{tr}\{Q(\beta_0)\Sigma^{-1}\}\Big\{H\textstyle\sum_{h=1}^{H}a_h^2(\Sigma-a_h\Sigma\beta_0\beta_0^{\mathsf{T}}\Sigma)\\
&\qquad -\textstyle\sum_{h,j=1,\dots,H}a_ha_j(\Sigma-a_h\Sigma\beta_0\beta_0^{\mathsf{T}}\Sigma)\Sigma^{-1}(\Sigma-a_j\Sigma\beta_0\beta_0^{\mathsf{T}}\Sigma)\Big\}\\
&= \mathrm{tr}\{Q(\beta_0)\Sigma^{-1}\}\{(H\textstyle\sum_{h=1}^{H}a_h^2)\Sigma-(H\sum_{h=1}^{H}a_h^3)\Psi-(\sum_{h=1}^{H}a_h)^2\Sigma\\
&\qquad +2(\textstyle\sum_{h=1}^{H}a_h^2)(\sum_{h=1}^{H}a_h)\Psi-(\sum_{h=1}^{H}a_h^2)^2(\beta_0^{\mathsf{T}}\Sigma\beta_0)\Psi\}\\
&\equiv \mathrm{tr}\{Q(\beta_0)\Sigma^{-1}\}(a^{\mathsf{T}}\Delta_Ha)\Sigma+c'\Psi.
\end{aligned}
$$

This completes the proof.

□

## B.3 Justification of the special case

We now prove the optimality of $w=(1,0,\dots,0)^{\mathsf{T}}$ in minimizing (12) in the case $\beta=(1,0,\dots,0)^{\mathsf{T}}$, $c_1>0$, and $\Sigma$ conveys compound symmetry. Recall that $w$ must fall in $\mathcal{W}_p=\{(w_1,\dots,w_p)^{\mathsf{T}}\in\mathbb{R}^p:\sum_{i=1}^{p}w_i=1,w_1\geq 0,\dots,w_p\geq 0\}$. Write (12) as

$$
w^{\mathsf{T}}\Omega_1w/(w^{\mathsf{T}}\Omega_2w)\equiv f(w)/g(w),
$$

where the form of $\Omega_1$ and $\Omega_2$ can be found in the proof of Proposition 2. In particular, $(1, a^2, \ldots, a^2)^\top$ is the unique eigenvector of $\Omega_2$ that corresponds to a nonzero eigenvalue. By simple algebra, for any $a$ that makes $\Sigma$ positive semi-definite, i.e., $a \in (-1/(p-1), 1)$, $(1, a^2, \ldots, a^2)^\top$ is not an eigenvector of $\Omega_1$. Hence, there always exists nonzero $\delta \in \mathbb{R}^p$ such that $(1, a^2, \ldots, a^2)^\top \delta = 0$ and $(1, a^2, \ldots, a^2)^\top \Omega_1 \delta \neq 0$. This means, for any $w \in \mathbb{R}^p$, if we move $w$ in its local neighborhood along with direction $\delta$, then $g(w)$ is invariant but $w^\top W_1 w$ will have a monotone change. Therefore, any minimizer of $f(w)/g(w)$ must be on the edge of $\mathcal{W}$, i.e., at least one of $w_i$'s being 0 or 1. Suppose we restrict $w_i = 0$ for some $i > 1$, then we can remove this component and resolve the exactly same problem in $\mathcal{W}_{p-1}$, and derive the same result that the minimum must be reached on the edge of $\mathcal{W}_{p-1}$. By removing the next $w_i = 0$ for $i > 2$ and repeat the process, we will finally reach $w_1 = 1$. Therefore, the minimum of $f(w)/g(w)$ is reached only if $w_1 = 1$ or $w_1 = 0$ or $w_i = 1$ for some $i > 2$. By simple algebra, we have

$$\frac{f(w)}{g(w)} = \begin{cases} \operatorname{tr}\{Q(\beta_0)\Sigma^{-1}\} & \text{if } w_1 = 1 \\ a^{-4}[\operatorname{tr}\{Q(\beta_0)\Sigma^{-1}\}a^2\{1 + a\sum_{i\neq j} w_i w_j\} + a^2 \sum_{i=2}^p w_i^2] & \text{if } w_1 = 0 \\ a^{-4}[\operatorname{tr}\{Q(\beta_0)\Sigma^{-1}\}a^2 + a^2] & \text{if } w_i = 1 \text{ for } i > 1, \end{cases}$$

where the latter two can be easily verified to be strictly greater than $\operatorname{tr}\{Q(\beta_0)\Sigma^{-1}\}$, based on $a \in (-(p-1)^{-1}, 1)$ due to the non-singularity of $\Sigma$ and $\sum_{i\neq j} w_i w_j \in [(p-2)/(p-1), 1]$ when $w_1 = 0$. Hence, $f(w)/g(w)$ is minimized at $w = (1, 0, \ldots, 0)^\top$.

## B.4 Proof of Theorem 1

By the theoretical consistency of the predictor augmentation estimator [28], $\operatorname{rank}(\mathrm{M_F})$ is truly estimated for all $\mathrm{F} \in \mathcal{G}$ with the probability converging to one, which means $P(\widehat{\mathcal{F}} = \mathcal{F}) \to 1$. To ease the presentation, we assume $\widehat{\mathcal{F}} = \mathcal{F}$ in the rest of the proof if no ambiguity is caused. To start with, consider the case that $\lim_{n\to\infty} \mathcal{D}(\mathcal{S}(\widehat{\beta}_{\mathrm{F}}), \mathcal{S}_{Y|X})$ is uniquely minimized by G, under which it suffices to show $\widehat{\mathrm{G}} = \mathrm{G}$ with probability converging to one. Recall the functional form of $g(\cdot)$ in (13) of the main text. By Lemma 5, we have, for any $\mathrm{F} \subset \mathcal{I}_{\mathrm{M}}$ and almost surely $P_S$,

$$g(\mathrm{F}) = \lim_{n\to\infty} \mathcal{D}(\mathcal{S}(\widehat{\beta}_{\mathrm{F}}), \mathcal{S}_{Y|X}) + n^{-c}\mathcal{C}(\mathrm{F}) + O_P(n^{-1/2}c_n^{3/2}), \tag{48}$$

where both remainder terms are $o_P(1)$. Since G uniquely minimizes $\lim_{n\to\infty} \mathcal{D}(\mathcal{S}(\widehat{\beta}_{\mathrm{F}}), \mathcal{S}_{Y|X})$ and $\widehat{G}$ minimizes $g(\cdot)$, we have $\widehat{\mathrm{G}} = \mathrm{G}$ with probability converging to one. Now suppose $\lim_{n\to\infty} \mathcal{D}(\mathcal{S}(\widehat{\beta}_{\mathrm{F}}), \mathcal{S}_{Y|X})$ is minimized by each $\mathrm{F} \in \widetilde{\mathcal{F}}$, where $\widetilde{\mathcal{F}}$ includes other elements than G. Using the same argument as above, we still have $\widehat{G} \in \widetilde{\mathcal{F}}$ with probability converging to one. By (48), for any F in $\widetilde{\mathcal{F}}$ with $\mathcal{C}(\mathrm{F}) > \mathcal{C}(\mathrm{G})$, we have

$$g(\mathrm{F}) - g(\mathrm{G}) = n^{-c}\{\mathcal{C}(\mathrm{F}) - \mathcal{C}(\mathrm{G})\} + O_P(n^{-1/2}c_n^{3/2})$$

almost surely $P_S$, which means $g(\mathrm{F}) > g(\mathrm{G})$ with probability converging to one. Since $\widehat{G}$ minimizes $g(\cdot)$, $\widehat{G}$ must coincide with G with probability converging to one. This completes the proof.

## B.5 Proof of Proposition 3

We first prove (A6) for the sliced average variance estimation and directional regression. For $i = 1, \ldots, p$ and $h = 1, \ldots, H$, the corresponding column of $\mathrm{M_{SAVE}}$ is $\Sigma^{-1}(\Sigma - \Sigma_h)_i$. Let $e_i$ be the $i$th column of $I_p$. We have

$$\|\Sigma^{-1}(\Sigma - \Sigma_h)_i\|_{\mathrm{FB}} = \|e_i - \Sigma^{-1}\Sigma_h e_i\|_{\mathrm{FB}} \leq 1 + \|\Sigma^{-1}\Sigma_h e_i\|_{\mathrm{FB}} \leq 1 + \tau_{\max}(\Sigma^{-1}\Sigma_h), \tag{49}$$

which is $O(1)$ uniformly for $i$ and $h$ under Condition $(i)$. Since $\mathrm{M}_{\mathrm{DR}} = (\mathrm{M}_{\mathrm{SIR}}, \mathrm{M}_{\mathrm{SAVE}} + (\Sigma^{-1}\mu_h\mu_h^{\mathsf{T}})_{h=1,\ldots,H})$, and, by simple algebra, $\tau_{\max}(\Sigma^{-1}\mu_h\mu_h^{\mathsf{T}}) = \mu_h^{\mathsf{T}}\Sigma^{-1}\mu_h$ and thus is uniformly bounded for $h = 1, \ldots, H$ under Condition $(ii)$, the same argument as (49) applies for directional regression.

We next prove (A7) for the sliced average variance estimation and directional regression. By (35) in Lemma 6, (37) in Lemma 7, and the union bound, there exist $c_2, c_3 > 0$ such that

$$P\{\|\widehat{\Sigma} - \Sigma\|_\infty + \|\widehat{\Sigma}_{\mathrm{F}} - \Sigma_{\mathrm{F}}\|_\infty \geq c_2\sqrt{H\log(p)/n}\} \leq c_3 H\mathcal{C}(\mathrm{F})p^{-3}. \tag{50}$$

Since F is a small fixed set, $\mathcal{C}(\mathrm{F})$ is uniformly upper bounded by some positive integer $K$. In addition, by (A5), we can replace $Hp^{-3}$ with $p^{-1}(pH)^{-1}$ in (50) above, which then is (A7) for the sliced average variance estimation with $\nu = 1$. Under Condition $(iii)$, Equation (A.6) in [31] implies that, with probability greater than $1 - 4Hp^{-3}$,

$$\max_{1\leq h\leq H, 1\leq i\leq p} |e_i^{\mathsf{T}}(\widehat{\mu}_h - \mu_h)| \leq c_8\sqrt{H\log(p)/n}$$

for some positive scalar $c_8$. Together with (A5) and (50), this readily implies (A7) for directional regression with $\nu = 1$. □

## B.6 Proof of Theorem 2

By the definition of F being a small fixed set, which is mentioned at the beginning of Subsection 4.1 in the main text, we assume $\mathcal{C}(\mathrm{F})$ to be uniformly upper bounded by some fixed integer $K > 0$, as in the proof of Proposition 3.

We first prove statement $(i)$. For the most generality of the proof, we use the general $\lambda_n v_j$, and abuse the notation to denote the corresponding minimizer of $\mathcal{L}_{\mathrm{F}}(\cdot)$ by $\widehat{\mathrm{M}}_{\mathrm{F}}$. Since $\mathcal{L}_{\mathrm{F}}(\widehat{\mathrm{M}}_{\mathrm{F}}) \leq \mathcal{L}_{\mathrm{F}}(\mathrm{M}_{\mathrm{F}})$, by simple algebra, we have

$$\begin{aligned}
&-\mathrm{tr}\{(\widehat{\Lambda}_{\mathrm{F}} - \Lambda_{\mathrm{F}})^{\mathsf{T}}(\widehat{\mathrm{M}}_{\mathrm{F}} - \mathrm{M}_{\mathrm{F}})\} + \mathrm{tr}\{\mathrm{M}_{\mathrm{F}}^{\mathsf{T}}(\widehat{\Sigma} - \Sigma)(\widehat{\mathrm{M}}_{\mathrm{F}} - \mathrm{M}_{\mathrm{F}})\} \\
&\quad + \mathrm{tr}\{(\widehat{\mathrm{M}}_{\mathrm{F}} - \mathrm{M}_{\mathrm{F}})^{\mathsf{T}}\widehat{\Sigma}(\widehat{\mathrm{M}}_{\mathrm{F}} - \mathrm{M}_{\mathrm{F}})\}/2 + \lambda_n \textstyle\sum_{j=1}^p v_j\|\widehat{\mathrm{M}}_{\mathrm{F}}^j\|_{\mathrm{FB}} \\
&\equiv A_1 + A_2 + A_3/2 + \lambda_n \textstyle\sum_{j=1}^p v_j\|\widehat{\mathrm{M}}_{\mathrm{F}}^j\|_{\mathrm{FB}} \\
&\leq \lambda_n \textstyle\sum_{j=1}^p v_j\|\mathrm{M}_{\mathrm{F}}^j\|_{\mathrm{FB}}.
\end{aligned} \tag{51}$$

We next justify the bound of $A_1$ and $A_2$ with respect to $\sum_{i=1}^p \|\widehat{\mathrm{M}}_{\mathrm{F}}^i - \mathrm{M}_{\mathrm{F}}^i\|_{\mathrm{FB}}$, which, together with conditions for $\lambda_n v_j$ (see (57) below) and a modification of $A_3$, reformulate (51) as a regulation of $\|\widehat{\mathrm{M}}_{\mathrm{F}} - \mathrm{M}_{\mathrm{F}}\|_{\mathrm{FB}}$. For convenience, we abuse the notations to let S be $\mathcal{A}(\mathrm{M}_{\mathrm{F}})$ and let $\widehat{\delta}_i$ be $\widehat{\mathrm{M}}_{\mathrm{F}}^i - \mathrm{M}_{\mathrm{F}}^i$ for each $i = 1, \ldots, p$.

First, we have $\|\widehat{\Lambda}_{\mathrm{F}}^i - \Lambda_{\mathrm{F}}^i\|_{\mathrm{FB}} \leq \sqrt{\mathcal{C}(\mathrm{F})}\|\widehat{\Lambda}_{\mathrm{F}} - \Lambda_{\mathrm{F}}\|_\infty \leq \sqrt{K}\|\widehat{\Lambda}_{\mathrm{F}} - \Lambda_{\mathrm{F}}\|_\infty$ for each $i = 1, \ldots, p$. By (A7), this implies, with probability $1 - O(p^{-\nu}/\mathcal{C}(\mathcal{I}_{\mathrm{M}}))$, $\max_{i=1,\ldots,p}\|\widehat{\Lambda}_{\mathrm{F}}^i - \Lambda_{\mathrm{F}}^i\|_{\mathrm{FB}} \leq c_2\sqrt{HK\log(p)/n}$, which further implies

$$|A_1| \leq c_2\sqrt{HK\log(p)/n}\textstyle\sum_{i=1}^p \|\widehat{\delta}_i\|_{\mathrm{FB}}. \tag{52}$$

For $A_2$, note that by Lemma 6, we have

$$\|\widehat{\Sigma} - \Sigma\|_\infty \leq c_{(k)}\sqrt{\log(p)/n}$$

with probability $1 - O(p^{-k})$, where $k$ satisfies $p^{-k} = o(p^{-\nu}/\mathcal{C}(\mathcal{I}_{\mathrm{M}}))$ as permitted by the regularity (14). Together with (A6), this implies

$$|A_2| \leq \max_{i=1,\ldots,p}\|\mathrm{M}_{\mathrm{F}}^{\mathsf{T}}(\widehat{\Sigma} - \Sigma)_i\|_{\mathrm{FB}}\textstyle\sum_{i=1}^p \|\widehat{\delta}_i\|_{\mathrm{FB}}$$

$$\leq \sqrt{\mathcal{C}(\mathrm{F})s_{\mathrm{F}}}\|\widehat{\Sigma}-\Sigma\|_{\infty}(\max_{j\in\mathrm{F}}\|\mathrm{M}_i\|_{\mathrm{FB}})\sum_{i=1}^{p}\|\widehat{\delta}_i\|_{\mathrm{FB}}$$

$$\leq c_(8)c_{(k)}\sqrt{Ks_{\mathrm{F}}\log(p)/n}\sum_{i=1}^{p}\|\widehat{\delta}_i\|_{\mathrm{FB}}, \tag{53}$$

where $c_(8)$ denotes the upper bound of $\max\{\|\mathrm{M}_i\|_{\mathrm{FB}} : i \in \mathcal{I}_{\mathrm{M}}\}$ in (A6) and $s_{\mathrm{F}}$ denotes the cardinality of S, i.e., $\mathcal{A}(\mathrm{M}_{\mathrm{F}})$. Let

$$\widetilde{c}_n = c_2\sqrt{HK} + c_{(k)}c_{(8)}\sqrt{Ks_{\mathrm{F}}}, \tag{54}$$

which diverges with both $H$ and $s_{\mathrm{F}}$. By plugging (52) and (53) into (51), we have, with probability greater than $1 - c_6 p^{-\nu}/\mathcal{C}(\mathcal{I}_{\mathrm{M}})$ for some $c_6$,

$$A_3/2 + \lambda_n \sum_{i=1}^{p} v_j\|\widehat{\delta}_j\|_{\mathrm{FB}}$$

$$\leq \lambda_n \sum_{j\in\mathrm{S}} v_j\|\mathrm{M}_{\mathrm{F}}^{j}\|_{\mathrm{FB}} - \lambda_n \sum_{j=1}^{p} v_j\|\widehat{\mathrm{M}}_{\mathrm{F}}^{j}\|_{\mathrm{FB}} + \lambda_n \sum_{j=1}^{p} v_j\|\widehat{\delta}_j\|_{\mathrm{FB}}$$

$$+ \widetilde{c}_n\sqrt{\log(p)/n}\sum_{j=1}^{p}\|\widehat{\delta}_j\|_{\mathrm{FB}}$$

$$\leq 2\lambda_n \sum_{j\in\mathrm{S}} v_j\|\widehat{\delta}_j\|_{\mathrm{FB}} + \widetilde{c}_n\sqrt{\log(p)/n}\sum_{j=1}^{p}\|\widehat{\delta}_j\|_{\mathrm{FB}}, \tag{55}$$

where the last "$\leq$" is by the triangle inequality $\|\widehat{\delta}_j\|_{\mathrm{FB}} \geq |\,\|\widehat{\mathrm{M}}_{\mathrm{F}}^{j}\|_{\mathrm{FB}} - \|\mathrm{M}_{\mathrm{F}}^{j}\|_{\mathrm{FB}}\,|$ for each $j \in \mathrm{S}$ and that $\widehat{\mathrm{M}}_{\mathrm{F}}^{j} = \widehat{\delta}_j$ otherwise. The lower bound $1 - c_6 p^{-\nu}/\mathcal{C}(\mathcal{I}_{\mathrm{M}})$ for the probability above comes from the probability of $|A_1|$ being bounded (see (52) and (A7)) and that of $\|\widehat{\Sigma}-\Sigma\|_{\infty}$ being bounded (see Lemma 6), which are $1 - O(p^{-\nu}/\mathcal{C}(\mathcal{I}_{\mathrm{M}}))$ and $1 - O(p^{-k})$ with $p^{-k} = o(p^{-\nu}/\mathcal{C}(\mathcal{I}_{\mathrm{M}}))$, respectively, the former less than the latter. Clearly, (55) can be rewritten as

$$A_3/2 + \sum_{j\notin\mathrm{S}}(\lambda_n v_j - \widetilde{c}_n\sqrt{\log(p)/n})\|\widehat{\delta}_j\|_{\mathrm{FB}} \leq \sum_{j\in\mathrm{S}}(\lambda_n v_j + \widetilde{c}_n\sqrt{\log(p)/n})\|\widehat{\delta}_j\|_{\mathrm{FB}}. \tag{56}$$

If, for some $c > 1$ and some large enough $C' > 0$, we have

$$\lambda_n v_j \leq C'\widetilde{c}_n\sqrt{\log(p)/n} \text{ if } j \in \mathrm{S}, \quad \lambda_n v_j \geq c\widetilde{c}_n\sqrt{\log(p)/n} \text{ if } j \notin \mathrm{S}, \tag{57}$$

with probability greater than $1 - c_6 p^{-\nu}/\mathcal{C}(\mathcal{I}_{\mathrm{M}})$, where $c > 1$ will be used later, then we can replace $\lambda_n v_j - \widetilde{c}_n\sqrt{\log(p)/n}$ with zero and $\lambda_n v_j + \widetilde{c}_n\sqrt{\log(p)/n}$ with $(C'+1)\widetilde{c}_n\sqrt{\log(p)/n}$ in (56). This implies, with probability greater than $1 - c_6 p^{-\nu}/\mathcal{C}(\mathcal{I}_{\mathrm{M}})$,

$$|A_3| \leq 2(C'+1)\widetilde{c}_n\sqrt{\log(p)/n}\sum_{j\in\mathrm{S}}\|\widehat{\delta}_j\|_{\mathrm{FB}} \tag{58}$$

Let $B_3$ be $\mathrm{tr}\{(\widehat{\mathrm{M}}_{\mathrm{F}} - \mathrm{M}_{\mathrm{F}})^{\mathsf{T}}\Sigma(\widehat{\mathrm{M}}_{\mathrm{F}} - \mathrm{M}_{\mathrm{F}})\}$. We have

$$|A_3 - B_3| = |\mathrm{tr}\{(\widehat{\mathrm{M}}_{\mathrm{F}} - \mathrm{M}_{\mathrm{F}})^{\mathsf{T}}(\widehat{\Sigma}-\Sigma)(\widehat{\mathrm{M}}_{\mathrm{F}} - \mathrm{M}_{\mathrm{F}})\}|$$

$$\leq \sum_{i,j=1,\ldots,p}\|\widehat{\Sigma}-\Sigma\|_{\infty}\|\widehat{\delta}_i\|_{\mathrm{FB}}\|\widehat{\delta}_j\|_{\mathrm{FB}} = \|\widehat{\Sigma}-\Sigma\|_{\infty}(\sum_{j=1}^{p}\|\widehat{\delta}_j\|_{\mathrm{FB}})^2. \tag{59}$$

In addition, by (56), we have

$$\sum_{j\notin\mathrm{S}}\|\widehat{\delta}_j\|_{\mathrm{FB}} \leq \{(C'+1)/(c-1)\}\sum_{j\in\mathrm{S}}\|\widehat{\delta}_j\|_{\mathrm{FB}}, \tag{60}$$

which, together with (59), Lemma 6, and the Cauchy-Schwarz inequlaity, implies

$$|A_3 - B_3| \leq C''\|\widehat{\Sigma}-\Sigma\|_{\infty}(\sum_{j\in\mathrm{S}}\|\widehat{\delta}_j\|_{\mathrm{FB}})^2 \leq C''c_{(k)}s_{\mathrm{F}}\sqrt{\log(p)/n}\sum_{j\in\mathrm{S}}\|\widehat{\delta}_j\|_{\mathrm{FB}}^2 \tag{61}$$

with probability $1 - O(p^{-k})$, where $C'' = \{(C'+1)/(c-1)+1\}^2$. In conjunction with (58), we have

$$B_3 \leq |A_3| + |A_3 - B_3|$$

$$\leq 2(C'+1)\widetilde{c}_n\sqrt{\log(p)/n}\sum_{j\in S}\|\widehat{\delta}_j\|_{FB} + C''c_{(k)}s_F\sqrt{\log(p)/n}\sum_{j\in S}\|\widehat{\delta}_j\|^2_{FB} \tag{62}$$

with probability greater than $1-c_6p^{-\nu}/\mathcal{C}(\mathcal{I}_M)$. On the other hand, by construction, we have $B_3 \geq \tau_{\min}(\Sigma)\sum_{j=1}^p \|\widehat{\delta}_j\|^2_{FB}$, which means, with probability greater than $1-c_6p^{-\nu}/\mathcal{C}(\mathcal{I}_M)$,

$$\tau_{\min}(\Sigma)\sum_{j=1}^p\|\widehat{\delta}_j\|^2_{FB} \leq 2(C'+1)\widetilde{c}_n\sqrt{\log(p)/n}\sum_{j\in S}\|\widehat{\delta}_j\|_{FB} + C''c_{(k)}s_F\sqrt{\log(p)/n}\sum_{j\in S}\|\widehat{\delta}_j\|^2_{FB}. \tag{63}$$

Since $s_F\sqrt{\log(p)/n} = o(1)$ by (A5), this means, with probability greater than $1-c_6p^{-\nu}/\mathcal{C}(\mathcal{I}_M)$,

$$\begin{aligned}\tau_{\min}(\Sigma)\sum_{j=1}^p\|\widehat{\delta}_j\|^2_{FB} &\leq 3(C'+1)\widetilde{c}_n\sqrt{\log(p)/n}\sum_{j\in S}\|\widehat{\delta}_j\|_{FB}\\ &\leq 3(C'+1)\widetilde{c}_n\sqrt{s_F\log(p)/n}\Big(\sum_{j\in S}\|\widehat{\delta}_j\|^2_{FB}\Big)^{1/2}\\ &\leq 3(C'+1)\widetilde{c}_n\sqrt{s_F\log(p)/n}\Big(\sum_{j=1}^p\|\widehat{\delta}_j\|^2_{FB}\Big)^{1/2},\end{aligned} \tag{64}$$

where the second last inequality is derived by the Cauchy-Schwarz inequality. By (A5), this immediately implies, with probability greater than $1-c_6p^{-\nu}/\mathcal{C}(\mathcal{I}_M)$,

$$\begin{aligned}\|\widehat{M}_F - M_F\|_{FB} = \Big(\sum_{j=1}^p\|\widehat{\delta}_j\|^2_{FB}\Big)^{1/2} &\leq \tau^{-1}_{\min}(\Sigma)3(C'+1)\widetilde{c}_n\sqrt{s_F\log(p)/n}\\ &\leq C(H^{1/2}s_F^{1/2}+s_F)\{\log(p)/n\}^{1/2},\end{aligned} \tag{65}$$

which is the desired consistency of $\widehat{M}_F$.

We next verify (57) for $\lambda_n = \widetilde{\lambda}_n$ with $v_j = 1$ and $\lambda_n = \widehat{\lambda}_n$ with $v_j$ specified in (17), so that (65) can be applied to both $\widetilde{M}_F$ and $\widehat{M}_F^S$, which then proves statement (*i*) and part one of statement (*ii*). It is clear that (57) is satisfied for $\widetilde{\lambda}_n$ with $v_j = 1$, so the proof is omitted. To show that (57) is satisfied (probabilistically) for $\widehat{\lambda}_n$ with $v_j$ being specified in (17), by (S1), we have $\min_{j\in S}\|M_F^i\|_{FB} = \Omega(n^{-\phi})$; by (65) and (A5), we have $\max_{j\in S}\|\widetilde{M}_F^j - M_F^j\|_{FB} = O(n^{-\zeta}) = o(n^{-\phi})$ with probability greater than $1-c_6p^{-\nu}/\mathcal{C}(\mathcal{I}_M)$ for some $c_6$. These, together with the triangle inequality for $\|\cdot\|_{FB}$, imply $\min_{j\in S}\|\widetilde{M}_F^j\|_{FB} = \Omega(n^{-\phi})$ with probability greater than $1-c_6p^{-\nu}/\mathcal{C}(\mathcal{I}_M)$, which also means $\min_{j\in S}\|\widetilde{M}_F^j\|_{FB}\neq 0$ or equivalently $v_j = \|\widetilde{M}_F^j\|_{FB}^{-\rho}\neq 0$ for all $j\in S$ with probability greater than $1-c_6p^{-\nu}/\mathcal{C}(\mathcal{I}_M)$. Hence, we have, with probability greater than $1-c_6p^{-\nu}/\mathcal{C}(\mathcal{I}_M)$,

$$\max_{j\in S}\widehat{\lambda}_n v_j = \widehat{\lambda}_n/\min_{j\in S}\|\widetilde{M}_F^j\|^\rho_{FB} = c_4c_5n^{-\rho\phi}\widetilde{c}_n\sqrt{\log(p)/n}O(n^{\rho\phi}) \leq C'\widetilde{c}_n\sqrt{\log(p)/n} \tag{66}$$

for some large $C'$. Since (65) implies $\max_{j\notin S}\|\widetilde{M}_F^j\|_{FB} = O_P(\widetilde{c}_n\sqrt{s_F\log(p)/n})$, which is $O_P(n^{-\zeta})$ by (A5), we have, with probability greater than $1-c_6p^{-\nu}/\mathcal{C}(\mathcal{I}_M)$,

$$\begin{aligned}\min_{j\notin S}\widehat{\lambda}_n v_j &\geq c_4c_5n^{-\rho\phi}\widetilde{c}_n\sqrt{\log(p)/n}\min\{K,(\max_{j\notin S}\|\widetilde{M}_F^j\|_{FB})^{-\rho}\}\\ &= c_4c_5n^{-\rho\phi}\Omega(n^{\rho\zeta})\widetilde{c}_n\sqrt{\log(p)/n} \geq c\widetilde{c}_n\sqrt{\log(p)/n},\end{aligned} \tag{67}$$

where the last inequality is due to $\phi<\zeta$ as assumed in (S1), which makes $c_4c_5\Omega(n^{\rho(\zeta-\phi)})$ a diverging sequence of positive scalars. Hence, (65) also applies to $\widehat{M}_F^S$.

We next show the variable consistency part in statement (*ii*). Similarly to the above, we have, with probability converging to one, $\min_{j\in S}\|\widehat{M}_F^j\|_{FB}\neq 0$ or equivalently $\mathcal{A}(\widehat{M}_F^S)\supseteq S$. Assume that for some $j\notin S$, $\widehat{M}_F^j\neq \mathbf{0}$. Then, by the Karush-Kush-Tucker(KKT) conditions, we have

$$\|(\widehat{\Lambda}_F - \widehat{\Sigma}\widehat{M}_F^S)^j\|_{FB} = \widehat{\lambda}_n v_j, \tag{68}$$

where $v_j$ is specified in (17). Write

$$\widehat{\Lambda}_{\rm F} - \widehat{\Sigma}\widehat{\rm M}_{\rm F}^{\rm S} = (\widehat{\Lambda}_{\rm F} - \Lambda_{\rm F}) - (\widehat{\Sigma} - \Sigma)(\widehat{\rm M}_{\rm F}^{\rm S} - {\rm M}_{\rm F}) - \Sigma(\widehat{\rm M}_{\rm F}^{\rm S} - {\rm M}_{\rm F}) + (\widehat{\Sigma} - \Sigma){\rm M}_{\rm F}$$
$$\equiv A_{11} - A_{12} - A_{13} + A_{14}. \tag{69}$$

By (A7), we have $\|A_{11}^j\|_{\rm FB} = O[\{Hs_{\rm F}\log(p)/n\}^{1/2}]$ with probability $1 - O(p^{-\nu}/\mathcal{C}(\mathcal{I}_{\rm M}))$. By Lemma 6 and (65), we have $\|A_{12}^j\|_{\rm FB} = O\{(H^{1/2}s_{\rm F}^{1/2} + s_{\rm F})\log(p)/n\}$ with probability $1 - O(p^{-\nu}/\mathcal{C}(\mathcal{I}_{\rm M}))$. Since $A_{13}^j = \sum_{i=1}^p \sigma_{ji}(\widehat{\rm M}_{\rm F}^{\rm S} - {\rm M}_{\rm F})^i$, we have $\|A_{13}^j\|_{\rm FB} \leq \max_{j=1,\ldots,p}|\sigma_{ji}|\sum_{i=1}^p\|(\widehat{\rm M}_{\rm F}^{\rm S} - {\rm M}_{\rm F})^i\|_{\rm FB}$. By (A4), we have $\max_{i=1,\ldots,p}\sigma_{ii} = O(1)$. Since $|\sigma_{ij}| \leq \max\{\sigma_{ii},\sigma_{jj}\}$, we have $\max_{i=1,\ldots,p}|\sigma_{ji}| = O(1)$. In addition, by plugging (67) into (56), we have

$$\textstyle\sum_{j\notin {\rm S}}\|(\widehat{\rm M}_{\rm F}^{\rm S} - {\rm M}_{\rm F})^j\|_{\rm FB} = O_P(\sum_{j\in{\rm S}}\|(\widehat{\rm M}_{\rm F}^{\rm S} - {\rm M}_{\rm F})^j\|_{\rm FB});$$

by the Cauchy-Schwarz inequality, we have

$$\textstyle(\sum_{j\in{\rm S}}\|(\widehat{\rm M}_{\rm F}^{\rm S} - {\rm M}_{\rm F})^j\|_{\rm FB})^2 \leq s_{\rm F}\sum_{j\in{\rm S}}\|(\widehat{\rm M}_{\rm F}^{\rm S} - {\rm M}_{\rm F})^j\|_{\rm FB}^2.$$

These, together with (64), imply

$$\textstyle\|A_{13}^j\|_{\rm FB} \leq O_P(\sum_{j=1}^p\|(\widehat{\rm M}_{\rm F}^{\rm S} - {\rm M}_{\rm F})^j\|_{\rm FB}) = O_P(\sum_{j\in{\rm S}}\|(\widehat{\rm M}_{\rm F}^{\rm S} - {\rm M}_{\rm F})^j\|_{\rm FB})$$
$$\textstyle= O_P[\{s_{\rm F}\sum_{j\in{\rm S}}\|(\widehat{\rm M}_{\rm F}^{\rm S} - {\rm M}_{\rm F})^j\|_{\rm FB}^2\}^{1/2}] = O_P[(H + s_{\rm F})^{1/2}s_{\rm F}\{\log(p)/n\}^{1/2}],$$

where the extra $(H + s_{\rm F})^{1/2}$ is inherited from $\widetilde{c}_n$ defined in (54). Since ${\rm M}_{\rm F}$ only has $s_{\rm F}$ nonzero rows, we have

$$\textstyle\|A_{14}^j\|_{\rm FB} = \|\sum_{i\in{\rm S}}(\widehat{\sigma}_{ji} - \sigma_{ji}){\rm M}_{\rm F}^i\|_{\rm FB} \leq \|\widehat{\Sigma} - \Sigma\|_\infty\sqrt{s_{\rm F}K}\|{\rm M}_{\rm F}\|_\infty = O_P[\{s_{\rm F}\log(p)/n\}^{1/2}].$$

By plugging these results into (68), we have

$$\widehat{\lambda}_n v_j = O_P[(H + s_{\rm F})^{1/2}s_{\rm F}\{\log(p)/n\}^{1/2}] = O_P(n^{-\zeta}s_{\rm F}^{1/2}).$$

Since $\rho(\zeta-\phi) > (\log_n s_{\rm F})/2 - \zeta$, this means $\widehat{\lambda}_n v_j = o_P(n^{\rho(\zeta-\phi)})$. However, from (67), we also have $\widehat{\lambda}_n v_j = \Omega_P(n^{\rho(\zeta-\phi)})$. This contradiction implies $\widehat{\rm M}_{\rm F}^j = \mathbf{0}$ for all $j \notin {\rm S}$. This completes the proof. □

*Remark.* Note that (A5) implies $s = o(n^{(1-2\zeta)/2})$. Thus, if $\zeta > 1/6$, then we have

$$(\log_n s_{\rm F})/2 - \zeta \leq (\log_n s)/2 - \zeta \leq 1/4 - 3\zeta/2 < 0,$$

which means $\rho > \{(\log_n s_{\rm F})/2 - \zeta\}/(\zeta - \phi)$ as long as $\rho > 0$. Thus, the additional requirement on $\rho$ in Theorem 2(*ii*) can be omitted in this case.

## B.7 Proof of Theorem 3

For clarity, we use ${\rm F}(k)$ to denote F at the $k$th iteration. Let $Q(\cdot)$ be the same as defined in Proposition 2 above. We first prove the statement for the maximal element of $\mathcal{E}_k$. By the construction of $\mathcal{E}_k$, it suffices to show that, with probability converging to one, for any $0 \leq k < d$,

$$\max_{i\notin{\rm F}(k)}\|Q(\widetilde{\rm M}_{{\rm F}(k)})\widetilde{\rm M}_i\|_{\rm FB} = \Omega_P(n^{-\xi}). \tag{70}$$

Here, we abuse the notation to let $Q(\widetilde{\mathrm{M}}_{\mathrm{F}(0)}) = I_p$. We will prove (70) by the induction of both (70) and

$$\tau_{\min}(\mathrm{M}_{\mathrm{F}(k+1)}) = \Omega_P(n^{-\xi}), \tag{71}$$

regarded as a pair of conjugate statements. In (71), $\Omega_P(\cdot)$ is used instead of $\Omega(\cdot)$ due to the randomness of $\mathrm{M}_{\mathrm{F}(k+1)}$ inherited from the randomness of $\mathrm{F}(k+1)$, although M is nonrandom. First, let $k = 0$, and let $\mathrm{M}_{[1]}$ be such that $\|\mathrm{M}_{[1]}\|_{\mathrm{FB}} = \max_{i\in\mathcal{I}_{\mathrm{M}}} \|\mathrm{M}_i\|_{\mathrm{FB}}$. Since $\|\mathrm{M}_i\|^2_{\mathrm{FB}} \geq \mathrm{M}_i^{\mathsf{T}}\beta\beta^{\mathsf{T}}\mathrm{M}_i = (\beta^{\mathsf{T}}\mathrm{M}_i)^2$ for any $\beta \in \mathcal{S}(\mathrm{M})$ and for each $i \in \mathcal{I}_{\mathrm{M}}$, by (S2), we have $\|\mathrm{M}_{[1]}\|_{\mathrm{FB}} = \Omega(n^{-\xi})$. Together with (18) and Condition (A5), we have

$$\|\widetilde{\mathrm{M}}_{\mathrm{F}(1)}\|_{\mathrm{FB}} \geq \|\widetilde{\mathrm{M}}_{[1]}\|_{\mathrm{FB}} \geq \|\mathrm{M}_{[1]}\|_{\mathrm{FB}} + O_P(n^{-\zeta}) = \Omega_P(n^{-\xi}). \tag{72}$$

Thus (70) holds for $k = 0$, which also completes the proof if $d = 1$. Suppose $d > 1$, which means $\xi \leq \zeta/2$ in (S2). By (18) and (72), we have

$$\|\mathrm{M}_{\mathrm{F}(1)}\|_{\mathrm{FB}} \geq \|\widetilde{\mathrm{M}}_{\mathrm{F}(1)}\|_{\mathrm{FB}} + O_P(n^{-\zeta}) = \Omega_P(n^{-\xi}),$$

which implies (71) for $k = 0$. Suppose that, for some $h < d - 1$, both (70) and (71) hold for $k = 0, \ldots, h$. We next prove (70) for $k = h + 1$ and then prove (71) for $k = h + 1$. Let $\mathrm{M}_{[h+2]}$ be such that $\|Q(\mathrm{M}_{\mathrm{F}(h+1)})\mathrm{M}_{[h+2]}\|_{\mathrm{FB}} = \max_{i\in\mathcal{I}_{\mathrm{M}}} \|Q(\mathrm{M}_{\mathrm{F}(h+1)})\mathrm{M}_i\|_{\mathrm{FB}}$. Since $h+1 < d$, there must exist $\beta \in \mathcal{S}(\mathrm{M})$ such that $\beta^{\mathsf{T}}\mathrm{M}_{\mathrm{F}(h+1)} = \mathbf{0}$, which, by (S2), means

$$\|Q(\mathrm{M}_{\mathrm{F}(h+1)})\mathrm{M}_{[h+2]}\|_{\mathrm{FB}} \geq |\beta^{\mathsf{T}}Q(\mathrm{M}_{\mathrm{F}(h+1)})\mathrm{M}_{[h+2]}| = |\beta^{\mathsf{T}}\mathrm{M}_{[h+2]}| = \Omega(n^{-\xi}). \tag{73}$$

In addition, recall Wedin's Theorem and its corollaries reviewed in Subsection A.5 above. By setting $\widehat{\mathrm{M}}$ and M in (40) to be $\mathrm{M}_{\mathrm{F}(h+1)}$ and $\widetilde{\mathrm{M}}_{\mathrm{F}(h+1)}$, respectively, we have

$$\|\Pi(\widetilde{\mathrm{M}}_{\mathrm{F}(h+1)}) - \Pi(\mathrm{M}_{\mathrm{F}(h+1)})\|_{\mathrm{FB}} \leq \sqrt{2}\tau^{-1}_{\min}(\mathrm{M}_{\mathrm{F}(h+1)})\|\widetilde{\mathrm{M}}_{\mathrm{F}(h+1)} - \mathrm{M}_{\mathrm{F}(h+1)}\|_{\mathrm{FB}}. \tag{74}$$

By plugging (71) for $k = h$ and $\|\widetilde{\mathrm{M}}_{\mathrm{F}(h+1)} - \mathrm{M}_{\mathrm{F}(h+1)}\|_{\mathrm{FB}} = O_P(n^{-\zeta})$, the latter implied by (18) and Condition (A5), (74) indicates

$$\|\Pi(\widetilde{\mathrm{M}}_{\mathrm{F}(h+1)}) - \Pi(\mathrm{M}_{\mathrm{F}(h+1)})\|_{\mathrm{FB}} = O_P(n^{-\zeta+\xi}). \tag{75}$$

Together with Theorem 2, we have, uniformly for $i \notin \mathrm{F}(h+1)$,

$$\begin{aligned}
&\|Q(\widetilde{\mathrm{M}}_{\mathrm{F}(h+1)})\widetilde{\mathrm{M}}_i - Q(\mathrm{M}_{\mathrm{F}(h+1)})\mathrm{M}_i\|_{\mathrm{FB}} \\
&\quad= \|(\widetilde{\mathrm{M}}_i - \mathrm{M}_i) - \{\Pi(\widetilde{\mathrm{M}}_{\mathrm{F}(h+1)}) - \Pi(\mathrm{M}_{\mathrm{F}(h+1)})\}\widetilde{\mathrm{M}}_i + \Pi(\mathrm{M}_{\mathrm{F}(h+1)})(\widetilde{\mathrm{M}}_i - \mathrm{M}_i)\|_{\mathrm{FB}} \\
&\quad\leq \|\widetilde{\mathrm{M}}_i - \mathrm{M}_i\|_{\mathrm{FB}} + \|\Pi(\widetilde{\mathrm{M}}_{\mathrm{F}(h+1)}) - \Pi(\mathrm{M}_{\mathrm{F}(h+1)})\|_{\mathrm{FB}}\|\widetilde{\mathrm{M}}_i\|_{\mathrm{FB}} + \|\widetilde{\mathrm{M}}_i - \mathrm{M}_i\|_{\mathrm{FB}} \\
&\quad= O_P(n^{-\zeta}) + O_P(n^{-\zeta+\xi})O_P(1 + n^{-\zeta}) + O_P(n^{-\zeta}) \\
&\quad= O_P(n^{-\zeta+\xi}).
\end{aligned} \tag{76}$$

Putting (73) and (76) together, we have

$$\begin{aligned}
&\max_{i\notin\mathrm{F}(h+1)} \|Q(\widetilde{\mathrm{M}}_{\mathrm{F}(h+1)})\widetilde{\mathrm{M}}_i\|_{\mathrm{FB}} \geq \|Q(\widetilde{\mathrm{M}}_{\mathrm{F}(h+1)})\widetilde{\mathrm{M}}_{[h+2]}\|_{\mathrm{FB}} \\
&\quad= \|Q(\mathrm{M}_{\mathrm{F}(h+1)})\mathrm{M}_{[h+2]}\|_{\mathrm{FB}} + O_P(n^{-\zeta+\xi}) = \Omega_P(n^{-\xi}) + O_P(n^{-\zeta+\xi}) = \Omega_P(n^{-\xi}),
\end{aligned} \tag{77}$$

where the last equality is due to $\xi \leq \zeta/2$ imposed in (S2). This proves (70) for $k = h + 1$. Note that the proof of (77) requires (75), which requires (71) for $k = h$. This is the reason that we include (71) in the induction as a conjugate to (70). Next, we use (70) for $k = 0, \dots, h+1$, where the case of $k = h+1$ is proved in (77), to prove (71) for $k = h+1$. To this end, we introduce

$$\Upsilon^{(h+2)} \equiv (\mathrm{M}_{\mathrm{S}(1)}, Q(\mathrm{M}_{\mathrm{F}(1)})\mathrm{M}_{\mathrm{S}(2)}, ..., Q(\mathrm{M}_{\mathrm{F}(h+1)})\mathrm{M}_{\mathrm{S}(k+2)}), \tag{78}$$

which is an invertible column transformation of $\mathrm{M}_{\mathrm{F}(h+2)}$. Since $\Upsilon^{(h+2)}$ and $\mathrm{M}_{\mathrm{F}(h+2)}$ span the same column space, we have $\mathrm{M}_{\mathrm{F}(h+2)} = \Pi(\Upsilon^{(h+2)})\mathrm{M}_{\mathrm{F}(h+2)}$, which means

$$\begin{aligned}\tau_{\min}(\mathrm{M}_{\mathrm{F}(h+2)}) &= \tau_{\min}\left[\Upsilon^{(h+2)}\{(\Upsilon^{(h+2)})^{\mathsf{T}}\Upsilon^{(h+2)}\}^{-1}(\Upsilon^{(h+2)})^{\mathsf{T}}\mathrm{M}_{\mathrm{F}(h+2)}\right] \\ &\geq \tau_{\min}(\Upsilon^{(h+2)})\tau_{\min}\left[\{(\Upsilon^{(h+2)})^{\mathsf{T}}\Upsilon^{(h+2)}\}^{-1}(\Upsilon^{(h+2)})^{\mathsf{T}}\mathrm{M}_{\mathrm{F}(h+2)}\right].\end{aligned} \tag{79}$$

Based on (79), we next prove (71) for $k = h+1$ by showing

$$\tau_{\min}(\Upsilon^{(h+2)}) = \Omega_P(n^{-\xi}), \tag{80}$$

$$\tau_{\min}\left[\{(\Upsilon^{(h+2)})^{\mathsf{T}}\Upsilon^{(h+2)}\}^{-1}(\Upsilon^{(h+2)})^{\mathsf{T}}\mathrm{M}_{\mathrm{F}(h+2)}\right] = \Omega_P(1), \tag{81}$$

which then would complete the induction for both (70) and (71) and prove (70) for $k = 0, \dots, d$, i.e., the first part of the theorem.

Proof of (80) Since the columns of $\Upsilon^{(h+2)}$ are orthogonal to each other, $(\Upsilon^{(h+2)})^{\mathsf{T}}\Upsilon^{(h+2)}$ must be a diagonal matrix, and we have

$$\tau_{\min}(\Upsilon^{(h+2)}) = \min\{\|Q(\mathrm{M}_{\mathrm{F}(k)})\mathrm{M}_{\mathrm{S}(k+1)}\|_{\mathrm{FB}} : k = 0, ..., h+1\}, \tag{82}$$

where we again abuse $Q(\mathrm{M}_{\mathrm{F}(0)})$ to denote $I_p$. To tackle $\|Q(\mathrm{M}_{\mathrm{F}(k)})\mathrm{M}_{\mathrm{S}(k+1)}\|_{\mathrm{FB}}$, recall that $\mathrm{S}(k+1)$ is the newly selected element in $\mathrm{F}(k+1)$, i.e., $\mathrm{F}(k+1) = \mathrm{F}(k) \cup \{\mathrm{S}(k+1)\}$, which means

$$\|Q(\widetilde{\mathrm{M}}_{\mathrm{F}(k)})\widetilde{\mathrm{M}}_{\mathrm{S}(k+1)}\|_{\mathrm{FB}} = \max_{j \notin \mathrm{F}(k)} \|Q(\widetilde{\mathrm{M}}_{\mathrm{F}(k)})\widetilde{\mathrm{M}}_j\|_{\mathrm{FB}}. \tag{83}$$

By (70) for $k = 0, \dots, h+1$ and (76), we have

$$\begin{aligned}\|Q(\mathrm{M}_{\mathrm{F}(k)})\mathrm{M}_{\mathrm{S}(k+1)}\|_{\mathrm{FB}} &\geq \|Q(\widetilde{\mathrm{M}}_{\mathrm{F}(k)})\widetilde{\mathrm{M}}_{\mathrm{S}(k+1)}\|_{\mathrm{FB}} + O_P(n^{-\zeta+\xi}) \\ &= \Omega_P(n^{-\xi}) + O_P(n^{-\zeta+\xi}) = \Omega_P(n^{-\xi})\end{aligned} \tag{84}$$

uniformly for $k = 0, \dots, h+1$, which, by (82), implies (80).

Proof of (81) Intuitively, (81) is about the "magnitude" of the column transformation from $\Upsilon^{(h+1)}$ to $\mathrm{M}_{\mathrm{F}(h+1)}$. We next verify it by Lemma 8 in Subsection A.4 above. Let $i, j$ be arbitrary integers in $\{1, \dots, h+2\}$. Denote the $i$th column of $\Upsilon^{(h+2)}$ by $\Upsilon_i^{(h+2)}$. Since $(\Upsilon_i^{(h+2)})^{\mathsf{T}}\mathrm{M}_{\mathrm{S}(j)} = \mathrm{M}_{\mathrm{S}(i)}^{\mathsf{T}}Q(\mathrm{M}_{\mathrm{F}(i-1)})\mathrm{M}_{\mathrm{S}(j)}$, which is 0 if $i > j$, $(\Upsilon^{(h+2)})^{\mathsf{T}}\mathrm{M}_{\mathrm{F}(h+2)}$ is an upper-triangular matrix. Together with that $(\Upsilon^{(h+2)})^{\mathsf{T}}\Upsilon^{(h+2)}$ is a diagonal matrix as mentioned above, $\{(\Upsilon^{(h+2)})^{\mathsf{T}}\Upsilon^{(h+2)}\}^{-1}(\Upsilon^{(h+2)})^{\mathsf{T}}\mathrm{M}_{\mathrm{F}(h+2)}$ in (81) is also an upper-triangular matrix. In addition, as $(\Upsilon_i^{(h+2)})^{\mathsf{T}}\mathrm{M}_{\mathrm{S}(i)} = (\Upsilon_i^{(h+2)})^{\mathsf{T}}\Upsilon_i^{(h+2)}$, each diagonal element of $\{(\Upsilon^{(h+2)})^{\mathsf{T}}\Upsilon^{(h+2)}\}^{-1}(\Upsilon^{(h+2)})^{\mathsf{T}}\mathrm{M}_{\mathrm{F}(h+2)}$ is exactly equal to one. Thus, by Lemma 8, (81) holds as long as the upper-half off-diagonal elements of $\{(\Upsilon^{(h+2)})^{\mathsf{T}}\Upsilon^{(h+2)}\}^{-1}(\Upsilon^{(h+2)})^{\mathsf{T}}\mathrm{M}_{\mathrm{F}(h+2)}$ are uniformly $O_P(1)$, that is,

$$(\Upsilon_i^{(h+2)})^{\mathsf{T}}\mathrm{M}_{\mathrm{S}(j)}/(\Upsilon_i^{(h+2)})^{\mathsf{T}}(\Upsilon_i^{(h+2)}) = O_P(1) \text{ uniformly for } i < j. \tag{85}$$

Recall that (71) holds for $k = 0, \ldots, h$ at this stage and (76) is implied by (71) for $k = h$. Thus, with the same argument, (76) also holds uniformly for $k = 0, \ldots, h$ if we replace $h$ with $k$ in (76). Together with (83), $\zeta < \xi/2$, and (70) for $k = 0, \ldots, h+1$, this implies

$$\begin{aligned}
&|(\Upsilon_i^{(h+2)})^\top \mathrm{M}_{\mathrm{S}(j)}/(\Upsilon_i^{(h+2)})^\top(\Upsilon_i^{(h+2)})| \\
&\quad = |\mathrm{M}_{\mathrm{S}(i)}^\top Q(\mathrm{M}_{\mathrm{F}(i-1)})\mathrm{M}_{\mathrm{S}(j)}/\mathrm{M}_{\mathrm{S}(i)}^\top Q(\mathrm{M}_{\mathrm{F}(i-1)})\mathrm{M}_{\mathrm{S}(i)}| \\
&\quad = |\widetilde{\mathrm{M}}_{\mathrm{S}(i)}^\top Q(\widetilde{\mathrm{M}}_{\mathrm{F}(i-1)})\widetilde{\mathrm{M}}_{\mathrm{S}(j)}/\widetilde{\mathrm{M}}_{\mathrm{S}(i)}^\top Q(\widetilde{\mathrm{M}}_{\mathrm{F}(i-1)})\widetilde{\mathrm{M}}_{\mathrm{S}(i)}| + o_P(1) \\
&\quad \leq 1 + o_P(1)
\end{aligned} \tag{86}$$

uniformly for $i, j = 1, \ldots, h+2$ with $i < j$. Thus, (81) holds.

We next prove $\mathcal{S}(\mathrm{M}_\mathrm{R}) = \mathcal{S}_{Y|X}$ with probability converging to one. Recall

$$\Upsilon^{(d)} \equiv (\mathrm{M}_{\mathrm{S}(1)}, Q(\mathrm{M}_{\mathrm{F}(1)})\mathrm{M}_{\mathrm{S}(2)}, \ldots, Q(\mathrm{M}_{\mathrm{F}(d-1)})\mathrm{M}_{\mathrm{S}(d)})$$

with $\mathcal{S}(\Upsilon^{(d)}) = \mathcal{S}(\mathrm{M}_{\mathrm{F}(d)})$. By (84), we have $\|Q(\mathrm{M}_{\mathrm{F}(h)})\mathrm{M}_{\mathrm{S}(h+1)}\|_{\mathrm{FB}} = \Omega_P(n^{-\xi})$ uniformly for $h = 1, \ldots, d-1$, which means that $\Upsilon^{(d)}$ has full column rank with probability converging to one. Since $\mathcal{S}_{Y|X}$ is $d$-dimensional and

$$\mathcal{S}(\Upsilon^{(d)}) = \mathcal{S}(\mathrm{M}_{\mathrm{F}(d)}) \subseteq \mathcal{S}(\mathrm{M}_\mathrm{R}) \subseteq \mathcal{S}(\mathrm{M}) = \mathcal{S}_{Y|X},$$

where the last equality is induced by the coverage condition, $\Upsilon$ having full column rank must imply $\mathcal{S}(\mathrm{M}_\mathrm{R}) = \mathcal{S}_{Y|X}$. Thus, we have $\mathcal{S}(\mathrm{M}_\mathrm{R}) = \mathcal{S}_{Y|X}$ with probability converging to one. This completes the proof. □

## B.8 Proof of Proposition 4

Recall that $\beta_0$ is an orthonormal basis of $\mathcal{S}(\mathrm{M})$. Thus, we have $\mathrm{M}_j = \Pi(\beta_0)\mathrm{M}_j = \beta_0\beta_0^\top \mathrm{M}_j$ for each $j \in \mathcal{I}_\mathrm{M}$ and subsequently $\mathrm{M}_j^i = \beta_0^i\beta_0^\top \mathrm{M}_j$ for each $i \in A$. Let $u_i = \beta_0^i/\|\beta_0^i\|_{\mathrm{FB}} \in \mathbb{R}^{1\times d}$. Since $u_i\beta_0^\top\beta_0 u_i^\top = 1$, we can regard $\beta_0 u_i^\top$ as some $\beta \in \mathcal{S}_{Y|X}$ with $\beta^\top\beta = 1$ under the coverage condition (2). Let $c_u = \min\{\|\beta_0^i\|_{\mathrm{FB}} : i \in A\}$. Together with (S2) and (19), these imply

$$\begin{aligned}
\min\{\max\{|\mathrm{M}_j^i| : j \in \mathcal{I}_\mathrm{M}\} : i \in A\} &= \min\{\max\{|\beta_0^i\beta_0^\top \mathrm{M}_j| : j \in \mathcal{I}_\mathrm{M}\} : i \in A\} \\
&\geq \min\{\max\{c_u \cdot |u_i\beta_0^\top \mathrm{M}_j| : j \in \mathcal{I}_\mathrm{M}\} : i \in A\} \\
&= c_u \cdot \min\{\max\{|u_i\beta_0^\top \mathrm{M}_j| : j \in \mathcal{I}_\mathrm{M}\} : i \in A\} \\
&\geq c_u \cdot \min\{\max\{|\beta^\top \mathrm{M}_j| : j \in \mathcal{I}_\mathrm{M}\} : \beta \in \mathcal{S}_{Y|X}, \beta^\top\beta = 1\} \\
&= \Omega(n^{-\omega}) \cdot \Omega(n^{-\xi}) = \Omega(n^{-(\xi+\omega)}) = \Omega(n^{-\zeta}).
\end{aligned}$$

This completes the proof. □

## B.9 Proof of Theorem 4

We will first prove statements $(i)$ and $(iii)$, and then prove statement $(ii)$. For simplicity of notations, we remove the superscript S from $\widehat{\mathrm{M}}_\mathrm{R}^\mathrm{S}$ and simply denote it by $\widehat{\mathrm{M}}_\mathrm{R}$ throughout the proof.

*Proof of (i) and (iii)* In short, these two statements are modifications of Theorem 2$(ii)$ where F is replaced with the random index set R and $\mathcal{A}(\mathrm{M}_\mathrm{F})$ is replaced with $A$. Because the proof of Theorem 2 involves $\mathrm{M}_\mathrm{F}$ only through Conditions (A7) and (S1), the same proof can be applied to deliver statements $(i)$ and $(\dddot{u})$ for $\widehat{\mathrm{M}}_{\mathrm{R}}$ as long as, first, the event $\mathcal{S}(\mathrm{M}_{\mathrm{R}}) = \mathcal{S}_{Y|X}$ holds with probability converging to one, which has been justified in Theorem 3, and, second, both (A7) and (S1) are satisfied for R after being adjusted for its randomness; that is,

$$P(\|\widehat{\Lambda}_{\mathrm{R}} - \Lambda_{\mathrm{R}}\|_{\infty} > c_2\{H\log(p)/n\}^{1/2}) \leq c_3 p^{-\nu}, \tag{87}$$

$$\min\{\|\mathrm{M}^i_{\mathrm{R}}\|_{\mathrm{FB}} : i \in A\} = \Omega_P(n^{-\phi}). \tag{88}$$

Note that the probability is $c_3p^{-\nu}$ in (87) rather than $c_3p^{-\nu}/\mathcal{C}(\mathcal{I}_{\mathrm{M}})$ in (A7), and that it is $\Omega_P(\cdot)$ used in (88) rather than $\Omega(\cdot)$ used in (S1). The latter is again due to the randomness of R, which makes the event $A(\mathrm{M}_{\mathrm{R}}) = A$ uncertain (but still with probability converging to one) and makes $\min\{\|\mathrm{M}^i_{\mathrm{R}}\|_{\mathrm{FB}} : i \in A(\mathrm{M}_{\mathrm{R}})\}$, the original term that appears in (S1) if we replace F with R, a random variable itself. Accordingly, given (87) and (88), we need to adjust the convergence rates of the probabilities of relevant events when parallelizing the proof of Theorem 2$(\ddot{u})$ for statements $(i)$ and $(\dddot{u})$ here. These include:

- the probability of $|A_1|$ in (52) (with F replaced by R) being bounded changed to $1 - O(p^{-\nu})$,
- the probability of $\|\widehat{\mathrm{M}}_{\mathrm{R}} - \mathrm{M}_{\mathrm{R}}\|$ in (65) (again with F replaced by R) being bounded changed to $1 - o_P(1)$,
- all the probabilities induced from these two.

The details of the parallelized proof are omitted to avoid repetitive presentation. We next prove (87) and (88) sequentially, which is all that suffices. For simplicity, we will omit the phrase "with probability converging to one" whenever appropriate.

*Proof of (87)* This is straightforward as, first, regardless of the nature of randomness of R, we always have

$$\|\widehat{\Lambda}_{\mathrm{R}} - \Lambda_{\mathrm{R}}\|_{\infty} \leq \max\{\|\widehat{\Lambda}_i - \Lambda_i\|_{\infty} : i \in \mathcal{I}_{\mathrm{M}}\},$$

and, second, by letting F in (A7) be each singleton $\{i\} \subset \mathcal{I}_{\mathrm{M}}$, we have

$$\begin{aligned} P(\max\{\|\widehat{\Lambda}_i - \Lambda_i\|_{\infty} : i \in \mathcal{I}_{\mathrm{M}}\} > c_2\{H\log(p)/n\}^{1/2}) \\ \leq P(\bigcup_{i\in\mathcal{I}_{\mathrm{M}}} \{\|\widehat{\Lambda}_i - \Lambda_i\|_{\infty} > c_2\{H\log(p)/n\}^{1/2}\}) \\ \leq \textstyle\sum_{i\in\mathcal{I}_{\mathrm{M}}} P(\|\widehat{\Lambda}_i - \Lambda_i\|_{\infty} > c_2\{H\log(p)/n\}^{1/2}) \\ \leq \mathcal{C}(\mathcal{I}_{\mathrm{M}})c_3p^{-\nu}/\mathcal{C}(\mathcal{I}_{\mathrm{M}}) = c_3p^{-\nu}. \end{aligned}$$

*Proof of (88)* Using the same notations as in the proof of Theorem 3, let $\mathrm{F}(k)$ be the index set of selected columns of $\widehat{\mathrm{M}}$ at the $k$th iteration. As $\mathrm{F}(1) \subset \ldots \subset \mathrm{F}(d) \subseteq \mathrm{R}$, we will prove the stronger statement

$$\min\{\|\mathrm{M}^i_{\mathrm{F}(d)}\|_{\mathrm{FB}} : i \in A\} = \Omega_P(n^{-\phi}). \tag{89}$$

To achieve this, we will first show that, under (S3),

$$\min\{\|\Upsilon^i\|_{\mathrm{FB}} : i \in A\} = \Omega_P(n^{-\phi}), \tag{90}$$

where $\Upsilon$ denotes $\Upsilon^{(d)}$ defined in (78) for simplicity and $\Upsilon^i$ denotes the $i$th row of $\Upsilon$. Then, we will regulate the "magnitude" of the column transformation from $\Upsilon$ to $\mathrm{M}_{\mathrm{F}(d)}$ similarly to (81). These together readily imply (89).

Under (S3), for each $i \in A$, there exists $\sigma(i) \in \mathcal{I}_{\mathrm{M}}$ such that $|\mathrm{M}^{i}_{\sigma(i)}| = \Omega(n^{-\phi})$ uniformly over $A$. As $\Upsilon$ spans $\mathcal{S}_{Y|X}$ (see the end of the proof of Theorem 3 above), each $\mathrm{M}_{\sigma(i)}$ must fall in $\mathcal{S}(\Upsilon)$. Thus, we have $\mathrm{M}_{\sigma(i)} = \Upsilon\Xi_{\sigma(i)}$ where $\Xi_{\sigma(i)} = (\Upsilon^{\mathsf{T}}\Upsilon)^{-1}\Upsilon^{\mathsf{T}}\mathrm{M}_{\sigma(i)}$. This means $|\mathrm{M}^{i}_{\sigma(i)}| \le \|\Upsilon^{i}\|_{\mathrm{FB}}\|\Xi_{\sigma(i)}\|_{\mathrm{FB}}$, which, together with $|\mathrm{M}^{i}_{\sigma(i)}| = \Omega(n^{-\phi})$, implies

$$\begin{aligned}\min\{\|\Upsilon^{i}\|_{\mathrm{FB}} : i \in A\} &= \min\{|\mathrm{M}^{i}_{\sigma(i)}|/\|\Xi_{\sigma(i)}\|_{\mathrm{FB}} : i \in A\}\\ &\ge \min\{|\mathrm{M}^{i}_{\sigma(i)}| : i \in A\}\ \min\{\|\Xi_{\sigma(i)}\|^{-1}_{\mathrm{FB}} : i \in A\}\\ &= \Omega(n^{-\phi})\min\{\|\Xi_{\sigma(i)}\|^{-1}_{\mathrm{FB}} : i \in A\}.\end{aligned}$$

Hence, (90) holds if $\min\{\|\Xi_{\sigma(i)}\|^{-1}_{\mathrm{FB}} : i \in A\} = \Omega_P(1)$ or equivalently

$$\max\{\|\Xi_{\sigma(i)}\|_{\mathrm{FB}} : i \in A\} = O_P(1). \tag{91}$$

Since the columns of $\Upsilon$ are orthogonal to each other, for $h = 1, \dots, d$, the $h$th element of $\Xi_{\sigma(i)}$ is

$$\Xi^{h}_{\sigma(i)} = \Upsilon^{\mathsf{T}}_{h}\mathrm{M}_{\sigma(i)}/(\Upsilon^{\mathsf{T}}_{h}\Upsilon_{h}) = \mathrm{M}^{\mathsf{T}}_{\mathrm{S}(h)}Q(\mathrm{M}_{\mathrm{F}(h-1)})\mathrm{M}_{\sigma(i)}/(\Upsilon^{\mathsf{T}}_{h}\Upsilon_{h}),$$

where we abuse the notations to denote $I_p$ by $Q(\mathrm{M}_{\mathrm{F}(0)})$. If $\mathrm{M}_{\sigma(i)}$ falls in $\mathcal{S}(\mathrm{M}_{\mathrm{F}(h-1)})$, then $Q(\mathrm{M}_{\mathrm{F}(h-1)})\mathrm{M}_{\sigma(i)} = 0$ and thus $\Xi^{h}_{\sigma(i)} = 0$. Otherwise, using the same argument as in (86) in the proof of Theorem 3, we have $|\Xi^{h}_{\sigma(i)}| = 1 + o_P(1)$ uniformly for $i \in A$ and $h = 1, \dots, d$. These imply (91) and consequently (90).

To connect (90) with (89), note that, as $\Upsilon$ is an invertible column transformation of $\mathrm{M}_{\mathrm{F}(d)}$, we have $\mathrm{M}_{\mathrm{F}(d)} = \Upsilon\{(\Upsilon^{\mathsf{T}}\Upsilon)^{-1}\Upsilon^{\mathsf{T}}\mathrm{M}_{\mathrm{F}(d)}\}$. This means

$$\begin{aligned}\|\mathrm{M}^{i}_{\mathrm{F}(d)}\|_{\mathrm{FB}} &= \|(\Upsilon^{i}/\|\Upsilon^{i}\|_{\mathrm{FB}})(\Upsilon^{\mathsf{T}}\Upsilon)^{-1}\Upsilon^{\mathsf{T}}\mathrm{M}_{\mathrm{F}(d)}\|_{\mathrm{FB}}\,\|\Upsilon^{i}\|_{\mathrm{FB}}\\ &\ge \tau_{\min}\{(\Upsilon^{\mathsf{T}}\Upsilon)^{-1}\Upsilon^{\mathsf{T}}\mathrm{M}_{\mathrm{F}(d)}\}\,\|\Upsilon^{i}\|_{\mathrm{FB}}.\end{aligned} \tag{92}$$

uniformly for $i \in A$. Referring to (81), we have $\tau_{\min}\{(\Upsilon^{\mathsf{T}}\Upsilon)^{-1}\Upsilon^{\mathsf{T}}\mathrm{M}_{\mathrm{F}(d)}\} = \Omega_P(1)$, which, together with (90), implies (89). This completes the proof of statements (*i*) and (*iii*).

*Proof of (ii)* Recall Wedin's Theorem reviewed in Subsection A.5 above. If we regard $\widehat{\beta}^{\mathrm{S}}_{\mathrm{R}}$, $\beta_0$, $\widehat{\mathrm{M}}_{\mathrm{R}}$, and $\mathrm{M}_{\mathrm{R}}$ as $\widehat{U}_1$, $U_1$, $\widehat{\mathrm{R}}$, and R, respectively, then, since $\mathrm{M}_{\mathrm{R}}$ has rank $d$ and consequently $\tau_{d+1}(\mathrm{M}_{\mathrm{R}}) = 0$, we have

$$\|\Pi(\widehat{\beta}^{\mathrm{S}}_{\mathrm{R}}) - \Pi(\beta_0)\|_{\mathrm{FB}} \le \sqrt{2}\tau^{-1}_{d}(\widehat{\mathrm{M}}_{\mathrm{R}})\|\widehat{\mathrm{M}}_{\mathrm{R}} - \mathrm{M}_{\mathrm{R}}\|_{\mathrm{FB}},$$

as long as $\tau_d(\widehat{\mathrm{M}}_{\mathrm{R}}) > 0$. Hence, given statement (*i*) of the theorem, it now suffices to prove $\tau_d(\widehat{\mathrm{M}}_{\mathrm{R}}) = \Omega_P(n^{-\xi})$. Since $\mathrm{F}(d) \subseteq \mathrm{R}$, we have

$$v^{\mathsf{T}}\widehat{\mathrm{M}}_{\mathrm{R}}\widehat{\mathrm{M}}^{\mathsf{T}}_{\mathrm{R}}v = \textstyle\sum_{i\in\mathrm{R}} v^{\mathsf{T}}\widehat{\mathrm{M}}_{i}\widehat{\mathrm{M}}^{\mathsf{T}}_{i}v \ge \sum_{i\in\mathrm{F}(d)} v^{\mathsf{T}}\widehat{\mathrm{M}}_{\mathrm{F}(i)}\widehat{\mathrm{M}}^{\mathsf{T}}_{\mathrm{F}(i)}v \tag{93}$$

for any $v \in \mathbb{R}^{p}$, which means $\tau_d(\widehat{\mathrm{M}}_{\mathrm{R}}) \ge \tau_d(\widehat{\mathrm{M}}_{\mathrm{F}(d)}) = \tau_{\min}(\widehat{\mathrm{M}}_{\mathrm{F}(d)})$. Thus, it suffices to show $\tau_{\min}(\widehat{\mathrm{M}}_{\mathrm{F}(d)}) = \Omega_P(n^{-\xi})$. Referring to (71) in the proof of Theorem 3, we have $\tau_{\min}(\mathrm{M}_{\mathrm{F}(d)}) = \Omega_P(n^{-\xi})$. Together with Weyl's Theorem reviewed in Subsection A.5 above and $\|\widehat{\mathrm{M}}_{\mathrm{F}(d)} - \mathrm{M}_{\mathrm{F}(d)}\|_{\mathrm{FB}} = O_P(n^{-\zeta})$ justified in statement (*i*) of the theorem, this implies

$$\begin{aligned}\tau_{\min}(\widehat{\mathrm{M}}_{\mathrm{F}(d)}) &\ge \tau_{\min}(\mathrm{M}_{\mathrm{F}(d)}) - \|\widehat{\mathrm{M}}_{\mathrm{F}(d)} - \mathrm{M}_{\mathrm{F}(d)}\|_{\mathrm{FB}}\\ &= \Omega_P(n^{-\xi}) + O_P(n^{-\zeta}) = \Omega_P(n^{-\xi}),\end{aligned}$$

where the last equality comes from $\xi < \zeta$. This completes the proof of statement (*ii*).

## Appendix C Complementary simulation studies

We now conduct a run time comparison between the proposed sparsity-adapted column selection approach and the existing sparse SDR methods under high-dimensional settings. For simplicity, we only take $\mathrm{M} = (\mathrm{M}_{\mathrm{SIR}}, \mathrm{M}_{\mathrm{SAVE}})$ and omit the easier case $\mathrm{M} = \mathrm{M}_{\mathrm{SAVE}}$ for the proposed approach, and we use Model Ⅷ with $\Sigma = \Sigma_a$ in Subsection 7.2 of the main text as an example. In addition, we will evaluate the impact of the number of slices $H$ on the consistency of both the efficiency-adapted column selection approach under low-dimensional settings and the sparsity-adapted column selection approach under high-dimensional settings.

For simplicity, we set $n = 200$, $H = 5$ for $\mathrm{M}_{\mathrm{SIR}}$, and $H = 2$ for $\mathrm{M}_{\mathrm{SAVE}}$ as in Subsection 7.2 of the main text. Because the computational time of a sparse SDR method grows with the dimension $p$ of $X$, we set $p$ to be each of 600, 900, 1200, 1500, and 1800. For fairness of comparison, we set 10 candidate choices for each tuning parameter in each sparse SDR method. These include the multiplicative scalars for both the nuclear norm penalty and the group lasso penalty in the sparse SIR-based method proposed in [47], and that for each time of minimizing (20) in our approach, etc. We also set the cross validation for tuning these parameters to be uniformly five-fold. In favor of both the sparse SIR-based method and the sparse sliced average variance estimation proposed in [31], we additionally assume $d$ to be known in their implementation, whose estimation therein would require conducting sparse SDR for multiple times. For each sparse SDR method, we take the logarithm of its computational time measured in seconds, and record its average over 200 independent runs in Figure 2. Each run is conducted on an Intel Core i7-11800H CPU with 16 cores, operating at 2.3 GHz.

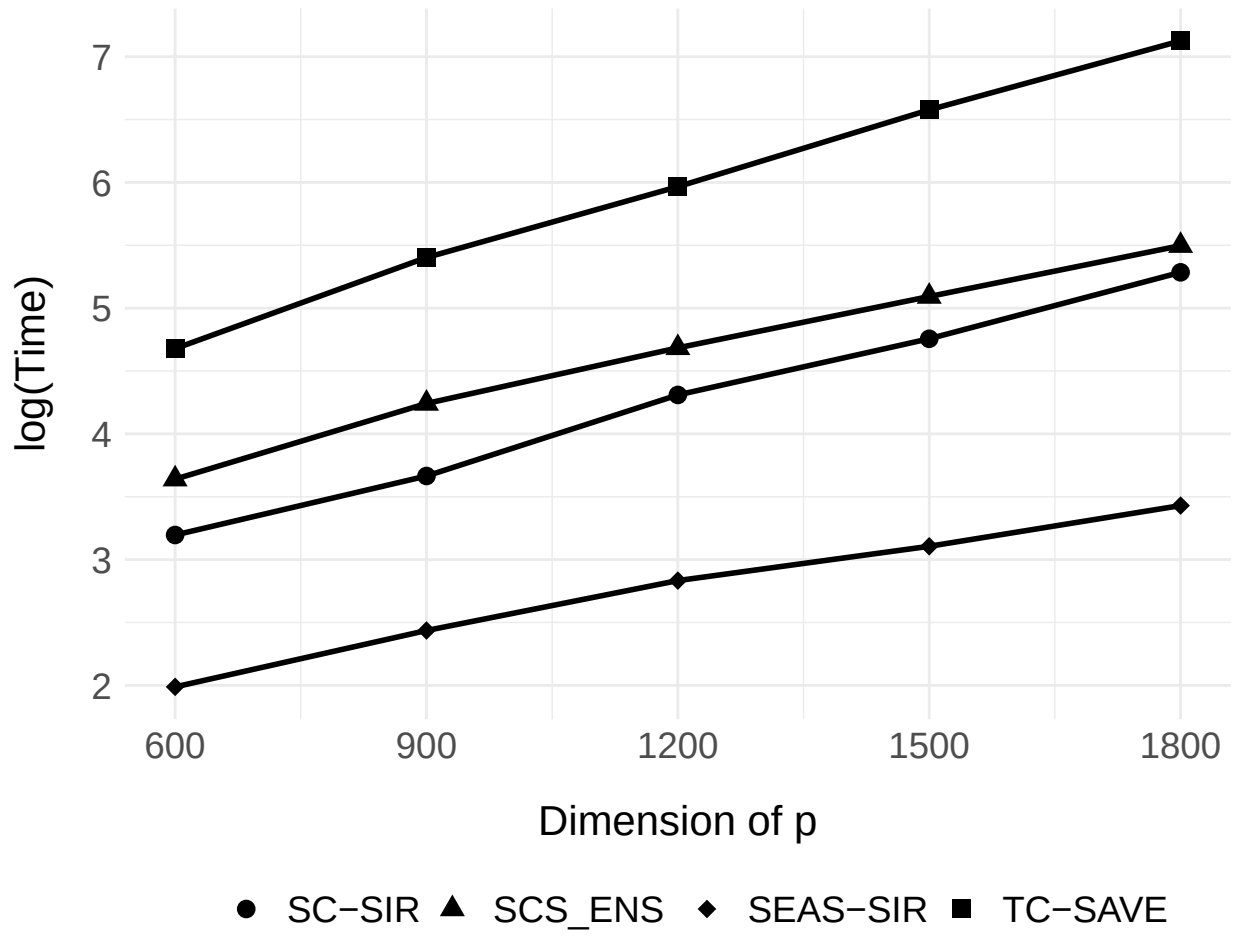

**Fig. 2.** The averages of log(computational time) of the sparse SDR methods, measured in log(seconds), over 200 independent runs. The working model is Model Ⅷ in Section 7 of the main text, with $\Sigma = \Sigma_a$, $H = 5$ for $\mathrm{M}_{\mathrm{SIR}}$, $H = 2$ for $\mathrm{M}_{\mathrm{SAVE}}$, $n = 200$, and $p$ being each of 600, 900, 1200, 1500, and 1800. The methods "SEAS-SIR", "SC-SIR", "TC-SAVE", and "SCS-ENS" stand for the sparse SIR-based method proposed in [47], that proposed in [31], the sparse sliced average variance estimation proposed in [31], and the proposed sparsity-adapted column selection approach applied to $\mathrm{M} = (\mathrm{M}_{\mathrm{SIR}}, \mathrm{M}_{\mathrm{SAVE}})$, respectively.

From Figure 2, the sparse SIR-based method proposed in [47] is clearly the fastest among all, which complies with the theoretical anticipation as it directly targets at the simple $M_{SIR}$ that only has five columns and it uses the well-studied penalized least squares estimation. With the initial column estimation and column selection as a prior stage, our column selection approach naturally requires more time than the sparse SIR-based method proposed in [47], namely about four times slower regardless of $p$. Interestingly, it is almost equally fast as the sparse SIR-based method proposed in [31]. In addition to the non-convex minimization that slows down the latter, this owes to three factors:

- the fast-screening mechanism [34] and the warm start mechanism [15] built in the `R` package `glmnet` that accelerate the minimization of (20),
- a simple forward column selection procedure that mainly involves calculating inner products of vectors in $\mathbb{R}^p$ and finding the maximum in a set of $O(pH)$ elements,
- the predictor augmentation estimator that quickly determines $d$.

For the same reasons, our approach is much faster than the sparse sliced average variance estimation proposed in [31]. Considering the large number of columns of $M_{SAVE}$ in comparison to $M_{SIR}$, these results suggest the computational effectiveness of the sparsity-adapted column selection approach as a generalization of the higher-order inverse regression methods under high-dimensional settings.

We next evaluate the impact of $H$ on the proposed efficiency-adapted and sparsity-adapted column selection approaches. First, we look into the efficiency-adapted column selection approach applied to $M = M_{SAVE}$, under the same model settings and the same measure of performance as in Subsection 7.1 of the main text. Instead of setting $H = 5$, we raise $H$ to each of 10 and 20. The results are summarized in Table 3 and Table 4, respectively.

**Table 3.** Performance of the efficiency-adapted column selection approach applied to the sliced average variance estimation when $H = 10$. The notations and the meanings of the numbers follow those in Table 1 of the main text.

| Model | $\Sigma_2$ | | | | | $\Sigma_8$ | | | | |
|---|---|---|---|---|---|---|---|---|---|---|
| | $p$ | Efficiency | # Col. | TSR | Opt | $p$ | Efficiency | # Col. | TSR | Opt |
| I | 6 | 0.83(0.009) | 10.1 | 1.00 | 0.83 | 6 | 0.94(0.009) | 19.1 | 0.81 | 0.89 |
| | 10 | 0.79(0.008) | 13.4 | 1.00 | 0.76 | 10 | 0.95(0.007) | 24.8 | 0.63 | 0.93 |
| II | 6 | 1.24(0.017) | 33.8 | 0.56 | 1.00 | 6 | 1.04(0.007) | 16.0 | 0.27 | 1.00 |
| | 10 | 1.26(0.014) | 32.7 | 0.33 | 1.00 | 10 | 1.01(0.001) | 21.2 | 0.21 | 1.00 |
| III | 6 | 0.89(0.008) | 26.8 | 1.00 | 0.87 | 6 | 0.95(0.008) | 19.1 | 0.79 | 0.90 |
| | 10 | 0.75(0.008) | 23.1 | 1.00 | 0.75 | 10 | 0.96(0.004) | 25.9 | 0.76 | 0.91 |
| IV | 6 | 0.87(0.011) | 24.6 | 1.00 | 0.86 | 6 | 0.89(0.011) | 22.8 | 0.79 | 0.78 |
| | 10 | 0.79(0.007) | 22.7 | 1.00 | 0.79 | 10 | 0.95(0.011) | 29.1 | 0.77 | 0.83 |

**Table 4.** Performance of the efficiency-adapted column selection approach applied to the sliced average variance estimation when $H = 20$. The notations and the meanings of the numbers follow those in Table 1 of the main text.

| Model | $\Sigma_2$ | | | | | $\Sigma_8$ | | | | |
|---|---|---|---|---|---|---|---|---|---|---|
| | $p$ | Efficiency | # Col. | TSR | Opt | $p$ | Efficiency | # Col. | TSR | Opt |
| I | 6 | 0.73(0.009) | 20.2 | 1.00 | 0.73 | 6 | 0.96(0.009) | 35.5 | 0.72 | 0.89 |
| | 10 | 0.67(0.009) | 24.5 | 1.00 | 0.66 | 10 | 0.93(0.007) | 32.4 | 0.76 | 0.92 |
| II | 6 | 1.49(0.009) | 33.2 | 0.28 | 1.00 | 6 | 1.01(0.001) | 27.8 | 0.23 | 1.00 |
| | 10 | 1.14(0.008) | 46.8 | 0.23 | 1.00 | 10 | 1.00(0.002) | 34.2 | 0.17 | 1.00 |
| III | 6 | 0.77(0.009) | 44.9 | 1.00 | 0.75 | 6 | 0.94(0.007) | 38.4 | 0.71 | 0.89 |
| | 10 | 0.70(0.007) | 43.2 | 1.00 | 0.70 | 10 | 0.96(0.004) | 46.6 | 0.63 | 0.92 |
| IV | 6 | 0.68(0.008) | 40.4 | 1.00 | 0.68 | 6 | 0.93(0.011) | 47.7 | 0.81 | 0.78 |
| | 10 | 0.56(0.009) | 43.3 | 1.00 | 0.55 | 10 | 0.96(0.011) | 44.8 | 0.65 | 0.86 |

Comparing the results in Tables 3 and 4 with those in Table 1 of the main text, the same phenomena can be observed for Models I, III, and IV. Referring to the model settings discussed in Subsection 7.1 of the main text, these suggest the robustness of the efficiency-adapted column selection approach to the choice of $H$ when the optimal submatrix of M have a small number of columns. For Model II where the ambient $\mathrm{M}_{\mathrm{SAVE}}$ is favorable over all of its submatrices, the proposed column selection approach becomes less effective as $H$ grows when $\Sigma = \Sigma_2$. This is caused by the practical need of precisely approximating $\varphi_0(\mathrm{G}) = \varphi_0(\mathcal{I}_{\mathrm{M}})$ for Model II and the ineffectiveness of the corresponding bootstrap estimation in (13) that is commonly recognized when the number of parameters ($\approx 0.5p^2H$) is large with respect to the sample size $n$. From an omitted simulation study, it is alleviated if we increase $n$ from 200 to 500.

For Models VII and VIII in Subsection 7.2 of the main text, we evaluate the sparsity-adapted column selection approach applied to both $\mathrm{M} = \mathrm{M}_{\mathrm{SAVE}}$ and $\mathrm{M} = (\mathrm{M}_{\mathrm{SIR}}, \mathrm{M}_{\mathrm{SAVE}})$, with $(n, p) = (200, 200)$ and $H$ in both $\mathrm{M}_{\mathrm{SIR}}$ and $\mathrm{M}_{\mathrm{SAVE}}$ raised to each of 5, 8, and 10. Since the sparse SIR-based method proposed in [47] outperforms that proposed in [31] in Table 2 of the main text, we only include the former as the benchmark of the comparison, in favor of which we again assume $d$ to be known in its implementation. Referring to the findings in the run time comparison above, we also omit the sparse sliced average variance estimation proposed in [31] due to its excessive computational cost as $H$ grows. The results based on 200 independent runs are summarized in Table 5, in the same format as Table 2 of the main text.

**Table 5.** Performance of the sparsity-adapted column selection approach when $H = 5$, 8, and 10. The notations and the meanings of the numbers follow those in Table 2 of the main text.

| Model | $H$ | Method | $\Sigma_a$ | | | | $\Sigma_b$ | | | |
|---|---|---|---|---|---|---|---|---|---|---|
| | | | Eff. | ACv. | ICv. | Col. | Eff. | ACv. | ICv. | Col. |
| VII | 5 | SEAS-SIR | 0.095(0.004) | 0.00 | 0.80 | – | 0.094(0.005) | 0.00 | 1.24 | – |
| | | SCS-SAVE | 0.105(0.005) | 0.00 | 0.07 | 3.35 | 0.103(0.005) | 0.00 | 0.04 | 3.64 |
| | | SCS-ENS | 0.086(0.004) | 0.00 | 0.02 | 4.54 | 0.094(0.004) | 0.00 | 0.06 | 4.42 |
| | 8 | SEAS-SIR | 0.102(0.005) | 0.00 | 1.49 | – | 0.095(0.005) | 0.00 | 2.61 | – |
| | | SCS-SAVE | 0.101(0.005) | 0.00 | 0.03 | 4.55 | 0.097(0.005) | 0.00 | 0.08 | 4.71 |
| | | SCS-ENS | 0.097(0.005) | 0.00 | 0.14 | 5.61 | 0.088(0.004) | 0.00 | 0.12 | 5.45 |
| | 10 | SEAS-SIR | 0.092(0.004) | 0.00 | 2.04 | – | 0.098(0.005) | 0.00 | 3.13 | – |
| | | SCS-SAVE | 0.112(0.005) | 0.00 | 0.04 | 5.14 | 0.106(0.005) | 0.00 | 0.05 | 4.85 |
| | | SCS-ENS | 0.093(0.004) | 0.00 | 0.01 | 6.82 | 0.093(0.004) | 0.00 | 0.03 | 6.52 |
| VIII | 5 | SEAS-SIR | 0.252(0.008) | 0.00 | 3.40 | – | 0.253(0.008) | 0.00 | 3.52 | – |
| | | SCS-SAVE | 0.532(0.024) | 0.64 | 0.82 | 4.78 | 0.549(0.026) | 0.78 | 0.44 | 5.49 |
| | | SCS-ENS | 0.240(0.008) | 0.02 | 0.45 | 7.77 | 0.234(0.008) | 0.03 | 0.42 | 7.85 |
| | 8 | SEAS-SIR | 0.222(0.007) | 0.00 | 7.58 | – | 0.228(0.008) | 0.00 | 6.82 | – |
| | | SCS-SAVE | 0.723(0.023) | 1.05 | 1.85 | 6.13 | 0.783(0.022) | 1.41 | 1.03 | 7.81 |
| | | SCS-ENS | 0.228(0.008) | 0.01 | 0.29 | 8.57 | 0.241(0.008) | 0.05 | 0.35 | 8.94 |
| | 10 | SEAS-SIR | 0.228(0.008) | 0.01 | 8.65 | – | 0.224(0.008) | 0.02 | 8.05 | – |
| | | SCS-SAVE | 0.820(0.023) | 0.93 | 1.24 | 7.43 | 0.923(0.018) | 2.28 | 1.62 | 8.05 |
| | | SCS-ENS | 0.231(0.008) | 0.04 | 0.24 | 9.63 | 0.253(0.013) | 0.06 | 0.24 | 9.78 |

Comparing the results in Table 5 with those in Table 2 of the main text, the sparse SIR-based method proposed in [47] is robust to the choice of $H$, and the same level of robustness is also enjoyed by the proposed sparsity-adapted column selection approach applied to $\mathrm{M} = (\mathrm{M}_{\mathrm{SIR}}, \mathrm{M}_{\mathrm{SAVE}})$. When applied to $\mathrm{M} = \mathrm{M}_{\mathrm{SAVE}}$, the robustness of the proposed column selection approach to $H$ depends on the model: it is equally robust as the sparse SIR-based method for Model VII but becomes inconsistent as $H$ grows for Model VIII. The latter is caused by the decay of the signal strength of $\mathrm{M}_{\mathrm{SAVE}}$ for large $H$, which practically violates Condition (S2) in the main text. To illustrate this, we evaluate the optimal submatrix of the non-sparse $\widehat{\mathrm{M}}_{\mathrm{SAVE}}$, i.e., $\widehat{\mathrm{M}}_{\mathrm{G}}$, for Model VIII under low-dimensional settings, where only the first five and the last five variables of $X$ are included in the data. The corresponding results in Table 6 shows its inconsistency for large $H$ even when the estimation error in $\widehat{\mathrm{M}}_{\mathrm{SAVE}}$ is practically negligible. These findings comply with the conclusion in [23], that is,

“The performance of SAVE (i.e., the sliced average variance estimation) is strongly affected by the choice of $c$ (i.e., $n/H$), but when $H$ is properly chosen, SAVE works very well. However, the range of $c$ that results in a good performance from SAVE is fairly narrow.“

To summarize, the robustness of the proposed sparsity-adapted column selection approach to $H$ synchronizes with the overall signal strength of the working M associated with $H$, which may vary with data. How to choose $H$ appropriately is an important issue in practice and deserves thorough studies in the future.

**Table 6.** Performance of the optimal submatrix $\widehat{\mathrm{M}}_{\mathrm{G}}$ of $\widehat{\mathrm{M}}_{\mathrm{SAVE}}$ for Model VIII, with $X$ replaced by its first five and last five variables. The numbers $a$ and $b$ in each cell $a(b)$ are respectively the sample mean and the sample standard deviation of the spectral loss defined in (10) of the main text, with F in (10) replaced by G and based on 200 runs.

| $H$ | $\Sigma_a$ | $\Sigma_b$ |
|---|---|---|
| 2 | 0.445(0.005) | 0.524(0.005) |
| 5 | 0.720(0.005) | 0.850(0.005) |
| 8 | 0.815(0.005) | 0.904(0.004) |
| 10 | 0.860(0.004) | 0.919(0.004) |